\documentclass[useAMS]{mn2e}
\usepackage{longtable}
\usepackage[tight]{subfigure}
\usepackage{graphicx}
\newcommand{\reduceme}{\mbox{R\raisebox{-0.35ex}{E}D\hspace{-0.05em}\raisebox{0.85ex}{uc}\hspace{-0.90em}\raisebox{-.35ex}{{m}}\hspace{0.05em}E}}

\title[Radial kinematics of BCGs]{Radial kinematics of brightest cluster galaxies}
\author[Loubser et al.]{S. I. Loubser$^{1}$\thanks{E-mail:
siloubser@uclan.ac.uk (SIL)}, A. E. Sansom$^{1}$, P. S\'{a}nchez-Bl\'{a}zquez$^{1}$, I. K.
Soechting$^{2}$, \newauthor{G. E. Bromage$^{1}$}\\
$^{1}$Centre for Astrophysics, University of Central Lancashire, Preston, PR1 2HE, UK\\
$^{2}$Oxford Astrophysics, Department of Physics, University of Oxford, Oxford, OX1 3RH, UK}
\begin{document}

\date{Accepted 2008 August 10. Received 2008 August 05; in original form 2008 February 01}

\pagerange{\pageref{firstpage}--\pageref{lastpage}} \pubyear{2008}

\maketitle

\label{firstpage}

\begin{abstract}
This is the first of a series of papers devoted to the investigation of a large sample of brightest cluster galaxies (BCGs), their kinematic and stellar population properties, and the relationships between those and the properties of the cluster. We have obtained high signal-to-noise ratio, long-slit spectra of these galaxies with Gemini and WHT with the primary purpose of investigating their stellar population properties. This paper describes the selection methods and criteria used to compile a new sample of galaxies, concentrating on BCGs previously classified as containing a halo (cD galaxies), together with the observations and data reduction. Here, we present the full sample of galaxies, and the measurement and interpretation of the radial velocity and velocity dispersion profiles of 41 BCGs. We find clear rotation curves for a number of these giant galaxies. In particular, we find rapid rotation ($>$ 100 km s$^{-1}$) for two BCGs, NGC6034 and NGC7768, indicating that it is unlikely that they formed through dissipationless mergers. Velocity substructure in the form of kinematically decoupled cores is detected in 12 galaxies, and we find five galaxies with velocity dispersion increasing with radius. The amount of rotation, the velocity substructure and the position of BCGs on the anisotropy-luminosity diagram are very similar to those of ``ordinary'' giant ellipticals in high density environments. 
\end{abstract}


\section{Introduction}
The galaxies in the centres of clusters are unique. They are usually the dominant, brightest and most massive, galaxies in their clusters. A wealth of imaging data has been accumulated for these intriguing objects (Malumuth $\&$ Kirshner 1985; Schombert 1986, 1987, 1988; Postman $\&$ Lauer 1995; Collins $\&$ Mann 1998; Brough et al.\ 2002; Laine et al.\ 2003), however the spectroscopic data is limited to very small samples or narrow wavelength coverage. 

Tonry (1984), Tonry (1985), Gorgas, Efstathiou $\&$ Aragon-Salamanca (1990), Fisher, Illingworth $\&$ Franx (1995a), Fisher, Franx $\&$ Illingworth (1995b), Cardiel, Gorgas $\&$ Aragon-Salamanca (1998) and Carter, Bridges $\&$ Hau (1999) each investigated 18 or less galaxies, and moreover measured only three or less indices. Brough et al.\ (2007) used a wide wavelength range but included only three brightest cluster galaxies (BCG). Von der Linden et al.\ (2007) studied BCGs from the Sloan Digital Sky Survey (SDSS), however the spatial information is lacking since fibers were used. 

Among BCGs there exists a special class of galaxies catalogued as cD galaxies, although there exists a great deal of confusion over the exact meaning of the classifications gE, D and cD. Matthews, Morgan $\&$ Schmidt (1964) outlined the following definition: ``D galaxies have an elliptical-like nucleus surrounded by an extensive envelope. The supergiant D galaxies observed near the centre of a number of Abell's rich clusters have diameters 3 -- 4 times as great as the ordinary lenticulars in the same clusters. These very large D galaxies observed in clusters are given the prefix ``c'', in a manner similar to the notation for supergiant stars in stellar spectroscopy''. Later, Schombert (1987) defined gE galaxies as distinct from other early-type galaxies by their large size; D galaxies as being gE galaxies with a shallow surface brightness profile slope; and cD galaxies as D galaxies with large extended stellar haloes. The last are also more diffuse than normal ellipticals (Schombert 1986). Because of the confusion surrounding the definition of D galaxies, and the fact that they are rarely regarded as a separate type of object in the modern literature, this study will only refer to cD and non-cD BCGs (i.e BCGs containing a halo or not). For BCGs, we adopt the definition to comply with recent literature (for example Von der Linden et al.\ 2007), where BCG refers to the central, dominant galaxy in a cluster. For a small fraction of clusters, the BCG might not strictly be the brightest galaxy in the cluster.

Approximately 20 per cent of rich clusters of galaxies contain a dominant central cD galaxy (Dressler 1984; Oegerle $\&$ Hill 2001), although they can be found in poor clusters as well (Giacintucci et al.\ 2007). Some clusters have more than one cD galaxy, but a cD galaxy is always the dominant member of a local subcluster. The surface-brightness profiles of cD galaxies are displaced above the de Vaucouleurs law (de Vaucouleurs 1948) at large radii. The break in the cD galaxy surface-brightness profile typically occurs between 24 and 26 mag/arcsec$^{2}$ in the V-band (Sarazin 1988). The interpretation of this deviation is that the galaxy is embedded in an extensive luminous stellar halo. Three main theories have been proposed over the last four decades to explain the properties of BCG, and in particular cD, galaxies. 

\paragraph*{Theory 1.}
In the first theory, BCG formation is caused by the presence of \textit{cooling flows} in clusters of galaxies (Cowie $\&$ Binney 1977). Cooling flow clusters are common in the local universe and BCGs are most often found at the centres of these systems (Edwards et al.\ 2007). If the central cluster density is high enough, intracluster gas can condense and form stars at the bottom of the potential well. Observations that support this idea are blue- and UV-colour excesses observed in the central galaxy of Abell 1795 (indicative of star formation) by McNamara et al.\ (1996) and molecular gas detected in ten out of 32 central cluster galaxies by Salom\'{e} $\&$ Combes (2003). Cardiel et al.\ (1998) obtained radial gradients for the D$_{4000}$ and Mg$_{2}$ spectral features in 11 central cluster galaxies. Their observations were consistent with an evolutionary sequence in which radio-triggered star formation bursts take place several times during the lifetime of the cooling flow in the centre of the cluster. However, McNamara $\&$ O'Connell (1992) find only small colour anomalies with small amplitudes, implying star formation rates that account for at most a few percent of the material that is cooling and accreting onto the central galaxy. 

More recently, \textit{XMM-Newton} observations showed that the X-ray gas in cluster centres does not cool significantly below a threshold temperature of $kT\sim1-2$ keV (Jord\'an et al.\ 2004, and references therein). The central cluster galaxies often host radio-loud AGN which may account for the necessary heating to counteract radiative cooling (von der Linden et al.\ 2007). Although BCGs are probably not completely formed in cooling flows, the flows play an important role in regulating the rate at which gas cools at the centres of groups and clusters. 

\paragraph*{Theory 2.}
The second theory was proposed by Merritt (1983) and suggests that the essential properties of BCGs, and in particular those with haloes, are determined when the clusters collapse (\textit{primordial origin}). Thereafter, frequent mergers of galaxies would be inhibited by the relatively high velocities between galaxies. Merritt (1983) argued that all galaxies had large haloes early in the life of the cluster. These haloes were then removed by the mean cluster tidal field during the initial collapse and returned to the cluster potential, except for the central member which remained unaffected because of its special position with respect to the cluster potential. 

\paragraph*{Theory 3.}
The third, and most widely accepted, theory is in the context of the $\Lambda$CDM cosmology and relates the formation of the central galaxy to mergings with or captures of less massive galaxies, and is known as \textit{``galactic cannibalism"}. It was first proposed by Ostriker $\&$ Tremaine (1975) and developed by Ostriker $\&$ Hausman (1977). 

The most complete quantitative prediction of the formation of BCGs in the now standard CDM model of structure formation is by De Lucia $\&$ Blaizot (2007). They used N-body and semi-analytic techniques to study the formation and evolution of BCGs and found that, in a model where cooling flows are suppressed at late times by AGN activity, the stars of BCGs are formed very early (50 per cent at $z \sim 5$ and 80 per cent at $z \sim 3$) and in many small galaxies. They also found that BCGs assemble late: half of their final mass is typically locked up in a single galaxy after $z \sim 0.5$ (illustrated in their figure 9). A very similar conclusion was reached by Romeo et al.\ (2008), who performed N-body and hydrodynamical simulations of the formation and evolution of galaxy groups and clusters in a $\Lambda$CDM cosmology to follow the build-up of two clusters and 12 groups. Observationally, Aragon-Salamanca et al. (1998) examined the K-band Hubble diagram for BCGs up to a redshift of $z$ = 1. They found that the BCGs had grown by a factor of two to four since $z$ = 1. Brough et al. (2002) found a similar result but discovered that the mass growth depended on the X-ray luminosity of the host cluster. They found that BCGs in high X-ray luminosity clusters showed no mass accretion since $z$ = 1 as opposed to BCGs in low X-ray luminosity clusters which grew by a factor of four. However, the recent near-infrared photometric study of 42 BCGs by Whiley et al.\ (2008) in the 0.2 $<$ $z$ $<$ 1 range contradicts this. They studied the colour and rest-frame $K$-band luminosity evolution of BCGs and found it to be in good agreement with population synthesis models of stellar populations which formed at $z \sim $ 2 and evolved passively thereafter.

Using the Millennium Simulation, De Lucia et al.\ (2006) studied how formation histories, ages and metallicities of elliptical galaxies depend on environment and on stellar mass. Their Figure 9 shows the effective number of progenitors of early-type galaxies as a function of galaxy stellar mass. The number of effective progenitors is less than two for galaxies up to stellar masses of $\simeq10^{11}$ M$_{\odot}$. This function suddenly increases up to the value of approximately five effective progenitors for the mass of a typical BCG.

A related theory, called \textit{tidal stripping}, was first proposed by Gallagher $\&$ Ostriker (1972). Cluster galaxies that pass near the gravitational centre of the cluster may be stripped of some of their material by the tidal forces from the cluster potential or the central galaxy potential. The stripped material falls to the centre of the potential well, and could contribute to the observed haloes of cD galaxies. The most massive galaxies surrounding the central galaxy would be preferentially depleted as they are most strongly affected by dynamical friction (Jord\'an et al.\ 2004). The difference between stripping and primordial origin is that stripping (and cD halo formation) begins after cluster collapse whereas primordial origin assumes that the tidal events occur before collapse, and that the cD halo is not a consequence of tidal stripping (Schombert 1988). 

The observation of multiple nuclei in central galaxies favours the cannibalism theory (Postman $\&$ Lauer 1995). Tonry (1984) observed the velocity and velocity dispersion profiles of NGC6166 and NGC7720 and their multiple nuclei. He found that the stellar velocity dispersion of the central galaxy demonstrated that the multiple nuclei are not following circular orbits. This was followed by a bigger sample of 14 multiple nuclei BCGs for which the redshifts and stellar velocity dispersions are presented in Tonry (1985). Yamada et al.\ (2002) showed that the BCG in a cluster at $z=1.26$ is composed of two distinct sub-units that are likely to fully merge on a time-scale of $10^{8}$ years. Jord\'an et al.\ (2004) studied the globular cluster systems in BCGs with confirmed haloes from Hubble Space Telescope observations. They concluded that the observed globular cluster metallicity distributions are consistent with those expected if BCGs galaxies form through cannibalism of numerous galaxies and protogalactic fragments that formed their stars and globular clusters before capture and disruption, although they state that the cannibalism scenario is not the only possible mechanism to explain these observations. The Jord\'an et al.\ (2004) globular cluster data also suggests that BCGs experienced their mergers prior to cluster virialisation, yet the presence of tidal streams suggest otherwise (Seigar, Graham $\&$ Jerjen 2007). 

Carter $\&$ Metcalfe (1980) and West (1989) showed that the major axis of BCGs tends to be aligned with the major axis of the cluster galaxy distribution. Recent studies of the Coma cluster (Torlina, De Propris $\&$ West 2007) show strong evidence that there are no other large-scale galaxy alignments other than for the BCGs. This also confirms that BCGs form via mechanism related to collimated infall of galaxies along the filaments and the growth of the cluster from the surrounding large scale structure (Boylan-Kolchin, Ma $\&$ Quataert 2006).  

In the CDM cosmology it is now understood that local massive cluster galaxies assemble late through the merging of smaller systems. In this picture, cooling flows are the main fuel for galaxy mass-growth at high redshift. This source is removed only at low redshifts in group or cluster environments, due to AGN feedback (De Lucia $\&$ Blaizot 2007).

Some of the outstanding issues are: whether BCGs have a different formation mechanism than elliptical galaxies; and if the formation of BCGs are controlled by environment. To address these issues the present project was initiated, and long-slit spectra of a large and statistically significant sample of BCGs were obtained. These data provide a set of spectral indices, covering a wide wavelength range. The luminous central galaxies will be contrasted with other early-type galaxies to look for relative differences in their evolution, using features sensitive to stellar population age and abundances of various elements. 

The first part of this project entails the determination of the stellar kinematics through derivation of velocity and velocity dispersion profiles. A forthcoming paper will be devoted to the measurement and analysis of line strengths for this sample of BCGs.

This paper is organised as follows. Section 2 details the sample and the selection criteria, followed by a description of the observations and data reductions in Section 3. The kinematical measurements are described in Section 4 and the individual galaxy kinematic profiles and notes are presented in Section 5. Section 6 discusses the kinematic properties of BCGs as a class, compared to other Hubble types. Conclusions and future work are given in Section 7. 

\section{Sample}
The initial intention of this project was to study cD galaxies in particular. However, the confusing regarding the classification of cD galaxies and the lack of deep photometry which would allow to conclusively distinguish between BCGs with and without a halo made the sample selection extremely challenging. Aiming at maximising the completeness of the cD sample and inclusion of any potential sub-populations we adopted rather broad selection criteria when restricting an intrinsic BCG sample. We expect to add deep photometric information in future publications which will allow to test the presence of a faint halo and discriminate between cD and non-cD BCGs. For the purpose of this publication we refer to our sample of 63 galaxies as a BCG sample and present kinematics results for 41 observed members. Two ordinary elliptical galaxies were also observed (one E and one E/SO) and their kinematics presented here. They will be used as a control sample in this study, where relevant.

The sample selection combines three methods making the best use of available information from literature and astronomical databases. The three search methods are described below.

\paragraph*{Method 1.}
Two well-known galaxy cluster classification systems that distinguish clusters containing a cD galaxy in the centre from other galaxy clusters, are those of Rood $\&$ Sastry (1971, hereafter RS) and Bautz $\&$ Morgan (1970, hereafter BM). The RS classification is based on the projected distribution of the brightest 10 members and the BM classification is based on brightness contrast between first- and second-ranked galaxies (i.e. the slope of the luminosity function at the bright end). 

Hoessel, Gunn $\&$ Thuan (1980) derived the BM types for all the nearby Abell clusters of galaxies, which made it possible to obtain a list of all nearby Abell clusters with BM types I and I-II (classified to contain a cD galaxy). Struble $\&$ Rood (1987) compiled a catalogue of the morphological properties of 2712 Abell clusters, derived from visual inspection of the KPNO photographic plates. This made it possible to obtain a list of Abell clusters with RS cluster classification type cD. The NASA/IPAC Extragalactic Database (NED)\footnote{http://nedwww.ipac.caltech.edu/.} was used to search for the brightest galaxies close to the centre of each cluster. If the brightest member was classified as either a cD or a D galaxy in NED (in the NED morphology or notes of previous observations), it was included in our list of BCGs. The RS and/or BM classifications are frequently used as a method to identify possible cD galaxies (for example Hill $\&$ Oegerle 1992; Baier $\&$ Wipper 1995; Giacintucci et al.\ 2007).

\paragraph*{Method 2.}
The second search was carried out by performing an all-sky search in the HyperLEDA\footnote{http://leda.univ-lyon1.fr/.} database for galaxies with the following properties: T-type between -3.7 and -4.3 (the de Vaucouleurs Third Reference Catalogue (RC3) classifies cD galaxies as T-type = --4); apparent B-magnitude brighter than 16; distance closer than 340 Mpc; and further than 15 degrees from the Galactic plane\footnote{An $H_{0}$ value of 75 km s$^{-1}$ Mpc$^{-1}$ is assumed throughout this work.}. It was found that T-type alone is not a reliable classification of galaxy types (in both the RC3 and HyperLeda catalogues), as various low luminosity galaxies are also categorised as galaxies with T-type = --4. This prompted an absolute magnitude cutoff of $M_{B}$ = --20 to be applied throughout the sample selection. The galaxies obtained with this search were added to our list of BCGs if they were classified as cD galaxies in NED. 

\paragraph*{Method 3.}
The third search was undertaken by choosing galaxies with published surface brightness profiles that had a slope consistent with cD galaxies and in the case of cD galaxies broke the de Vaucouleurs $r^{\frac{1}{4}}$ law at large radii. Galaxies were chosen from the series of papers by Schombert (1986; 1987; and 1988) and from Malumuth $\&$ Kirshner (1985). These are all optical photometric studies of the brightest cluster members of clusters. Galaxies with the surface brightness slopes and profiles conforming to the cD/D criteria set in Schombert (1987) were included in our list since that forms the basis of most definitions of cD galaxies.

\medskip The complete list of galaxies and their properties is given in Table \ref{table:cDs}. The following global criteria were applied to the sample: apparent B-magnitude brighter than 16; distance closer than 340 Mpc; and the absolute magnitude cutoff at $M_{B}$ = --20. Method 1 delivered 13 galaxies, 2 delivered 32 galaxies and method 3 contributed 29 galaxies (indicated in Table \ref{table:cDs}). With 11 of the merged list being duplicates present from more than one method, this provided an all-sky list of 63 galaxies. 

In summary: These 63 galaxies were classified as cD/D either in NED (in the morphological classification or in the notes of previous observations) and/or have profiles breaking the $r^{\frac{1}{4}}$ law. With the exception of NGC4946 and NGC6047, which are included in the sample as control galaxies but known to be an E and E/SO galaxy respectively, all galaxies have thus been previously classified in the literature to contain a halo, albeit very inhomogeneously. Despite the confusion regarding the halo classification, all these galaxies are the dominant central galaxies in clusters or groups. If selection method 1 and 3 above are assumed to be more robust methods of identifying a BCG with a halo, then 67 per cent of the BCG sample are confirmed cD galaxies. 

Ten galaxies in the BCG sample are in common with the Fisher et al.\ (1995a) sample of 13 BCGs. For six Abell clusters, the choice of BCG differs from that by Postman $\&$ Lauer (1995). These are: Abell 0189, 0194, 0262, 0548, 4038 and 2151. In each of the first five cases, the choice of BCG in this sample is brighter and more dominant than the BCG in the Postman $\&$ Lauer (1995) sample. In the case of Abell 2151, a comparison of the brightness of the BCG choices was complicated by the fact than the BCG chosen by Postman $\&$ Lauer (1995), NGC6041, is actually an interacting pair of galaxies. For this cluster, we followed previous authors (e.g. Aragon-Salamanca et al.\ 1998) who have chosen NGC6034 to be the BCG as it has a very prominent cD halo (Schombert 1987, 1988), and is close to an X-ray peak.

Some of our clusters are known to have substructure (Rines et al.\ 2002; Adami et al.\ 2005) and, in these cases, there may be more than one local X-ray maximum, which can host a dominant galaxy central to that substructure. Hence, according to the definition adopted here, there can be more than one BCG per cluster. In particular, three of the selected BCGs (NGC4889, NGC4874 and NGC4839) are hosted by the Coma cluster, all of them showing pronounced envelopes (Schombert 1987; Andreon et al.\ 1996). The first two are the dominant galaxies in the cluster (it is not clear which one is the original central galaxy of the main cluster, see e.g. Adami et al.\ 2005), while NGC4839 is the dominant galaxy of a group that appears to be infalling into the main cluster (Neumann et al.\ 2001). Two galaxies (NGC6173 and NGC6160) from the cluster Abell 2197 are also included in the sample. These two galaxies are believed to be the dominant galaxies of two main groups in the process of merging (Abell 2197E and Abell 2197W; Schombert 1987).

An internally consistent photometric study of this BCG sample will be done before investigating the properties of cD galaxies as a separate class. Liu et al.\ (2008) measured the extended envelopes of a sample of BCGs using Petrosian profiles (Patel et al.\ 2006). There is a distinct signature of a plateau in the Petrosian profiles of cD galaxies with an extended stellar halo which is not present for normal elliptical galaxies. However, they find that the deviation of the surface brightness profiles from a single S\'{e}rsic profile does not have any sharp transitions. Therefore, it is very difficult to unambiguously separate cD from non-cD BCGs based on surface brightness profiles alone (Liu et al.\ 2008), and confirming the presence or absence of cD haloes depends on choices of profiles fitted to the surface brightness as a function of radius.
 
In all the conclusions of this study, it will be clearly stated whether it applies to BCGs or includes the two normal elliptical galaxies.
The 43 galaxies discussed further in this first paper are indicated in the last column of Table \ref{table:cDs} (numbered 1 to 5 according to observing run), and range from $M_{B}=-23.54$ to $M_{B}=-20.71$ for the BCGs. 

\begin{table*}
\begin{scriptsize}
\begin{tabular}{l r r l l r@{$\pm$}l r c c c c}
\hline
Object & $\alpha$(J2000) & $\delta$(J2000) & Cluster & NED-type &\multicolumn{2}{c}{T-type} & \multicolumn{1}{c}{PA} & $m_{B}$ & $M_{B}$ & Ref & Run \\
Name &  &  &  & \multicolumn{2}{c}{} &  & \multicolumn{1}{c}{degrees} \\
\hline
 IC1565 &0	39	26.3& 06	44	03& A0076 & cD;E & -3.9 & 1.2 & -- & 14.41 &-21.96&2& \\
 UGC00579  & 0	56	16.2& -01	15	22& A0119 & cD;E & -4.8 & 0.4 & 41 &14.40  &-22.34 &3&6\\
 ESO541-013  &1	02	41.8& -21	52	55& A0133 & cD;E+3 pec & -4.0 & 0.7 & 16 &14.71 &-22.54 &2 &6\\
 PGC004072   & 1	08	50.8& -15	24	31& A0151 & D & -4.2 & 1.0 & 83 &14.75 &-22.37 &2&6\\
 IC1633     &1	09	55.6& -45	55	52& A2877 & cD;E+1 & -3.9 & 0.5 & 97 &12.57 & -22.60 &2, 3&3\\
 IC1634     & 1	11	02.9& 17	39	46& A0154 & MLT SYS & -4.1 & 1.0 & 153 &15.27 & -22.64 &1& \\
 IC1695     &1	25	07.6& 08	41	58& A0193 & cD;S? &\multicolumn{2}{c}{--} & 103 &14.87 & -21.98 &1& \\
 NGC0533    & 1	25	31.5& 01	45	33& A0189 & cD;E3: &-4.8 & 0.6 & 50 &12.45 & -22.25 &3&6\\
 NGC0541   & 1	25	44.3&-01	22	46 & A0194 & cD;SO-: & -3.9 & 0.9 & 69 &13.09 & -21.62 &2&6\\
 IC1733     & 1	50	42.9& 33	04	55& A0260 & cD;E: & -4.9 & 0.8 & -- &14.14 & -22.16  &1&\\
 NGC0708   &1	52	46.3& 36	09	12& A0262 & cD;E &-4.9 & 0.4 & 39 &13.63 &  -20.98 &3&\\
 UGC02232  & 2	46	03.9& 36	54	19& A0376 & cD;E/D &-3.9 & 1.3 & -- &15.50 & -21.64 &2& 2\\
 NGC1129   & 2	54	27.4& 41	34	46& AWM7 & E;BrClG &-4.7 & 1.2 & 73 &13.34 &  -21.67 &3&\\
 UGC02450  & 2	58	57.8& 13	34	59& A0401 & cD & 2.8 & 5.0 &  29 &15.24 & -23.50 &1& \\
 NGC1275   & 3	19	48.2& 41	30	42& A0426 & cD;pec;NLRG & -2.2 & 1.7 & 110 &12.55 & -22.69&3&\\
 NGC1399   & 3	38	29.0&-35	26	58 & RBS454 & cD;E1 pec & -4.5 & 0.5 & -- &10.42 & -20.81&3& 3\\
 ESO303-005& 4	13	58.8& -38	05	50& RBS512 & cD? &-4.0 & 0.5 & -- &15.38 & -21.51&2&3 \\
 MCG-02-12-039  & 4	33	37.8&-13	15	40 & A0496 & cD;E+? & -3.9 & 0.9 & 180 &13.95 & -22.50 &2, 3&3\\
 ESO202-043 & 4	37	47.6& -51	25	23& A S0479 & E+ &-3.8 & 0.8 & 133 &14.45 & -21.72 &2&6\\
 ESO552-020   & 4	54	52.3& -18	06	53& CID 28 & cD;E+ &-3.9 & 0.7 & 148 &13.54 & -22.51&2&3\\
 NGC1713    & 4	58	54.6& 00	29	20& CID 27 & cD;E+ & -4.3 & 0.6 & 39 &13.88 & -20.71&2&2\\
 UGC03197   & 4	59	55.8& 80	10	43& A0505 & cD &-4.9 & 0.9 & 85 &15.25 & -22.26 &3&\\
 ESO488-027 & 5	48	38.5& -25	28	44& A0548 & cD;E+1 &-3.8 & 0.6  & 68 &14.22 & -22.22 &2&3\\
 PGC025714  & 9	08	32.4& -09	37	47& A0754 & D &-4.2 & 1.4 & 122 &14.32 & -23.04&2&6\\
 PGC026269  &9	18	05.7& -12	05	44& A0780 & (R')SAO-:;BrClGSy3 & -2.6 & 1.0 & 133 &14.38 & -22.84&1&3\\
 NGC2832    & 9	19	46.9& 33	44	59& A0779 & E+2:;cD & -4.3 & 0.6 & 172 &12.79 & -22.38&2, 3&2\\
 UGC05515   & 10	13	38.3& 00	55	32& A0957 & E+pec: & -4.0 & 0.7 & 83 &14.49 & -22.28&2, 3&5\\
 PGC030223  & 10	20	26.6& -06	31	35& A0978 & D & -3.9 & 1.2 & 1 &15.40 & -21.85&2&6\\
 NGC3311    & 10	36	42.9& -27	31	37& A1060 & cD;E+2 & -3.3 & 1.3 &  -- &12.69 & -21.33&1&5\\
 NGC3842    & 11	44	02.2& 19	56	59& A1367 & E;BrClG &-4.8 & 0.7 &  177 &12.80 & -22.18&3&1\\
 PGC044257  & 12	57	11.4& -17	24	36& A1644 & cD;E+4 &-4.1 & 1.0 & 44 & 13.50 & -23.54&1, 2&5\\
 NGC4839    & 12	57	24.2& 27	29	54& A1656 & cD;SAO &-4.0 & 0.8 & 64 & 13.05 & -22.26&2, 3&4\\
 NGC4874    &12	59	35.5& 27	57	36& A1656 & cD;Di &-3.7 & 0.9 & 45 &12.75 & -22.50&2, 3&1\\
 NGC4884(9)    & 13	00	07.9& 27	58	41& A1656 & cD;E4;Db &-4.3 & 0.5 & 82 &12.48 & -22.56 &2&1\\
 NGC4946$^{\star}$    & 13	05	29.4& -43	35	28& A3526 & E+? &-3.8 & 0.8 & 135 &13.41 & -20.27& &5\\
 ESO444-046 &13	27	56.9& -31	29	44& A3558 & cD;E+4 &-3.7 & 1.0 &  161 &14.07 & -22.86 &2, 3&5\\
 LEDA094683 & 13 53 06.4 & 05 08 59 & A1809 & cD &-2.3 & 5.0 &46&15.30& -22.35& 1 & 5\\
 GSC555700266& 14 01 36.4 & -11 07 43& A1837 & cD &\multicolumn{2}{c}{--} &--&14.59&\multicolumn{1}{c}{--}&1&5\\
 NGC5539    &14	17	37.8 & 08	10	46& A1890 & cD & 0.7 & 5.0 & 39 & 14.92 & -22.62&1 & \\
 IC1101     & 15	10	56.1& 05	44	41& A2029 & cD;SO-: &-2.9 & 1.1 & 25 &15.10 & -23.05&3 &5\\
 UGC09799   & 15	16	44.6& 07	01	16& A2052 & cD;E & -4.7 & 1.3 & 30 & 14.34 & -21.84&3 &\\
 PGC054913  &15	23	05.3 & 08	36	33& A2063 & cD;S &-3.8 & 1.6 & -- & 14.71 & -21.46&2&\\
 UGC09958   & 15	39	39.1&21	46	58& A2107 & cD;SO-: &-2.9 & 0.7 & 98 & 14.76 & -22.00&3 &\\
 UGC10012   & 15	44	59.0& 36	06	36& A2124 & cD;E &-4.9 & 0.6 & 143 &14.54 & -23.15&3 &\\
 UGC10143   &16	02	17.3&15	58	28& A2147 & cD;E+ & -4.0 & 0.8 & 12 & 14.27 & -21.97&2, 3&1\\
 NGC6034    &16	03	32.1& 17	11	55& A2151 & E+ & -4.0 & 0.4 & 59 & 14.58 & -21.60&2, 3&1\\
 NGC6047$^{\star}$    & 16	05	09.1& 17	43	48& A2151 & E+ &-3.5 & 0.8 & 117 & 14.58 & -21.45& &1\\
 NGC6086    & 16	12	35.6&29	29	06& A2162 & cD;E &-4.8 & 0.6 & 3 & 13.81 & -22.19&3&4\\
 NGC6160    &16	27	41.1& 40	55	37& A2197 & cD;E & -4.8 & 0.6 & 72 & 14.17 & -21.70&3&4\\
 NGC6166    & 16	28	38.5& 39	33	04& A2199 & cD;E & -4.2 & 1.4 & 38 & 12.88 & -22.92&2, 3&1\\
 NGC6173    & 16	29	44.9&40	48	42& A2197 & cD;E &-4.8 & 0.6 & 138 &13.13  & -22.58&3&4\\
 NGC6269    &16	57	58.1& 27	51	16& AWM5 & cD;E &-4.7 & 0.7 & 80 & 13.38 &-22.98&3&4\\
 IC4765     &18	47	18.5 & -63	19	50& A S0805 & cD;E+4 &-3.9 & 0.5 & 123 &12.46 &-21.97&2&5\\
 NGC7012    & 21	06	45.8& -44	48	49& A S0921 & E+4 pec & -3.9 & 0.5 & 100 &13.93 & -21.79&2&3 \\
 ESO146-028 & 22	28	51.1& -60	52	55& RXCJ2228.8-6053 & E+3 & -3.8 & 0.8 & 154 &13.94 & -22.54&2&5\\
 ESO346-003 & 22	49	22.0& -37	28	20& A S1065 & E+2 pec: &-3.8 & 0.8 & 118 &14.00 &-21.59&2&5\\
 NGC7597    & 23	18	30.3& 18	41	20& A2572 & cD;S? &-2.0 & 0.8 & 133 &15.00 &  -21.52&1&  2 \\
 NGC7647    &23	23	57.4& 16	46	38& A2589 & cD;E &-4.7 & 0.8 & 174 &14.61 &-21.93&3&1\\
 NGC7649    &23	24	20.1&14	38	49& A2593 & cD;E &-4.6 & 1.1 & 78 &15.65 & -20.99&3&2\\
 PGC071807  & 23	35	01.5&27	22	20& A2622 & cD & \multicolumn{2}{c}{--} & 138 &15.28 & -22.05&1&4\\
 NGC7720    & 23	38	29.5& 27	01	51& A2634 & cD;E+pec: &-4.3 & 0.6 & -- &13.43 & -22.63&2&1 \\
 IC5358     & 23	47	45.0& -28	08	26& A4038 & cD;E+4 pec &-3.8 & 1.1 & 114 &13.64 & -22.00&2&3\\
 NGC7768    &23	50	58.3&27	08	51& A2666 & cD;E &-4.9 & 0.5 & 55 &13.28 & -22.39&3&1\\
 PGC072804  & 23	54	13.7& -10	25	09& A2670 & cD;XBONG &-4.4 & 2.4 & 76 &15.32 & -22.82 &1&5\\
 ESO349-010 &23	57	00.7& -34	45	33& A4059 & cD;E+4 &-3.7 & 1.2 & 155 &14.18 & -22.69&2&3\\
\hline
\end{tabular}
\end{scriptsize}
\caption[BCG Sample.]{The new sample of BCGs, including the two ordinary ellipticals (marked $\star$). Columns 2, 3, 6 -- 9 are from the HyperLeda catalogue, and 4 and 5 from NED. PA is the position angle of the galaxy major axis. Ref 1,2 and 3 correspond to the three search methods for objects (column 10). Run (column 11) 1=WHT, 2=Gemini N 2006B, 3=Gemini S 2006B, 4=Gemini N 2007A, 5=Gemini S 2007A, 6=Gemini S 2007B (The data from Run 6 are not presented in this paper). Besides the galaxies classified as cD/D in the NED morphology (column 5); NGC6034, UGC05515 and NGC1129 were classified as cD/D in Schombert (1987); NGC7012, ESO146-028, ESO346-003, ESO202-043, NGC3842, PGC026269 and IC1634 were classified as cD/D in the NED notes by previous observations.}
\label{table:cDs}
\end{table*}

\section{Observations and data reduction}
\subsection{WHT observations}

Spectroscopic observations of 10 of the 43 galaxies presented here (listed in Table \ref{table:WHTobjects}) were carried out during the period 23 to 26 June 2006, using the 4.2m William Herschel telescope (WHT) equipped with the ISIS double spectrograph mounted at the \textit{f}/11 Cassegrain focus. The Marconi2 CCD was employed on the red arm and the EEV12 on the blue. The spatial scale of both CCDs was 0.4 arcsec/pixel (with 2 $\times$ 2 binning) and the slit width was 1 arcsec. Comparison spectra were provided by Cu/Ar and Cu/Ne arcs, and flat-field illumination by a tungsten lamp. 

Two different dichroics were used. The 5300 \AA{} dichroic was used for three nights and delivered an unvignetted spectral range of 3900 -- 5460 \AA{} in the blue arm and 5730 -- 6960 \AA{} in the red arm. The 6100 \AA{} dichroic was used for one night, delivering an unvignetted spectral range of 4500 -- 6120 \AA{} in the blue arm and 7950 -- 9600 \AA{} in the red arm. The exposure times are listed in Table \ref{table:WHTobjects}. This combination made it possible to observe 29 absorption indices consisting of the original 21 Lick indices from CN to TiO$_{2}$ (Burstein et al.\ 1984; Faber et al.\ 1985; Worthey et al.\ 1994), the four H$\delta$ and H$\gamma$ indices contributed by  Worthey $\&$ Ottaviani (1997), the near-IR CaII triplet at $\sim$ 8600 \AA{} and the MgI index. This collection of 29 indices will be referred to as the Lick indices in this study. 

Dispersion in the blue arm was 0.90 \AA{}/pixel (grating R600B) and 0.88 \AA{}/pixel (grating R600R) in the red. The spectral resolution ranged between 3.3 and 4.0 \AA{}, depending on the dichroic and arm used. Seeing was typically better than 1 arcsecond, and all necessary calibration frames were obtained. In most cases, the slit was placed on the major axis of the galaxy. Exceptions occurred if there was another early-type galaxy (unrelated to the sample) in close proximity, in which case the slit was positioned to go through the centres of both galaxies, to allow a control sample of ellipticals with the same observational setup (two control sample ellipticals were observed but are not used in this present paper). Five spectrophotometric standard stars were also observed for flux calibration. 

A total of 22 Lick calibration stars (G and K type stars which are considered the main contributors to the visible light in early-type galaxies) were observed with the 5300 \AA{} dichroic and 10 with the 6100 \AA{} dichroic. A neutral density filter was used for the Lick stars with both dichroics. These stars are used as templates for velocity dispersion measurements as well as to transform the line-strength indices to the Lick system. Before and/or after every standard star or object spectrum, an arc spectrum and flat-field frame were obtained in the red and blue arm at each telescope pointing, to allow accurate wavelength calibration as well as removal of the response function of the dichroics. 

\begin{table*}
\centering
\begin{tabular}{l c c r r r c r c}
\hline Object & Exposure  & Exposure  & \multicolumn{1}{c}{Slit} & \multicolumn{1}{c}{Major Axis} & \multicolumn{1}{c}{$r_{\rm e}$} &  $\epsilon$ & \multicolumn{1}{c}{$a_{\rm e}$} & Fraction \\
 & (5300 \AA{}) & (6100 \AA{}) & \multicolumn{1}{c}{PA} & \multicolumn{1}{c}{MA} & & & & times \\
 & seconds & seconds & \multicolumn{1}{c}{degrees} & \multicolumn{1}{c}{degrees} & \multicolumn{1}{c}{arcsec} & & \multicolumn{1}{c}{arcsec} & $a_{\rm e}$ \\
\hline NGC3842 & 6$\times$1200 & 2$\times$900 & 5 & 177 & 16.4 & 0.29 & 16.3 &1.22\\
 NGC4874 & 6$\times$1200 & 3$\times$900 & 82 & 45 & 20.0 & 0.00 & 20.0 &0.60\\
 NGC4889 & 6$\times$1200 & 2$\times$900 & 80 & 82 & 25.7 & 0.34 & 25.7 &0.89\\
 NGC6034 & 6$\times$1200 & 3$\times$900 & 50 & 59 & 8.2 & 0.27 & 8.1 &1.23\\
 NGC6047 & 3$\times$1200 & -- & 88 & 117 & 7.4 & 0.27 & 7.0 &0.61\\
 NGC6166 & 6$\times$1200 & 3$\times$900 & 35 & 38 & 22.6 & 0.26 & 22.6 &0.22\\
 NGC7647 & 6$\times$1200 & 3$\times$900 & 174 & 174 & 11.2 & 0.49 & 11.2 &1.21\\
 NGC7720 & 6$\times$1200 & 3$\times$900 & 10 & -- & 9.3 & 0.19 & 9.3 &0.65\\
 NGC7768 & 3$\times$1200 & 3$\times$900 & 60 & 55 & 14.7 & 0.19 & 14.7 &0.82\\
 UGC10143 & 6$\times$1200 & 3$\times$900 & 12 & 12 & 10.5 & 0.41 & 10.5 &0.65\\
\hline
\end{tabular}
\caption[Exposure Times for WHT BCGs.]{Exposure times in each of the two dichroics and position angles for the WHT BCGs. The PA is given as deg E of N. The half-light radii ($r_{\rm e}$) were calculated from the 2MASS catalogue, except NGC6166 and NGC7720 for which they were calculated with the \textsc{vaucoul} task in the \textsc{reduceme} package. The last column lists the fraction of the effective half-light radii spanned by the radial profiles measured in this work. Table \ref{table:cDs} contains more detailed properties for the galaxies.}
\label{table:WHTobjects}
\end{table*}

\subsection{Gemini observations}

The Gemini Multi-object Spectrograph (GMOS) observations were performed in queue scheduling mode. The B600 grating was used (2.7 \AA{} resolution) with a 0.5 arcsec slit width to match the observations of the Lick stars described below. Dispersion was 0.914 \AA{}/pixel, and the spatial scale was 0.146 arcsec/pixel (with 2 $\times$ 2 binning). Ten galaxies were observed at Gemini South in semester 2006B (August 2006 to January 2007) and five galaxies at Gemini North in the same observing semester. A further 12 galaxies were observed at Gemini South and six at Gemini North in 2007A (February 2007 to August 2007).

Bryan Miller et al.\ (private communication) performed a series of observing programmes from 2002 to 2004 in which they acquired long-slit spectroscopic data of ``Lick Stars''. These are well suited as calibration data for our observing programmes, and our instrumental set-up corresponded to the set-up used for observations of the Lick stars. Spectral dithering was carried out with two central wavelengths at 5080 \AA{} and at 5120 \AA{} for the Lick stars as well as the galaxy data described here. This is done in order to obtain the full spectrum uninterrupted by the gaps between the three CCDs. The three CCDs mosaiced together delivered a spectral range from 3700 to 6500 \AA{}. Exposure times were 2 $\times$ 900 seconds at each position, which amounts to an hour (mostly along the major axis) for each galaxy, except in the case of ESO444-046 where only one 900 second exposure at each position was taken. Imaging sequences (3 $\times$ 20 seconds with the $g$ and $r$ filter) were performed to allow insight into the possible presence and location of dust lanes or gas relative to the slit position. Standard stars were observed for relative flux calibration. The observations were mostly executed in dark time. The seeing typically ranged from 0.6 to 1.2 arcsec, and very rarely went up to 2.2 arcsec. Calibration arc (CuAr) and flat field spectra were also frequently observed at the two different central wavelengths. General bias frames and imaging flat field frames were also obtained.

For the Gemini data, information on the slit placements during observations was used to reconstruct the slit position on images of the galaxies. The slit location on the targets was determined using the slit acquisition images and the target positions from the imaging sequences. The images were taken with the same telescope pointing, and used to measure the offsets of the slit position from the galaxy centre (Table \ref{table:2007Aobjects}). In the few cases where the same galaxy was observed on two different nights, this was done for both nights. Offsets from the galaxy centres are shown in Table \ref{table:2007Aobjects}. The majority of the galaxies were centred correctly or with very small offsets. There are no offsets indicated for the WHT observations, because these were carried out in visiting observer mode, and all offsets were zero.

\begin{table*}
\begin{tabular}{l r r r c r c c} 
\hline Object & \multicolumn{1}{c}{Slit} & \multicolumn{1}{c}{Major Axis} & \multicolumn{1}{c}{$r_{\rm e}$} & $\epsilon$ & \multicolumn{1}{c}{$a_{\rm e}$} & Fraction & Offset\\
       &  \multicolumn{1}{c}{PA}  & \multicolumn{1}{c}{MA} & arcsec & & arcsec & $a_{\rm e}$ & arcsec\\
\hline\multicolumn{8}{c}{2006B South} \\ 
 ESO303-005 & 55 & -- & 9.1 & 0.25 & 9.1 & 0.50 &0.70\\							
 ESO349-010 & 14 & 155 & 15.3 & 0.52 & 12.3 & 1.18 & 0.30\\								
 ESO488-027 & 88 & 68 & 10.3 & 0.15 & 10.1 & 0.94 & 2.20\\								
 ESO552-020 & 148 & 148 & 18.3 & 0.43 & 18.3 & 0.76 & 0.64\\			
 IC1633 & 97 & 97 & 23.9 & 0.17 & 23.9 & 0.75 & 0.00\\									 
 IC5358 & 40 & 114 & 17.4 & 0.60 & 8.3 & 1.44 & 0.26\\
 MCG-02-12-039 & 166 & 180 & 15.2 & 0.19 & 15.1 & 1.13 & 0.64\\
 NGC1399 & 222 & -- & 42.2 & 0.06 & 42.2 & 0.44 & 0.45\\
 NGC7012 & 289 & 100 & 15.8 & 0.44 & 15.6 & 0.96 & 0.00\\
 PGC026269 & 313 & 133 & 7.3 & 0.00 & 7.3 & 0.89 & 0.00\\
\multicolumn{8}{c}{2006B North} \\ 
 NGC1713 & 330 & 39 & 15.5 & 0.14 & 14.0 & 1.07 & 0.00\\
 NGC2832 & 226 & 172 & 21.2 & 0.17 & 19.6 & 1.12 & 0.50\\
 NGC7597 & 46 & 133 & 9.4 & 0.00 & 9.4 & 0.64 & 0.00\\
 NGC7649 & 78 & 78 & 12.8 & 0.31 & 12.8 & 0.82 & 0.20\\
 UGC02232 & 60 &  -- & 9.7 & 0.00 & 9.7 & 0.82 & 0.00\\
\multicolumn{8}{c}{2007A South} \\ 
ESO146-028 & 154 & 154 & 12.4 & 0.35 & 12.4 &0.48 &0.15\\					
ESO346-003 & 297 & 118 & 11.9 & 0.18 & 11.9 &0.29 &0.15\\				 
ESO444-046 & 196 & 161 & 16.0 & 0.29 & 14.9 &0.24 &0.45\\									
GSC555700266 & 204 & -- & 10.6 & 0.30 & 10.6 &0.47 &0.00\\
IC1101 & 204 & 25 & 12.2 & 0.50 & 12.2 &0.39 &0.20\\								
IC4765 & 287 & 123 & 28.0 & 0.46 & 27.1 & 0.13 &0.75\\									
LEDA094683 & 226 & 46 & 7.3 & 0.33 & 7.3 &0.54 &2.00\\
NGC3311 & 63 & -- & 26.6 & 0.17 & 26.6 &0.53 &0.40\\
NGC4946 & 134 & 135 & 17.0 & 0.20 & 17.0 &0.74 &0.00\\
PGC044257 & 267 & 44 & 15.3 & 0.63 & 10.5 &0.38 &0.00\\
PGC072804 & 76 & 76 & 7.6 & 0.16 & 7.6 &0.52 &0.00\\
UGC05515 & 293 & 83 & 12.0 & 0.13 & 11.8 &0.38 &0.00\\
\multicolumn{8}{c}{2007A North} \\
NGC4839 & 63 & 64 & 17.2 & 0.52 & 17.2 &0.61 &0.00\\
NGC6086  & 270 & 3 & 11.2 & 0.29 & 8.1 &0.51 &0.10\\
NGC6160 & 140 & 72 & 13.9 & 0.47 & 9.0 &0.41 &0.00\\
NGC6173 & 139 & 138 & 15.0 & 0.26 & 15.0 &0.47 &0.10\\
NGC6269 & 306 & 80 & 14.1 & 0.20 & 13.1 &0.50 &0.35\\
PGC071807 & 138 & 138 & 6.7 & 0.70 & 6.7 &0.71 &0.10\\
\hline
\end{tabular}
\caption{Galaxies observed with the Gemini North and South telescopes. Table \ref{table:cDs} contains more detailed properties for the galaxies. The PA is given as deg E of N. The half-light radii ($r_{\rm e}$) were calculated from the 2MASS catalogue, except ESO303-005 for which it was calculated with the \textsc{vaucoul} task in the \textsc{reduceme} package. The second last column lists the fraction of the effective half-light radii spanned by the radial profiles measured in this work, and the last column lists the offsets as described in the text.}
\label{table:2007Aobjects}
\end{table*}

\subsection{Data reduction}

All the WHT data reductions were performed with the \textsc{iraf}\footnote{\textsc{iraf} is distributed by the National Optical Astronomy Observatories, which are operated by the Association of Universities for Research in Astronomy Inc., under cooperative agreement with the National Science Foundation.}and \textsc{starlink}\footnote{\textsc{starlink} is an interactive reduction and analysis facility that was developed in the UK, funded by the PPARC/STFC Research Council.} reduction packages. The blue and red spectra and spectra taken using the 5300 and 6100 \AA{} dichroics were treated in similar parallel streams of data reduction. Overscan correction, ADU conversion, bias subtraction and trimming of the image edges were done first. Flat field spectra were normalised by correcting for the lamp response function and all star and galaxy spectra were appropriately flat-fielded. The dichroics introduced an intermediate frequency pattern which varied with the position of the telescope. This was successfully removed by flat-fielding, using the flat-field images taken directly before and/or after the science frames (i.e. at the same telescope pointing). It was necessary to make an illumination correction using normalised twilight flat field spectra to correct for spatial gradients in the star and galaxy spectra. Full 2D wavelength calibration was performed and star and galaxy spectra were transformed to equal linear spectral bin sizes.

All the flux calibration and Lick stars were sky-subtracted and S-distortion-corrected before 1D spectra were extracted. All 1D Lick star spectra were flux calibrated and corrected for atmospheric extinction.  

A cosmic ray cleaning algorithm was applied to all the 2D galaxy spectra before they were background subtracted and S-distortion corrected. For each individual galaxy frame, a sky region consisting of 40 rows was chosen on each side of the galaxy and averaged together to obtain a 1D sky spectrum. The sky spectrum was grown into a 2D spectrum and subtracted from the data frame. Flux calibration and extinction corrections were also applied to the 2D spectra. The galaxy frames were then co-added (also removing any remaining cosmic ray effects in the process) to form a single frame for each galaxy in each of the blue and red arms and for each of the two dichroics.

The basic Gemini data reductions were performed with the Gemini-specific GMOS data reduction package that can be added onto the standard \textsc{iraf} reduction package. Spectra with different central wavelengths (5080 and 5120 \AA{}) and spectra taken with GMOS-North and GMOS-South were treated in similar parallel streams of data reduction. The raw Lick star data were obtained from the Gemini Science Archive\footnote{http://www4.cadc-ccda.hia-iha.nrc-cnrc.gc.ca/gsa/}. The basic data reductions were conducted in a similar manner as for the WHT data. For each of the two central wavelength settings, a bad pixel spectrum was made in order to avoid interpolation between the gaps. When the spectra of the two different wavelength settings were combined, these bad pixel masks were used to ensure that only true data were averaged. 

Proctor et al.\ (2005) and Brough et al.\ (2007) also used GMOS data to measure Lick indices, and found negligible scattered light that influenced their results. Nevertheless, as a test of the effect of scattered light in the GMOS spectrograph on our data, bright skylines were used to extract flat field profiles that could be compared to the shape of the dome flats that were used. It was found that scattered light effects did not influence the regions where galaxy data were extracted or the regions used for background subtraction. 

An error analysis was completed to account for both Poisson (photonic) noise and systematic errors in the data. Variance frames were created from the data frames after bias subtraction and ADU conversion, and reduced in a way similar to the data frames, but the arithmetic manipulations were performed by incorporating the usual error propagation laws. Thus, the variance frames represent the square of the Poisson noise at the start of this reduction process, and all other systematic noise effects are added onto the variance frames in quadrature. 

Flat-fielding produced negligible errors with the exception of the WHT flat-field spectra taken in the upper wavelength region (red spectra of dichroic 6100 \AA{} on night four) which contained fringe patterns. This contributed about four per cent to the errors and these were appropriately propagated in the variance frame. Errors contributed by the illumination correction proved negligible. Wavelength calibration was accurate to within 0.1 \AA{} in all cases. After that, the variance frames were also S-distortion corrected, following the procedure used for the corresponding data frames. The sky subtraction process added to the uncertainty in the data and was properly taken into account by calculating the following for each variance frame:
\begin{equation}
[\Delta Sky (\lambda)]^{2} = [\frac{1}{\sqrt{N}}\ \Delta A (\lambda)]^{2}, 
\end{equation} 
where $\Delta Sky$ is the error contribution of the sky subtraction being taken into account, $N$ is the number of spatial cross sections averaged for the sky determination and $A$ is the Poisson error on the average one line sky spectrum. This 1D spectrum was then grown into a 2D spectrum with the proper dimensions and added onto the variance frames in all cases. 

Relative flux calibration was accurate to within five to 10 per cent, and the variance frames were also co-added so that they could be treated alongside the galaxy data in further analysis. For the Gemini data, variance frames were co-added for each galaxy by dealing with the gaps in the same manner as the data (described above) to avoid any interpolation. The fact that only $N$/2 frames were added in the regions where the gaps occur, instead of $N$ individual frames per galaxy in all the other regions, was taken into account in the variance frame combination procedure. At the end of this reduction process, each 2D galaxy data frame had an associated 2D variance array. 

As mentioned, some of the galaxies were observed at a position angle (PA) other than the major axis (MA), due to the two control elliptical galaxies in the WHT observations, and due to the availability of suitably bright guide stars in the Gemini observations. This had an effect on the relative half-light radii (r$_{\rm e}$) which had to be taken into account. Tables \ref{table:WHTobjects} and \ref{table:2007Aobjects} list the r$_{\rm e}$ along the semimajor axis of the galaxies, computed from the 2MASS $K$-band 20th magnitude arcsec$^{-2}$ isophotal radius using the formula by Jarrett et al.\ (2003):
$\log r_{\rm e} \sim \log r_{\rm K20} - 0.4$.
Using the ellipticities ($\epsilon$) of the galaxies (data from NED), the effective half-light radii at PA were computed according to
\begin{equation}
 a_{\rm e} = \frac{r_{\rm e}(1-\epsilon)}{1-\epsilon \ \mid cos(\mid PA - MA \mid)\mid}.
\end{equation} 
The fraction of $a_{\rm e}$ which the radial profiles measured in this work spans is also listed in Tables \ref{table:WHTobjects} and \ref{table:2007Aobjects}. For those galaxies for which a MA angle is not available in HyperLeda, the PA is assumed to be the same as the MA (as intended during the observation preparations). 

\section{Central kinematical measurements}

The central radial velocity and velocity dispersion values of the galaxies were derived for regions with the size of $a_{\rm e}/8$, and with the galaxy centres defined as the luminosity peaks. Thus, the central values are for an aperture of $1 \times a_{\rm e}/8$ arcsec$^{2}$ for the WHT data and $0.5 \times a_{\rm e}/8 $ arcsec$^{2}$ for the Gemini data. 

\begin{figure}
   \centering
   \includegraphics[scale=0.27]{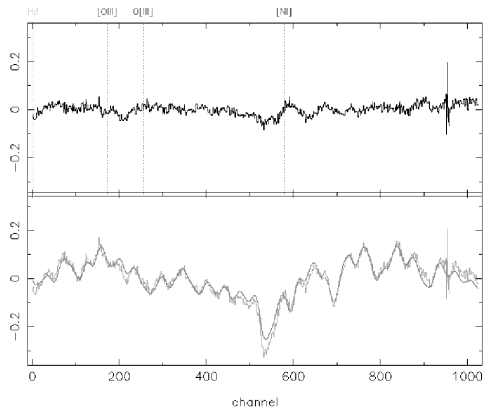}
   \caption[Example of the Calculation of the Derived Kinematics for NGC7768 with the \textsc{movel} algorithm.]{A typical example of the calculation of the derived kinematics for NGC7768 with the \textsc{movel} algorithm. In the lower panel, the smooth line corresponds to the optimal template spectrum, plotted together with the galaxy spectrum which has been shifted and broadened to the derived parameters. The upper panel shows the residuals of fit (in relative flux units), with the dotted lines indicating the positions of typical emission lines. This illustration spans 4900 to 5400 \AA{}.}
   \label{fig:WHTreduceme}
\end{figure}

\begin{figure}
   \centering
   \includegraphics[scale=0.30]{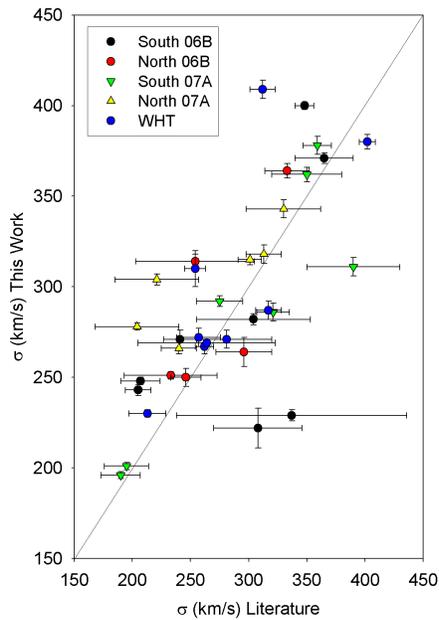}
\caption[Velocity Dispersion Measurements]{Velocity dispersion results compared to literature values. The latter values are the most recent ones from HyperLeda as described in the text.}
\label{fig:VelDisp}
\end{figure}

The Lick stars were shifted to zero radial velocity before using them as templates. Radial velocities and velocity dispersions of the galaxies were measured using the \textsc{movel} and \textsc{optema} algorithms described by Gonz\'{a}lez (1993) and integrated into the \reduceme\ data reduction package\footnote{\reduceme\ is an astronomical data reduction package, specially devoted to the analysis of long-slit spectroscopy data. It was developed by N. Cardiel and J. Gorgas (Cardiel 1999). \newline http://www.ucm.es/info/Astrof/software/reduceme/reduceme.html}. This routine is an improvement of the classic Fourier Quotient Technique described by Sargent et al.\ (1977). It is an iterative procedure in which a galaxy model is processed in parallel with the galaxy spectrum. In this way, a comparison between the input and recovered broadening functions for the model allows one to correct the galaxy power spectrum for any imperfections of the data handling procedures in Fourier Space. The procedure uses an algorithm that obtains the linear combination of individual templates that best matches the observed galaxy spectrum. The best template is scaled, shifted and broadened to match the galaxy kinematic parameters and is feeded to the \textsc{movel} algorithm. The output kinematic parameters are then used to create and improve composite templates and the process continues until it converges. This \textsc{optema} algorithm is able to overcome the typical template mismatch problem (see van der Marel $\&$ Franx 1993) by constructing an optimal template as a linear combination of stellar spectra of different spectral types and luminosity classes for each galaxy (Gonz\'{a}lez 1993; S\'{a}nchez-Bl\'{a}zquez et al.\ 2006). A new optimal template was constructed with each spatial bin in the kinematical measurements. 

Skyline residuals and emission lines were masked. Emission lines were detected in 13 galaxies: ESO349-010, MCG-02-12-039, NGC0541, NGC1713, NGC3311, NGC4874, NGC4946, NGC6166, NGC6173, NGC7012, NGC7649, NGC7720 and PGC044257. The detection thresholds and emission-line correction will be described in a future paper devoted to the stellar populations.

For the Gemini data, 20 stars were used to make the optimal template (excluding the stars not of G or K spectral type), and 22 stars for the WHT data. Figure \ref{fig:WHTreduceme} shows a typical fit between the observed central spectrum of a galaxy and the optimal template for that galaxy corrected by the derived kinematic parameters.\footnote{The derived recession velocities were automatically corrected for the standard special relativistic correction:
 $z=(1+\frac{V_{\rm nr}}{c})\gamma-1$ 
where $\gamma=\frac{c}{\sqrt{c^{2}-V^{2}}}$, and $V_{\rm nr}$ is the non-relativistic radial velocity and $c$ the speed of light.} 

The errors were computed through Monte Carlo numerical simulations and using the appropriate error spectra. In each simulation, gaussian noise was added to the galaxy spectra in accordance with the error spectra and the radial velocity and velocity dispersion derived. Errors in the radial velocities and velocity dispersions were calculated as the unbiased standard deviation of the different solutions. The errors computed in this manner are expected to incorporate all the uncertainties, from the basic reduction process (which is accounted for in the error spectra) to the final measurement of the parameters. 

When comparing measured central velocity dispersion values with those from the literature, emphasis must be placed on the fact that published values are measurements from within different apertures, which can have a systematic effect on the values. Accurate comparison is further limited because of different effective spatial resolution, data of different quality and different analysis techniques. 

Figure \ref{fig:VelDisp} shows the rough comparison (not taking into account the above differences) between the velocity dispersion values from HyperLeda (if available) and those derived in this study. For most of the galaxies, more than one velocity dispersion values are listed in HyperLeda (in which case the most recent was used for this comparison). In the case of NGC1399, for example, published values range between 233 km s$^{-1}$ and 420 km s$^{-1}$, which shows the effect of different apertures and measurements have (this can be seen in the steep change in the measured velocity dispersion profile of NGC1399 in Figure \ref{fig:IC1633}). The absolute mean difference between the values measured here and the literature is $\vert \sigma(\rm{here}) - \sigma(\rm{lit,\ uncorrected}) \vert$ = 33 $\pm$ 29 km s$^{-1}$, which is acceptable for the purposes of this paper.

\begin{table}
\begin{scriptsize}
\begin{tabular}{l r@{$\pm$}l r@{$\pm$}l r@{$\pm$}l r@{$\pm$}l}
\hline Galaxy  & \multicolumn{2}{c}{$\sigma$} & \multicolumn{2}{c}{$V$} & \multicolumn{2}{c}{$V_{\rm max}$} & \multicolumn{2}{c}{log$(V_{\rm max}/\sigma_{0})^{\ast}$}\\
 & \multicolumn{2}{c}{km s$^{-1}$} & \multicolumn{2}{c}{km s$^{-1}$} & \multicolumn{2}{c}{km s$^{-1}$} & \multicolumn{2}{c}{ }\\
 & \multicolumn{2}{c}{Observed} &  \multicolumn{2}{c}{Observed} & \multicolumn{2}{c}{ } & \multicolumn{2}{c}{ } \\ 
\hline ESO146-028 & 299 & 3 & 12061 & 3 & 14 & 18 & -1.2 & 0.9\\
 ESO303-005 & 276 & 5 & 14526 & 4 & 23 & 17 & -0.8 & 0.4\\
 ESO346-003 & 226 & 4 & 8459 & 9 & 51 & 13 & -0.3 & 0.1\\ 
 ESO349-010 & 282 & 3 & 14420 & 3 & 46 & 19 & -0.8 & 0.4\\
 ESO444-046 & 292 & 3 & 13741 & 3 & 26 & 16 & -0.9 & 0.4\\
 ESO488-027 & 248 & 2 & 11754 & 4 & 53 & 12 & -0.3 & 0.1\\
 ESO552-020 & 229 & 3 & 9128 & 5 & 20 & 19 & -1.0 & 0.8\\
 GSC555700266 & 312 & 9 & 20059 & 6 & 49 & 14 & -0.6 & 0.2\\
 IC1101 & 378 & 5 & 22585 & 9 & 41 & 31 & -1.0 & 0.8\\
 IC1633 & 400 & 2 & 7061 & 2 & 20 & 17 & -1.0 & 0.4\\
 IC4765 & 286 & 5 & 4403 & 6 & 50 & 28 & -0.8 & 0.6\\
 IC5358 & 243 & 3 & 8539 & 9 & 55 & 13 & -0.7 & 0.3\\
 Leda094683 & 332 & 5 & 22748 & 5 & 32 & 26 & -0.9 & 0.6\\
 MCG-02-12-039 & 271 & 5 & 9648 & 3 & 59 & 28 & -0.3 & 0.3\\
 NGC1399 & 371 & 3 & 1406 & 2 & 24 & 11 & -0.6 & 0.1\\
 NGC1713 & 251 & 2 & 4472 & 2 & 30 & 11 & -0.5 & 0.1\\
 NGC2832 & 364 & 4 & 6841 & 4 & 64 & 22 & -0.4 & 0.2\\
 NGC3311 & 196 & 2 & 3709 & 2 & 25 & 9 & -0.6 & 0.2\\
 NGC3842 & 287 & 5 & 6211 & 2 & 11 & 13 & -1.2 & 0.8\\
 NGC4839 & 278 & 2 & 7380 & 2 & 44 & 14 & -0.8 & 0.3\\
 NGC4874 & 267 & 4 & 6953 & 2 & 15 & 12 & -0.4 & 0.1\\
 NGC4889 & 380 & 4 & 6266 & 5 & 40 & 19 & -0.8 & 0.3 \\
 NGC4946 & 201 & 2 & 3148 & 9 & 62 & 9 & -0.2 & 0.1\\
 NGC6034 & 325 & 4 & 10113 & 28 & 134 & 15 & -0.2 & 0.1\\
 NGC6047 & 230 & 2 & 9031 & 8 & 59 & 16 & -0.4 & 0.2\\
 NGC6086 & 318 & 5 & 9483 & 2 & 13 & 18 & -1.2 & 0.9\\
 NGC6160 & 266 & 3 & 9428 & 3 & 12 & 18 & -1.3 & 1.4\\
 NGC6166 & 310 & 10 & 9100 & 2 & 31 & 20 & -0.7 & 0.4\\
 NGC6173 & 304 & 3 & 8766 & 2 & 24 & 16 & -0.9 & 0.4\\
 NGC6269 & 343 & 5 & 10363 & 3 & 26 & 35 & -0.8 & 0.7\\
 NGC7012 & 240 & 3 & 8653 & 3 & 24 & 16 & -0.9 & 0.6\\
 NGC7597 & 264 & 8 & 10907 & 5 & 63 & 20 & 0.2 & 0.1\\
 NGC7647 & 271 & 5 & 11842 & 7 & 36 & 14 & -0.9 & 0.4\\
 NGC7649 & 250 & 5 & 12258 & 5 & 41 & 14 & -0.6 & 0.2\\
 NGC7720 & 409 & 5 & 9034 & 5 & 50 & 18 & -0.6 & 0.2\\
 NGC7768 & 272 & 5 & 7875 & 24 & 114 & 11 & -0.1 & 0.1\\
 PGC026269 & 222 & 11 & 15816 & 8 & 51 & 20 & 0.2 & 0.1\\
 PGC044257 & 247 & 9 & 14014 & 6 & 20 & 16 & -1.2 & 1.0\\
 PGC071807 & 315 & 3 & 17911 & 5 & 16 & 13 & -1.5 & 1.2\\
 PGC072804 & 311 & 5 & 22433 & 4 & 50 & 24 & -0.4 & 0.2\\
 UGC02232 & 314 & 4 & 14208 & 3 & 42 & 18 & 0.0 & 0.1\\
 UGC05515 & 362 & 4 & 13103 & 7 & 56 & 18 & -0.4 & 0.1\\
 UGC10143 & 262 & 2 & 10344 & 5 & 19 & 11 & -1.1 & 0.5\\
\hline
\end{tabular}
\end{scriptsize}
\caption[Central Velocity Dispersions and Radial Velocities of the cD Galaxies.]{Central velocity dispersions ($\sigma$) and radial velocities ($V$) of the galaxies. The measurement and determination of $V_{\rm max}$ and $(V_{\rm max}/\sigma_{0})^{\ast}$ are described in Section \ref{kinematicanalysis}.}
\label{table:GalVel}
\end{table} 

\section{Kinematics of individual galaxies}
\label{Sec:kinematics}

The galaxy spectra were binned in the spatial direction to ensure a minimum S/N of 20 per \AA{} in the H$\beta$ region of the spectrum for kinematical measurements as a function of radius. Thus, the spatial cross-sections are broader with increasing radius from the centre of the galaxy. The radial kinematics (radial velocities and velocity dispersions) were measured with the same \textsc{movel} and \textsc{optema} algorithms in the \reduceme\ package and with the same optimal Lick star templates as the central kinematics. The values of the kinematic parameters are evaluated at the luminosity-weighted centres of the spatial bins used to derive the parameters. 

Dark time at both Gemini telescopes is highly oversubscribed. To speed up the completion of this multi-semester project, we relaxed the observing conditions to accept grey time as well. However, as a consequence it was not possible to extract radial profiles to very large radii for those targets observed during grey time, but the central values are still accurate. Eight galaxies (three WHT and five Gemini) could be extracted to radii $>$ 1$a_{\rm e}$ and 12 galaxies (one WHT and 11 Gemini) could only be extracted to radii $<$ 0.5$a_{\rm e}$. The rotational velocity $V_{\rm max}$ was measured as half the difference between the peaks of the rotation curve.

\begin{figure*}
   \mbox{\subfigure[ESO146-028.]{\includegraphics[scale=0.23]{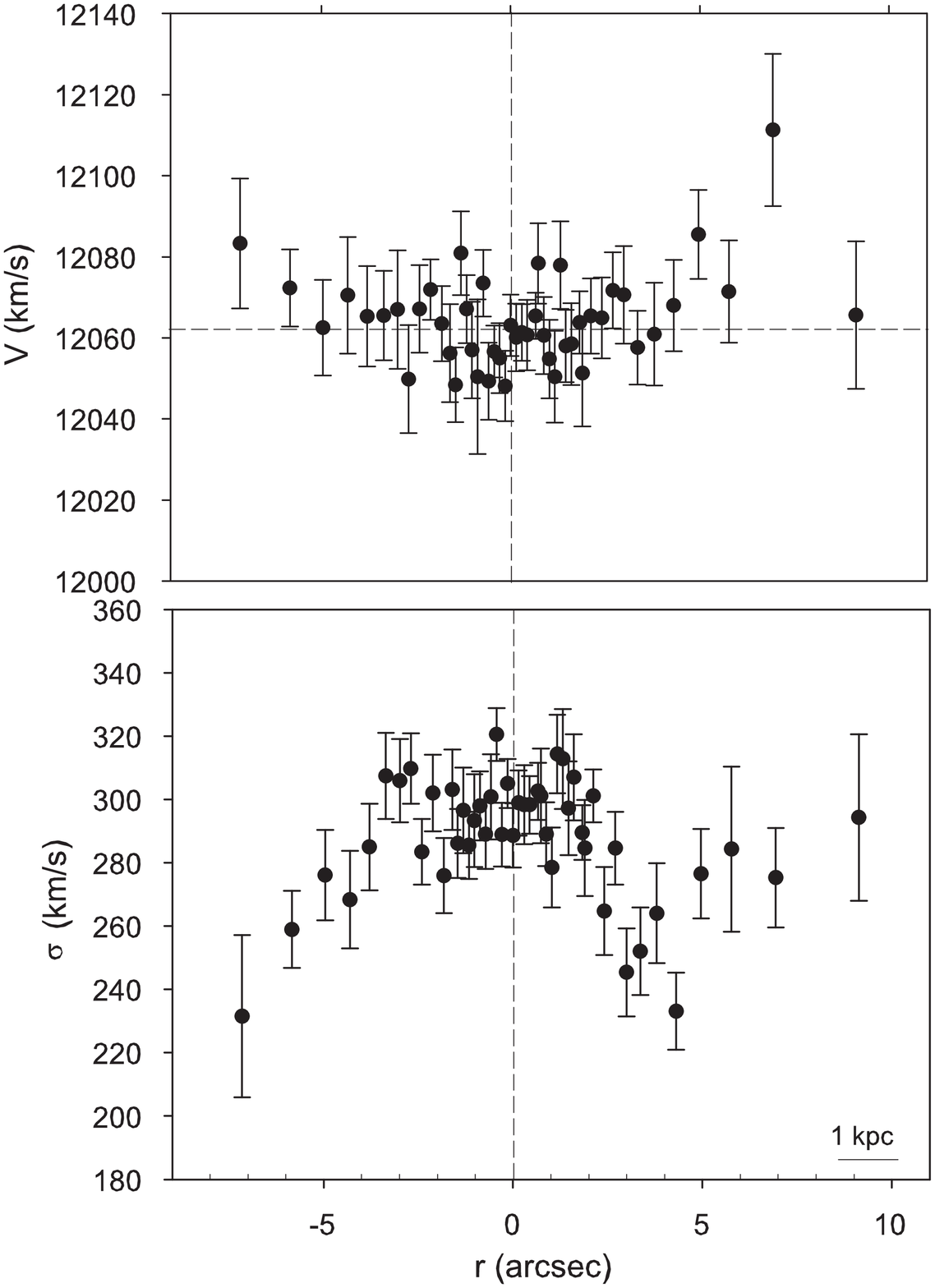}}\quad
         \subfigure[ESO303-005.]{\includegraphics[scale=0.23]{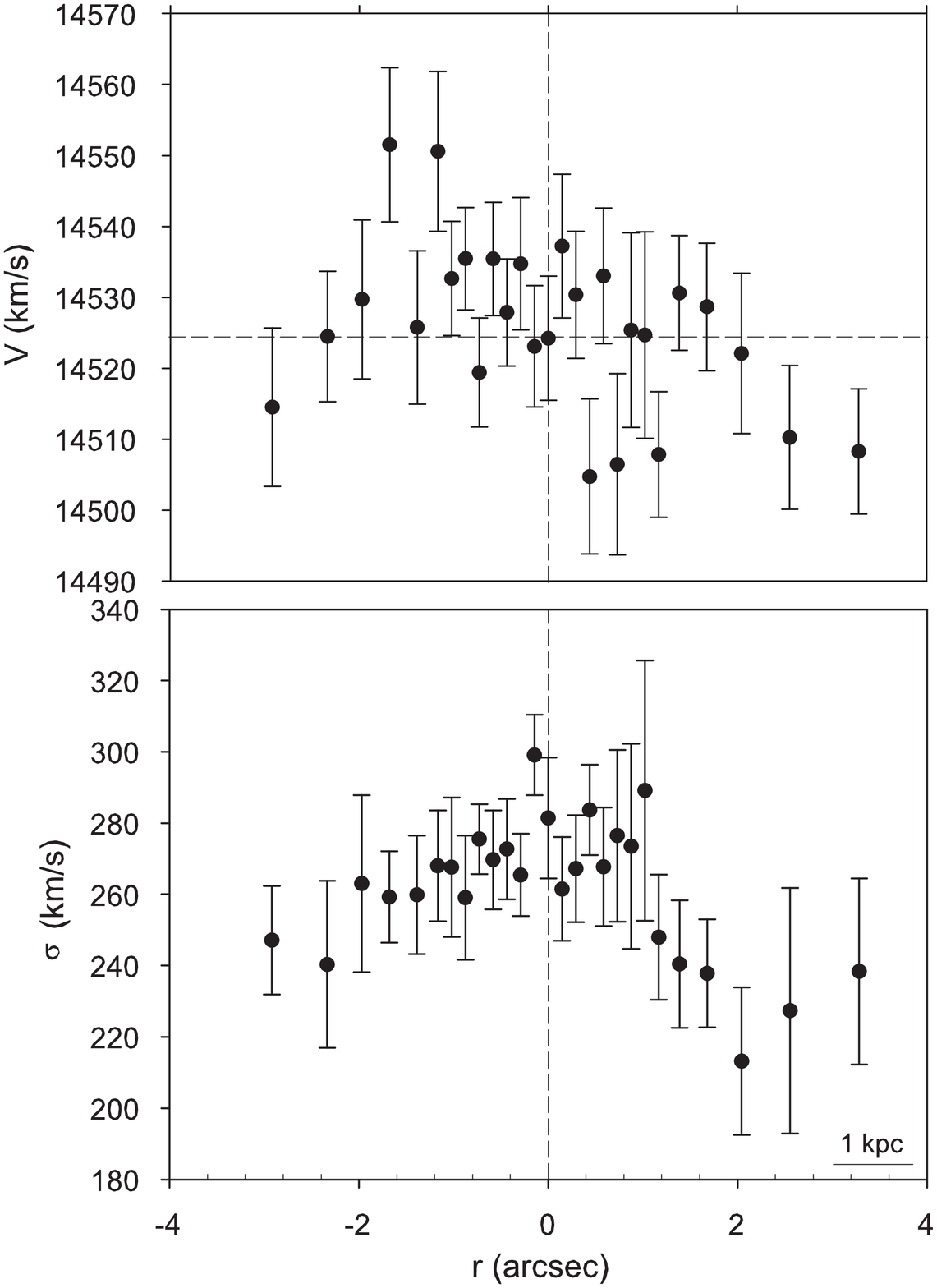}}\quad
         \subfigure[ESO346-003.]{\includegraphics[scale=0.23]{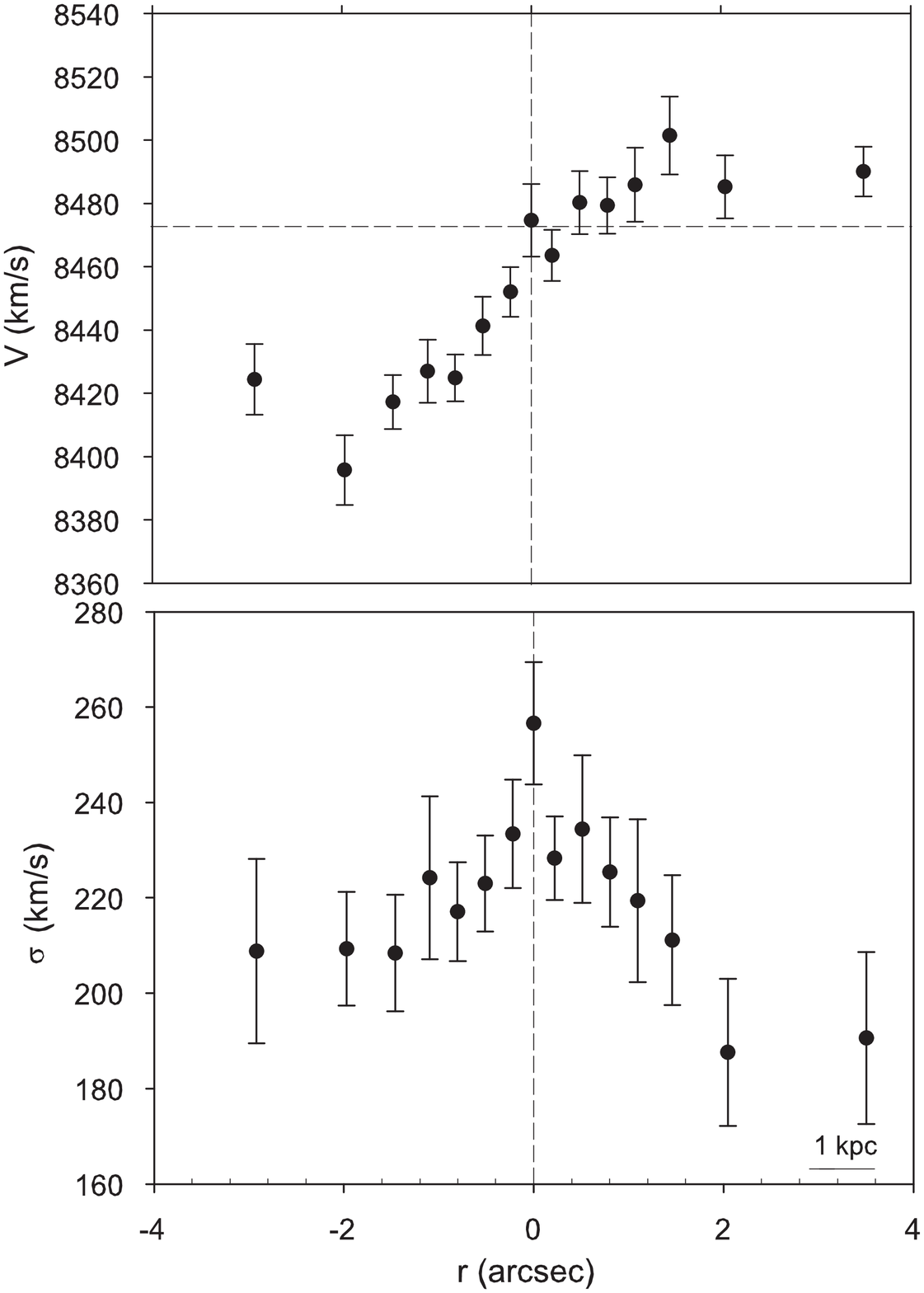}}}
   \mbox{\subfigure[ESO349-010.]{\includegraphics[scale=0.23]{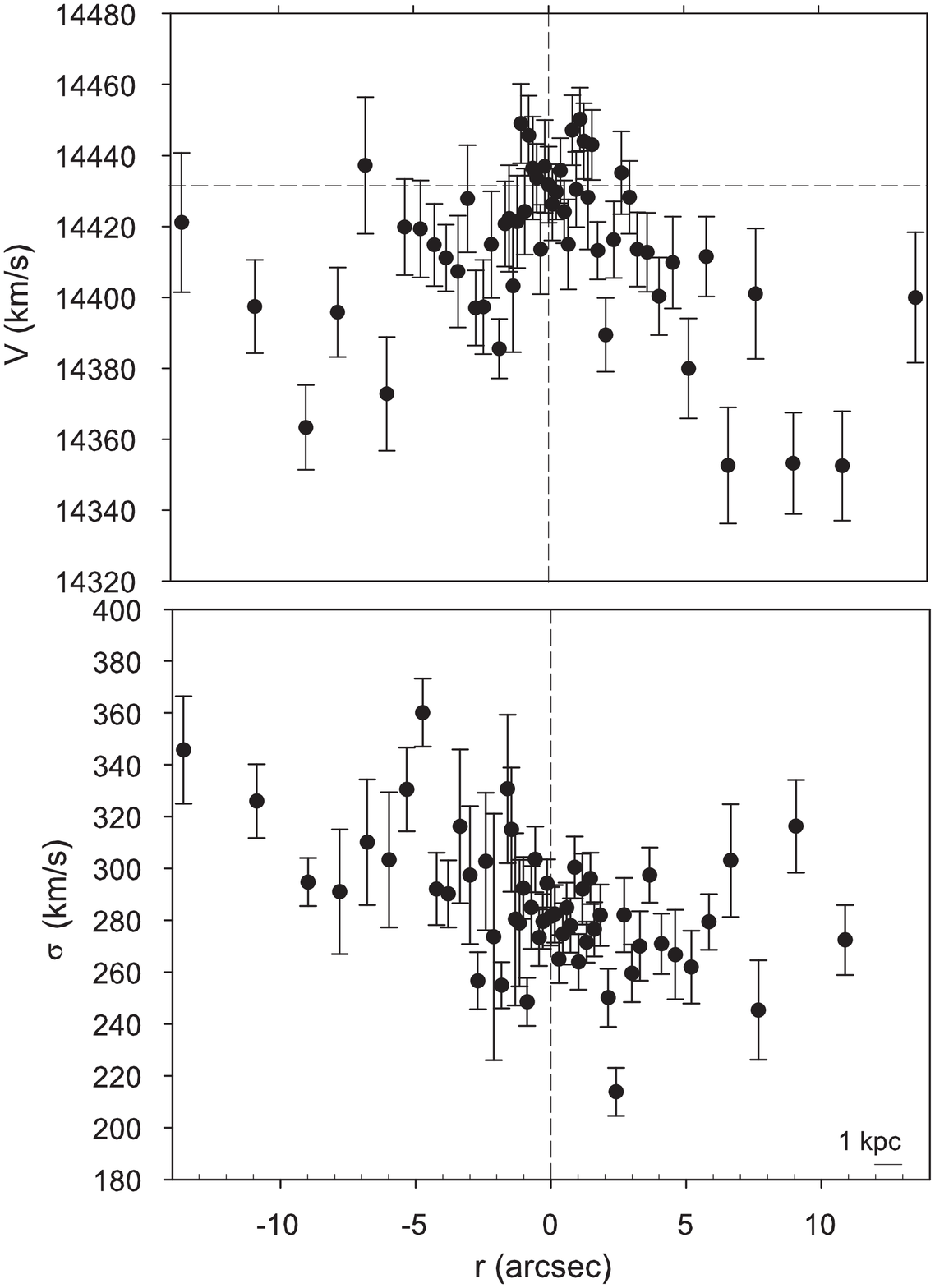}}\quad
         \subfigure[ESO444-046.]{\includegraphics[scale=0.23]{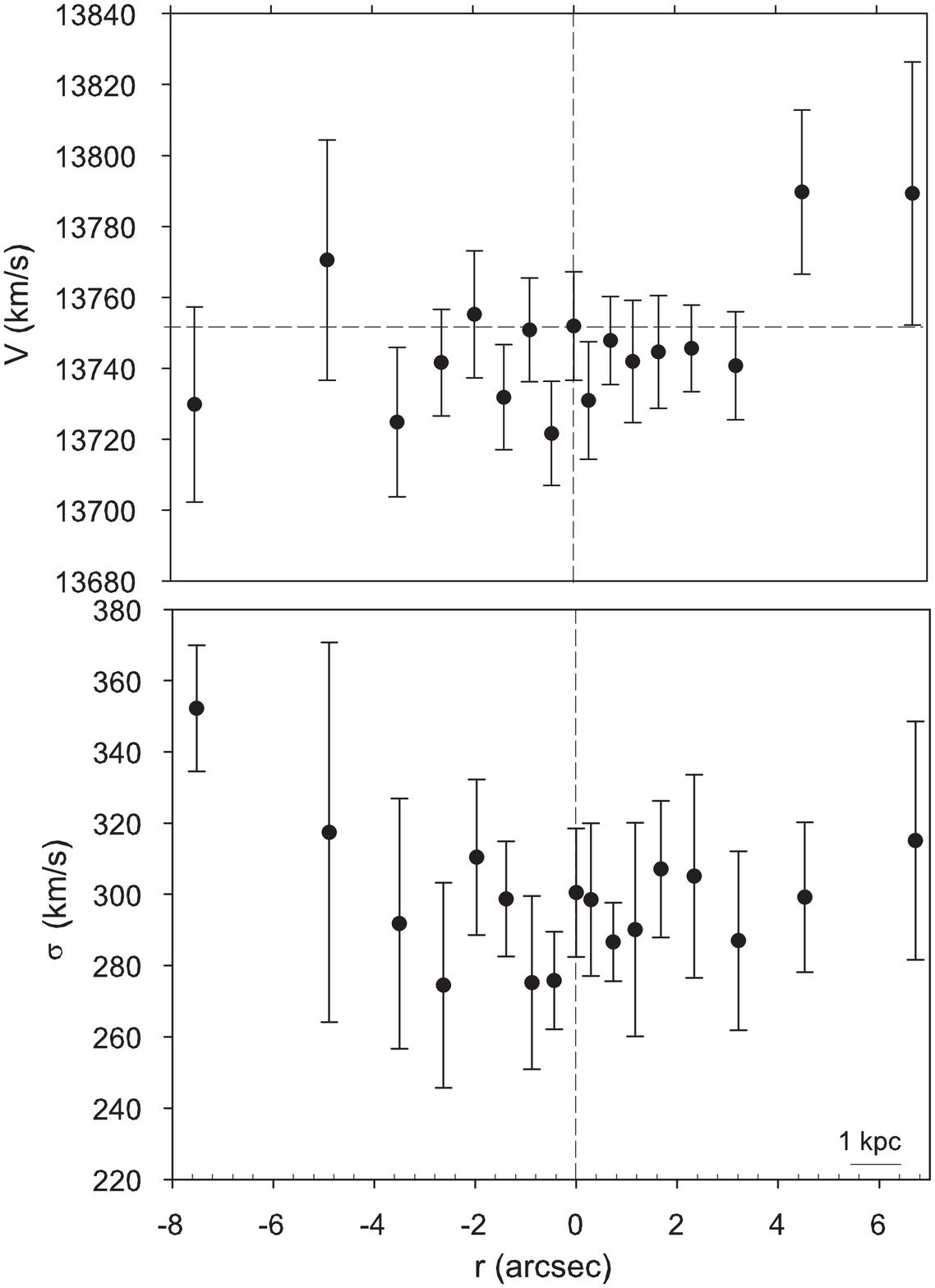}}\quad
         \subfigure[ESO488-027.]{\includegraphics[scale=0.23]{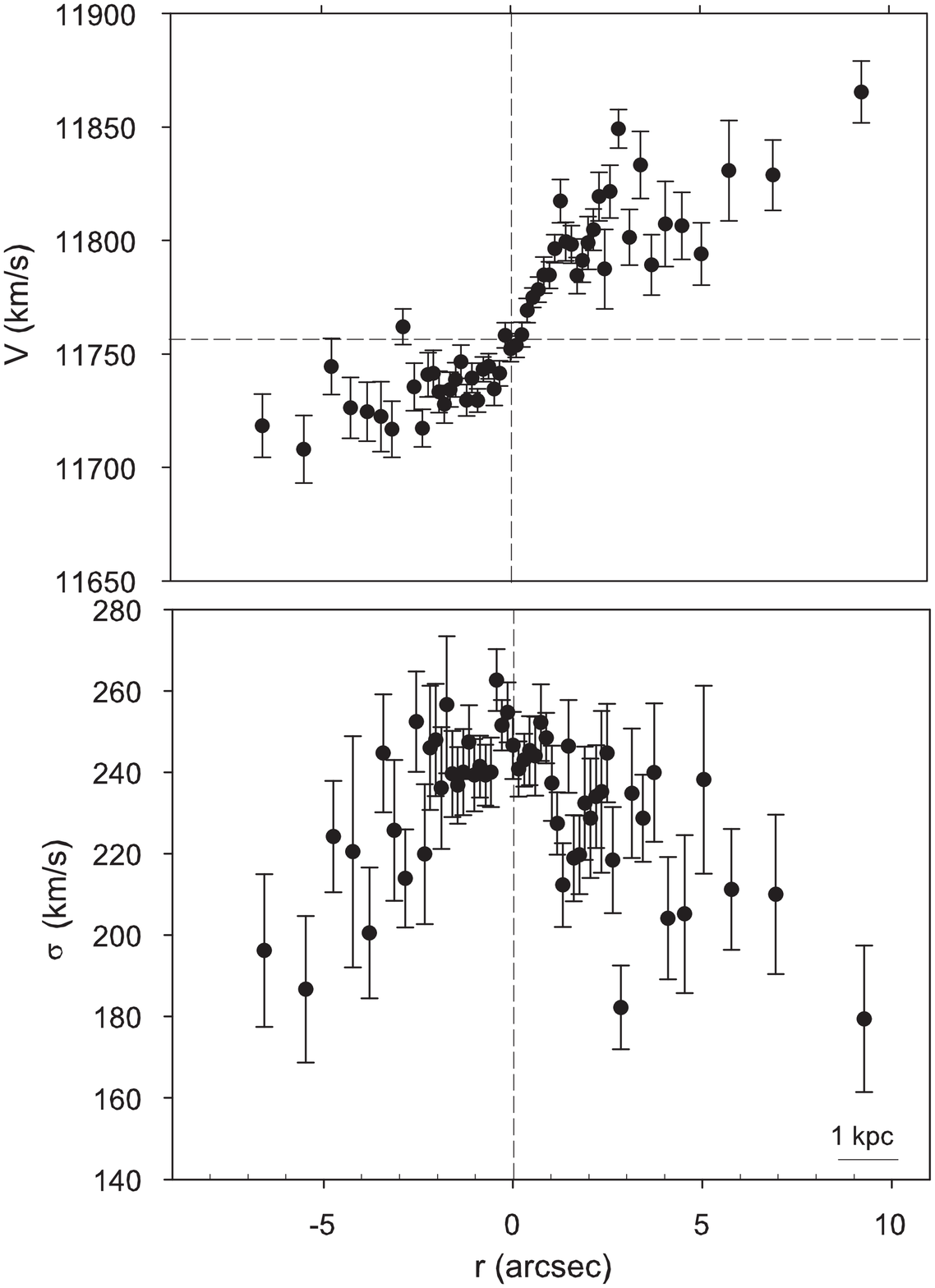}}}
   \mbox{\subfigure[ESO552-020.]{\includegraphics[scale=0.23]{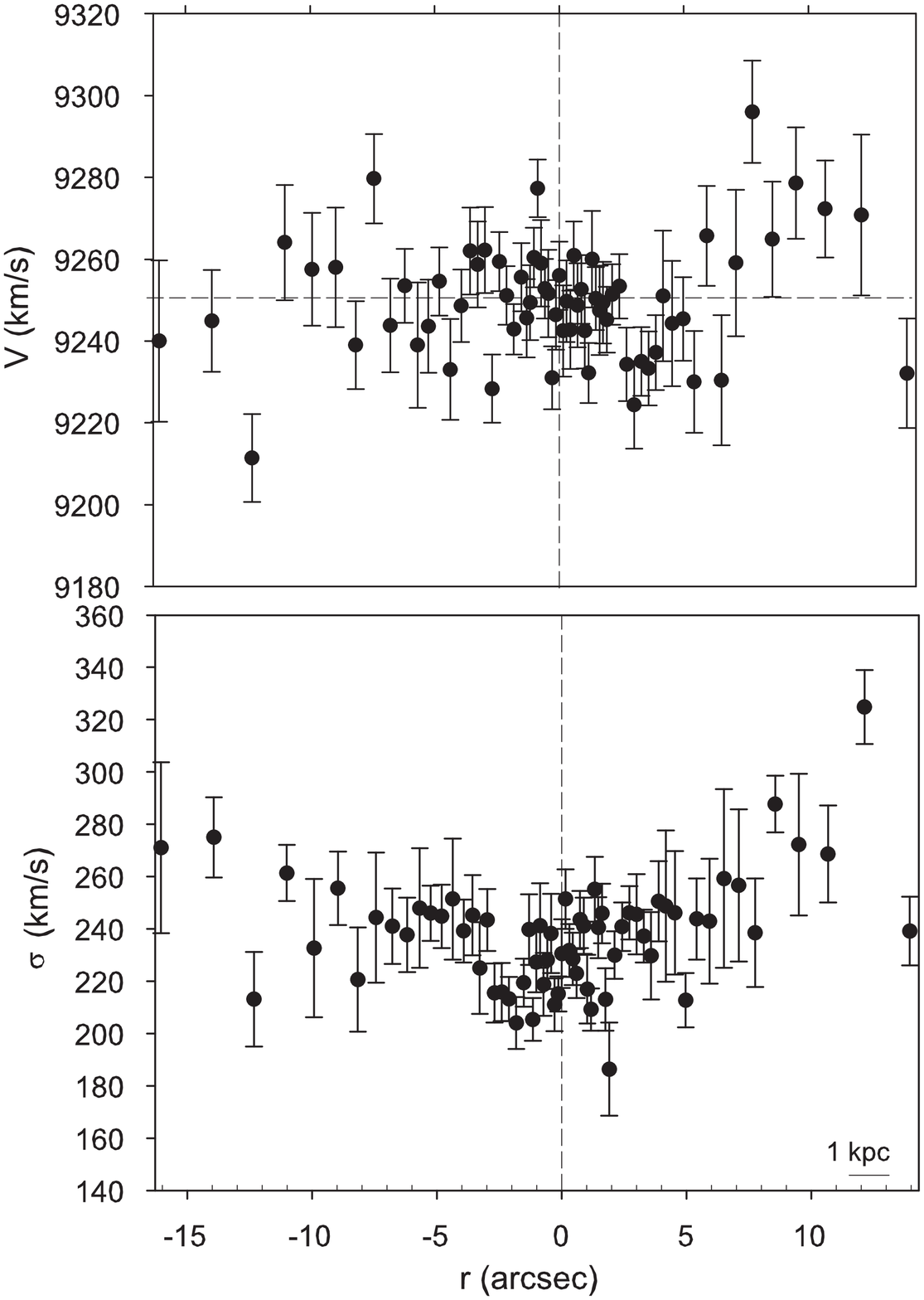}}\quad
         \subfigure[GSC555700266.]{\includegraphics[scale=0.23]{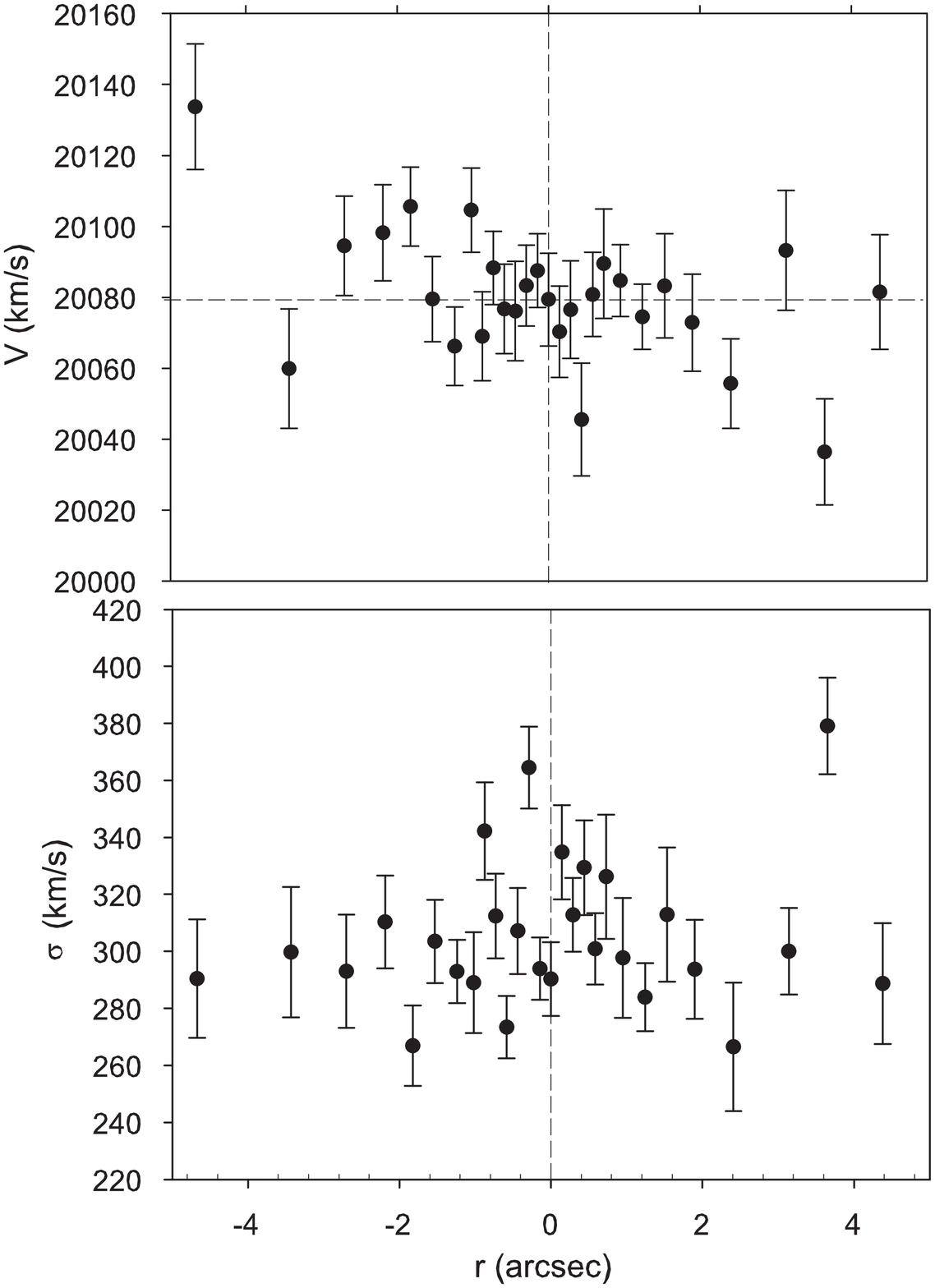}}\quad
         \subfigure[IC1101.]{\includegraphics[scale=0.23]{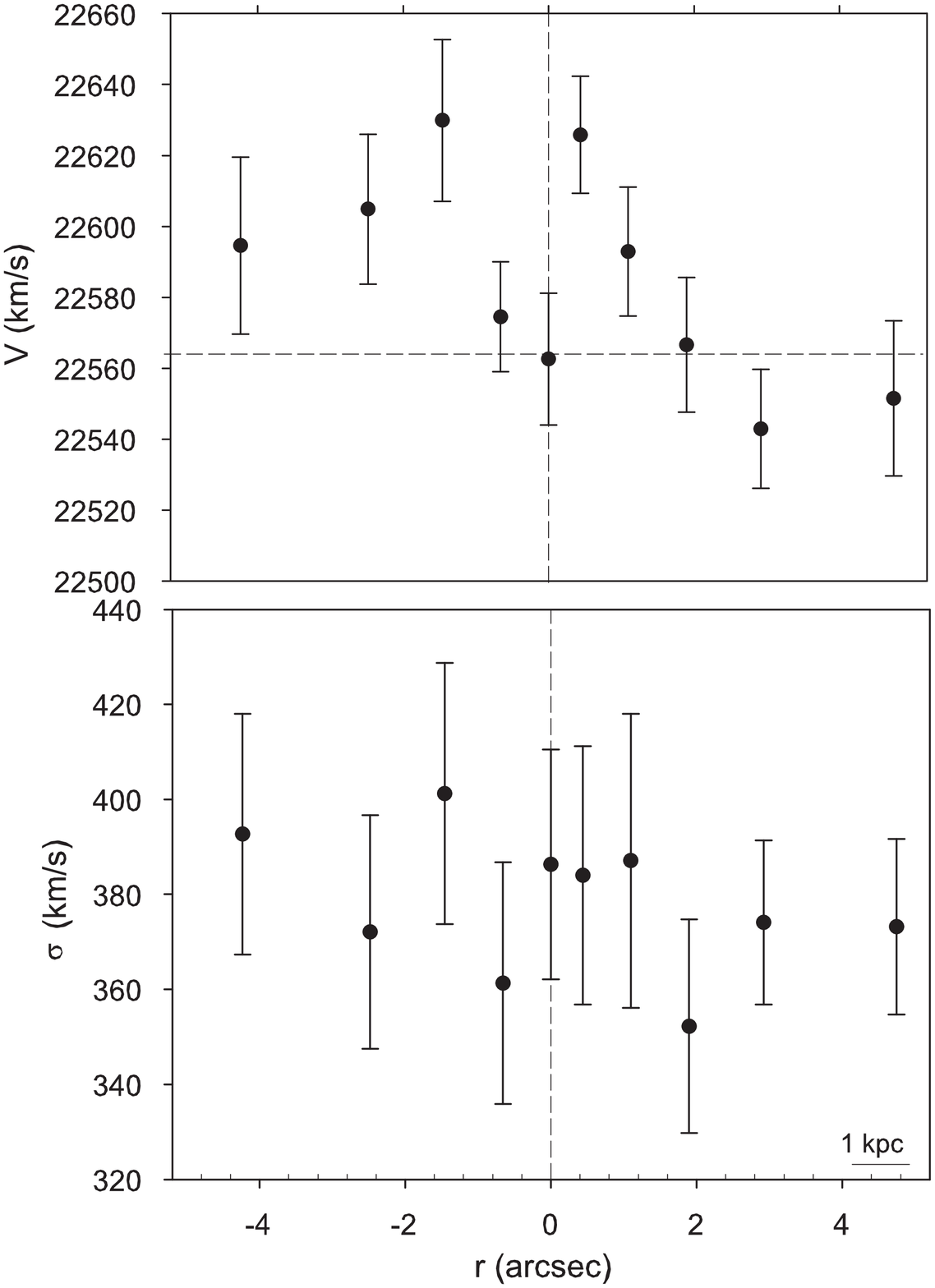}}}
\caption{Radial profiles of velocities (V) and velocity dispersion ($\sigma$). The vertical lines indicate the centre of the galaxy, and the horizontal line the radial velocity of the central bin.}
\label{fig:NGC3842}
\end{figure*}

\begin{figure*}
   \centering
   \mbox{\subfigure[IC1633.]{\includegraphics[scale=0.23]{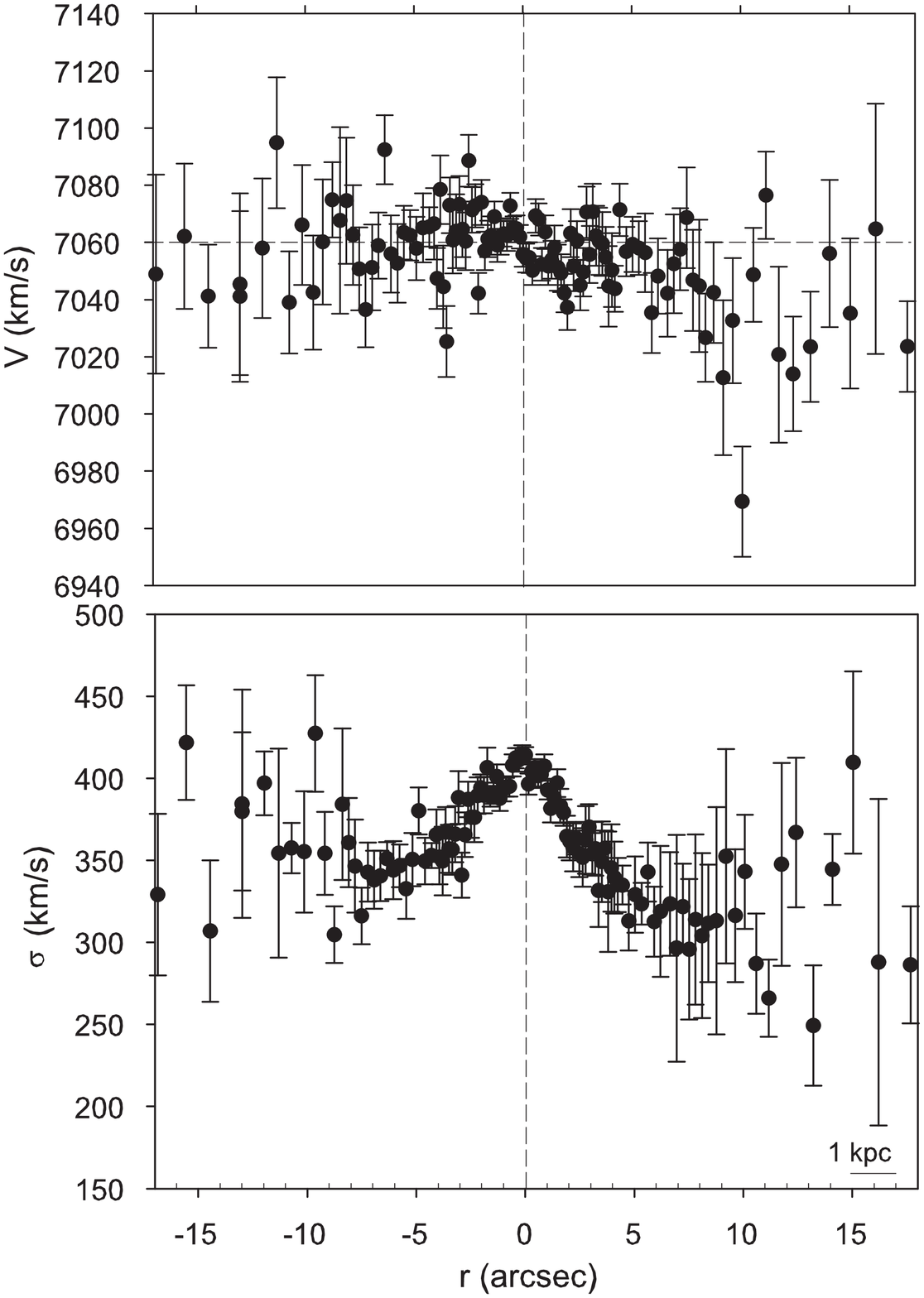}}\quad
         \subfigure[IC4765.]{\includegraphics[scale=0.23]{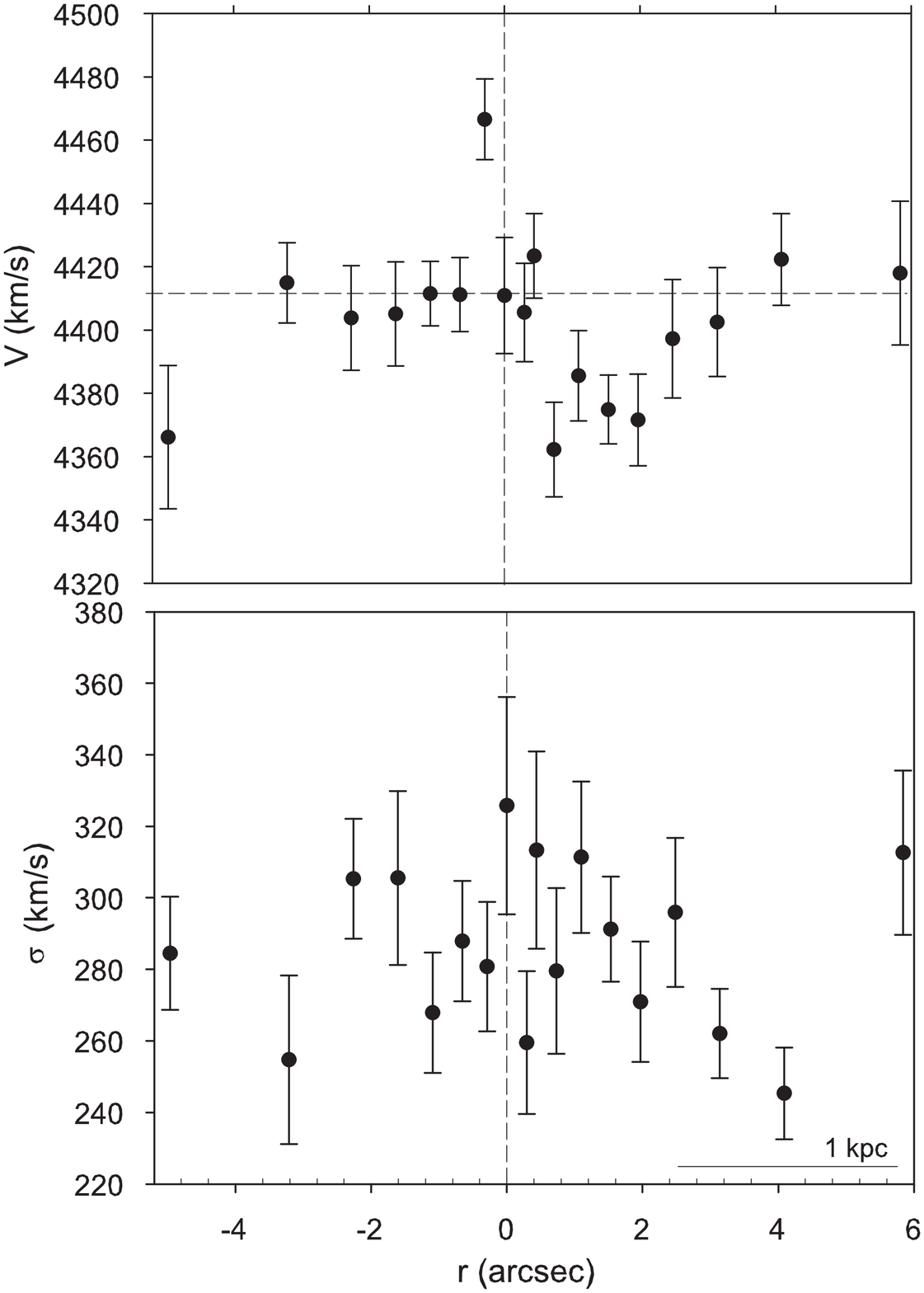}}\quad
         \subfigure[IC5358.]{\includegraphics[scale=0.23]{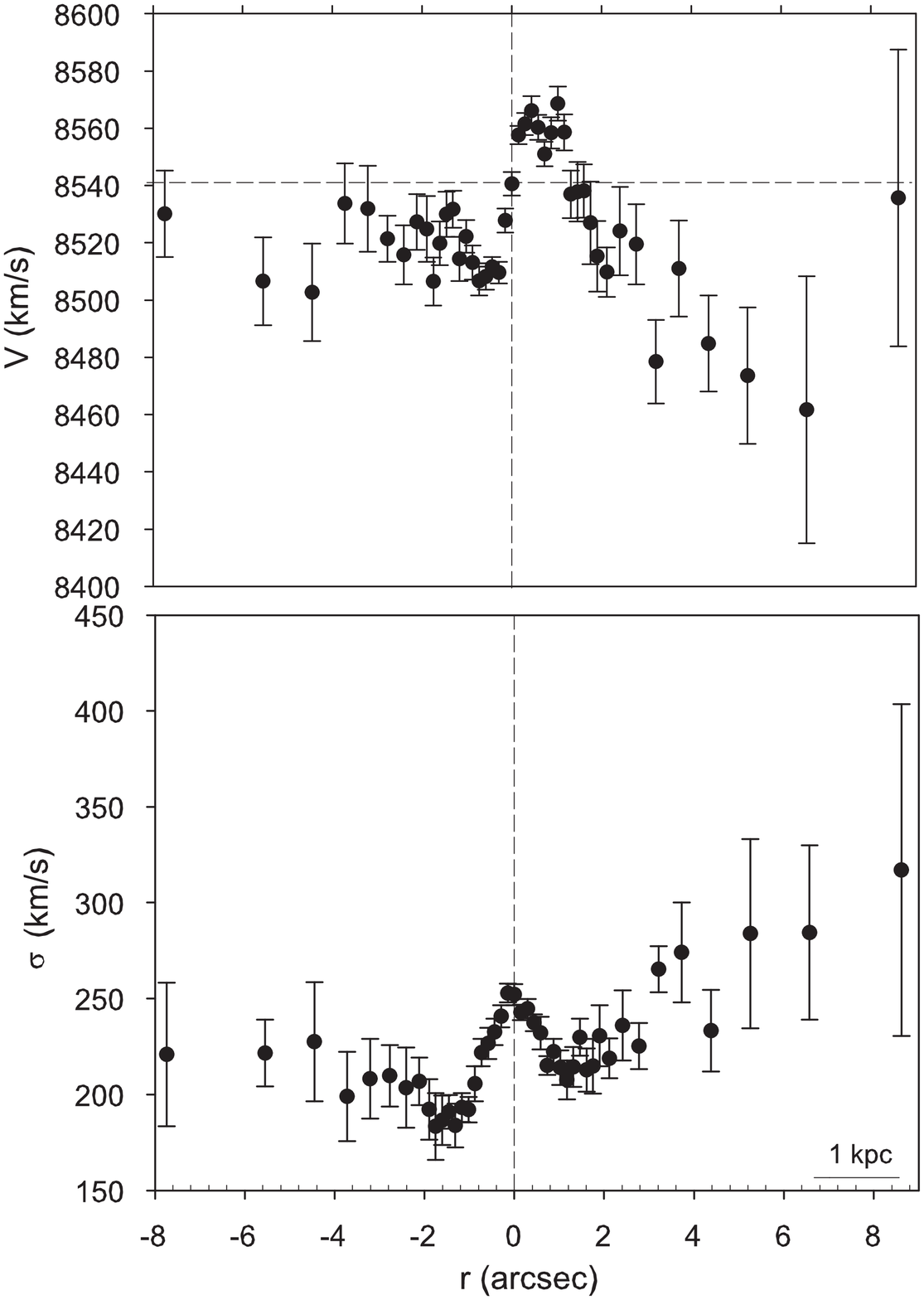}}}
   \mbox{\subfigure[Leda094683.]{\includegraphics[scale=0.23]{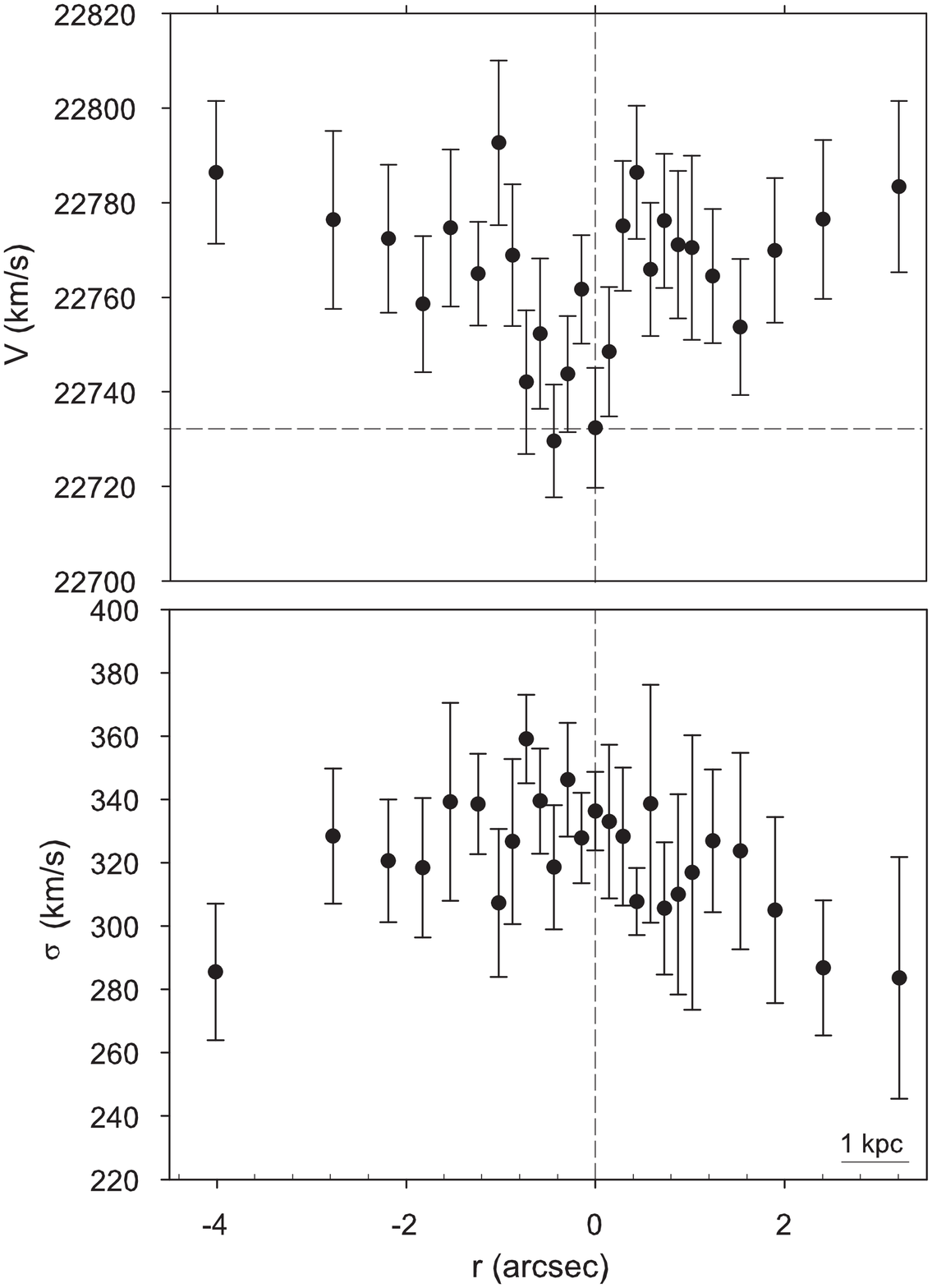}}\quad
         \subfigure[MCG-02-12-039.]{\includegraphics[scale=0.23]{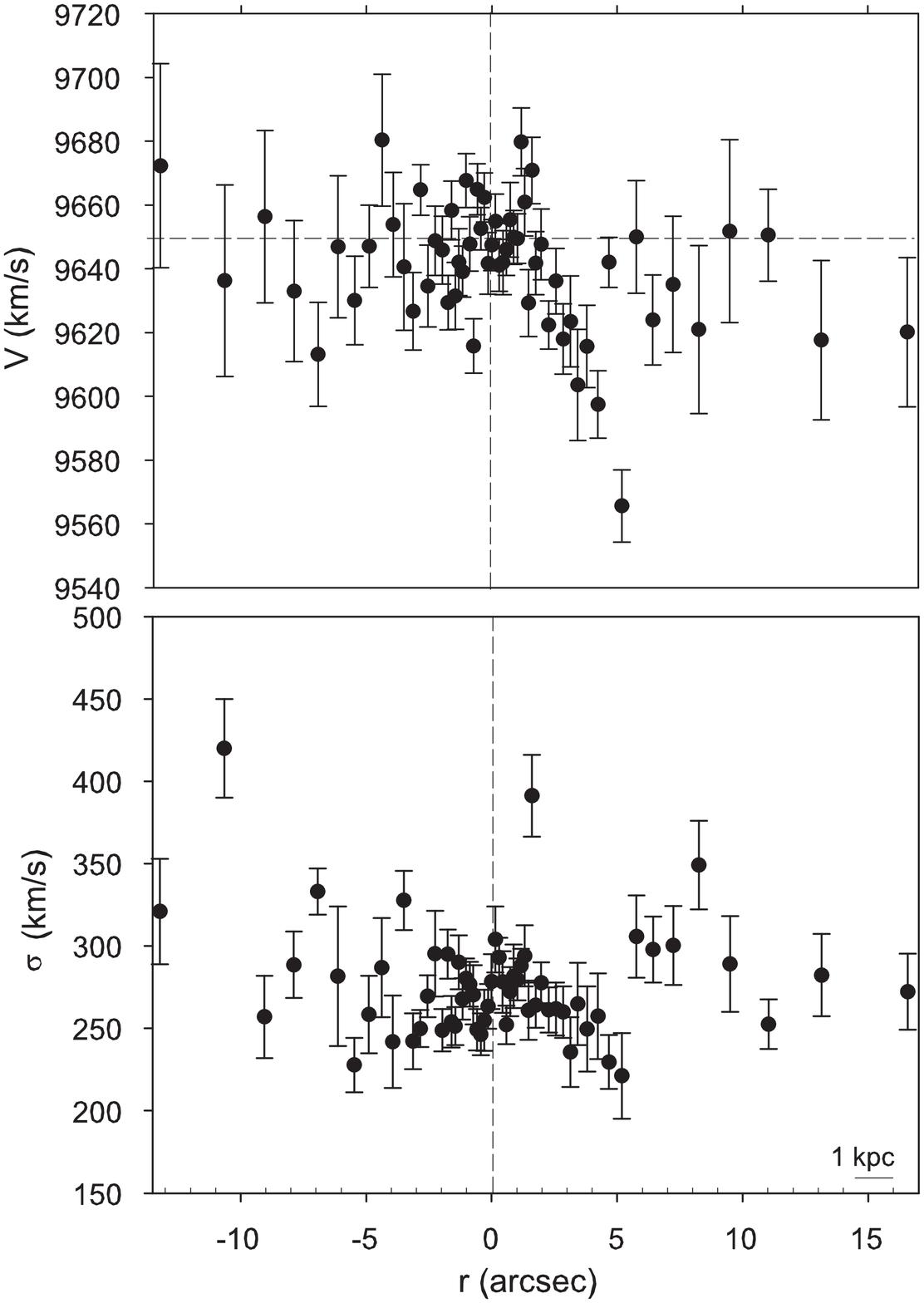}}\quad
         \subfigure[NGC1399.]{\includegraphics[scale=0.23]{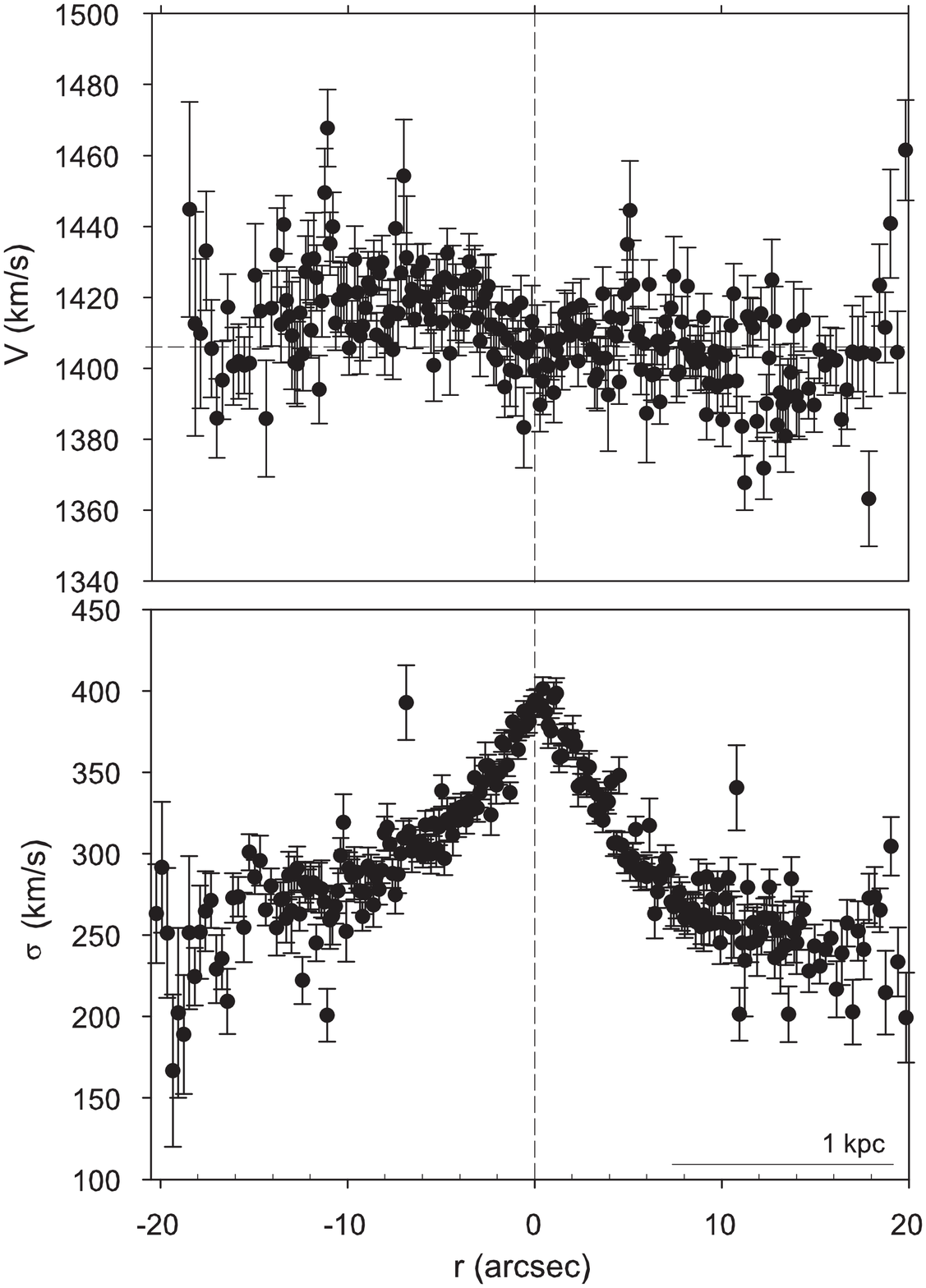}}}
   \mbox{\subfigure[NGC1713.]{\includegraphics[scale=0.23]{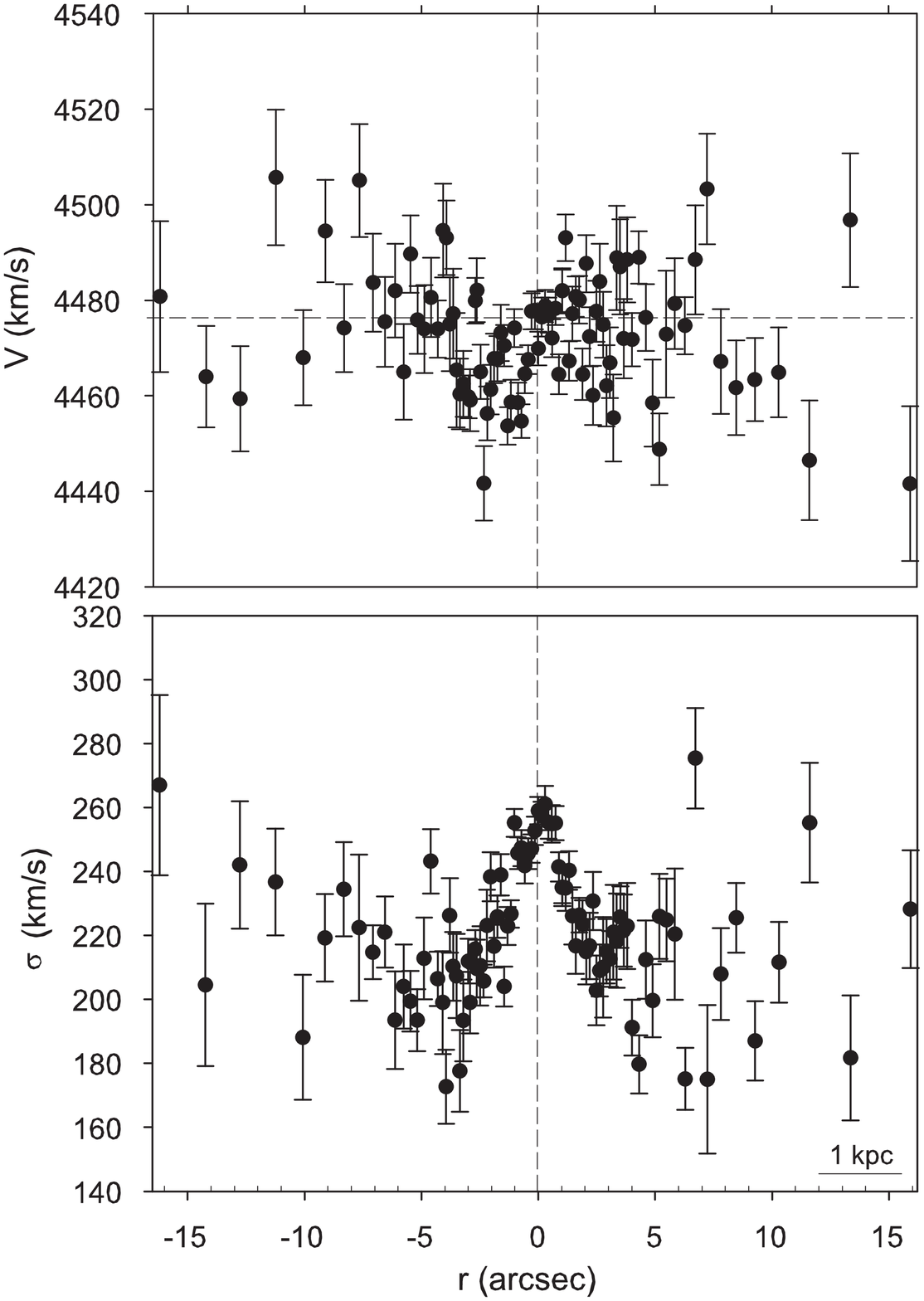}}\quad
         \subfigure[NGC2832.]{\includegraphics[scale=0.23]{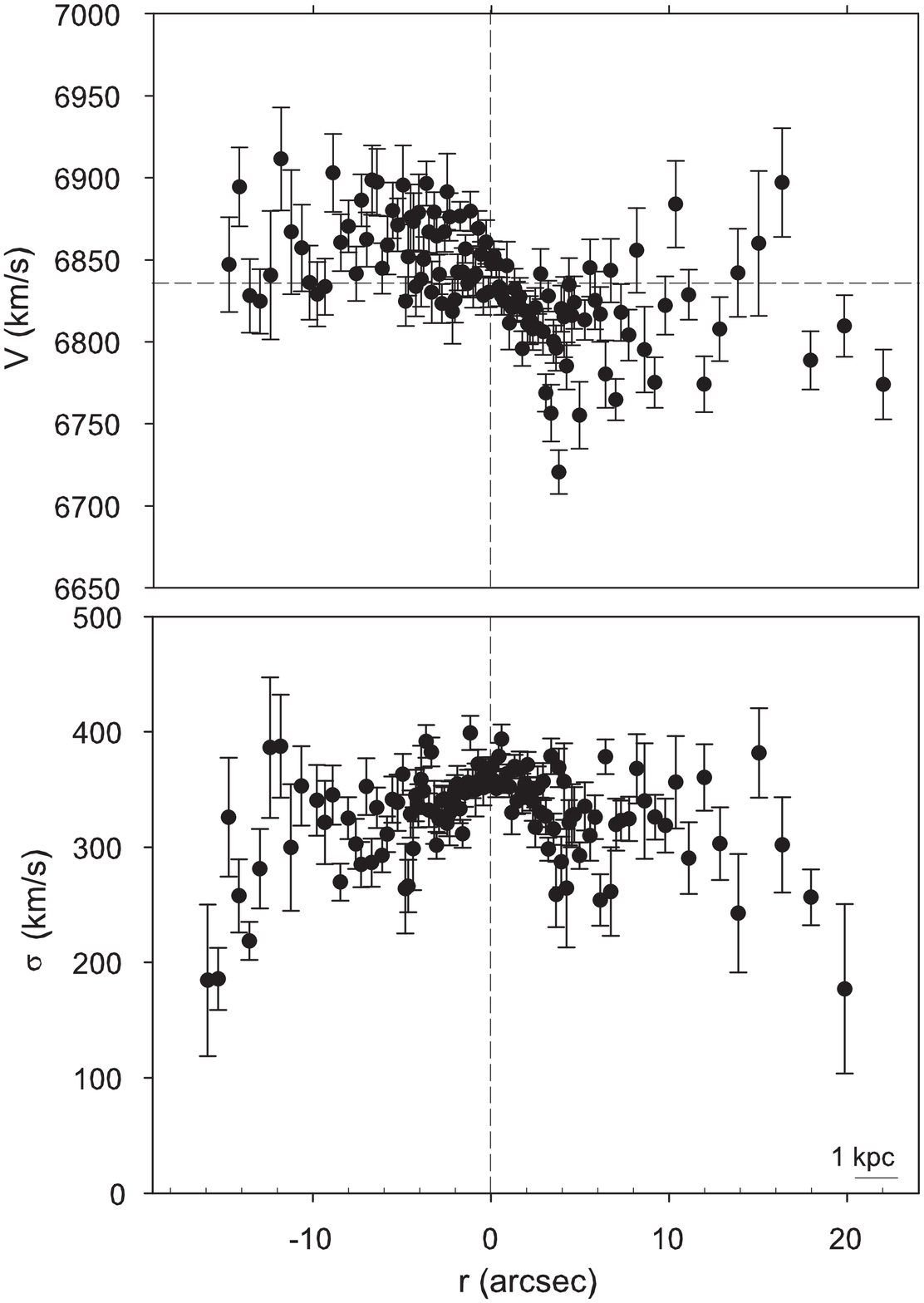}}\quad
         \subfigure[NGC3311.]{\includegraphics[scale=0.23]{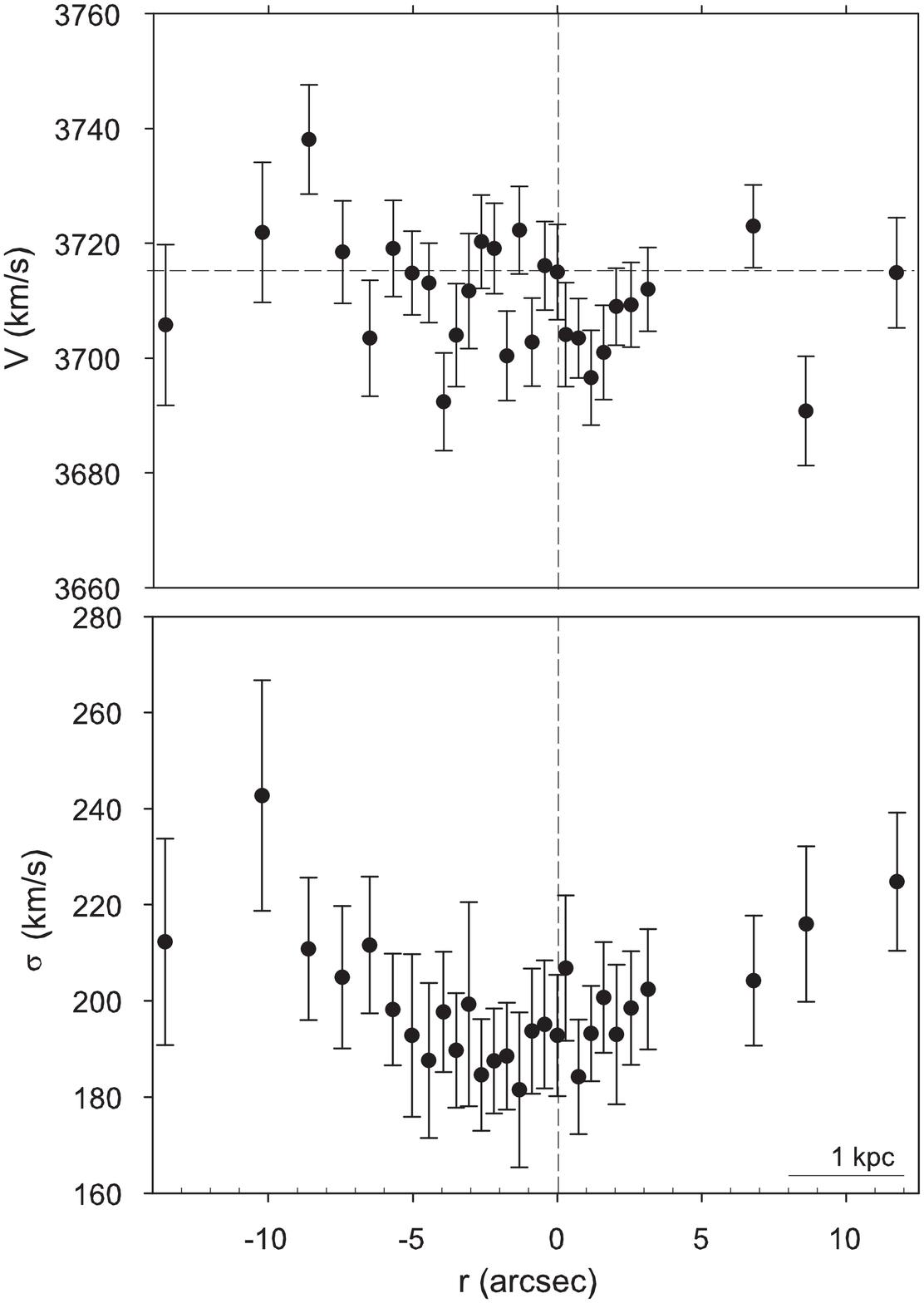}}}
\caption{Radial profiles of velocities (V) and velocity dispersion ($\sigma$). The dashed lines are as in Figure \ref{fig:NGC3842}.}
\label{fig:ESO146}
\end{figure*}

\begin{figure*}
   \centering
   \mbox{\subfigure[NGC3842.]{\includegraphics[scale=0.23]{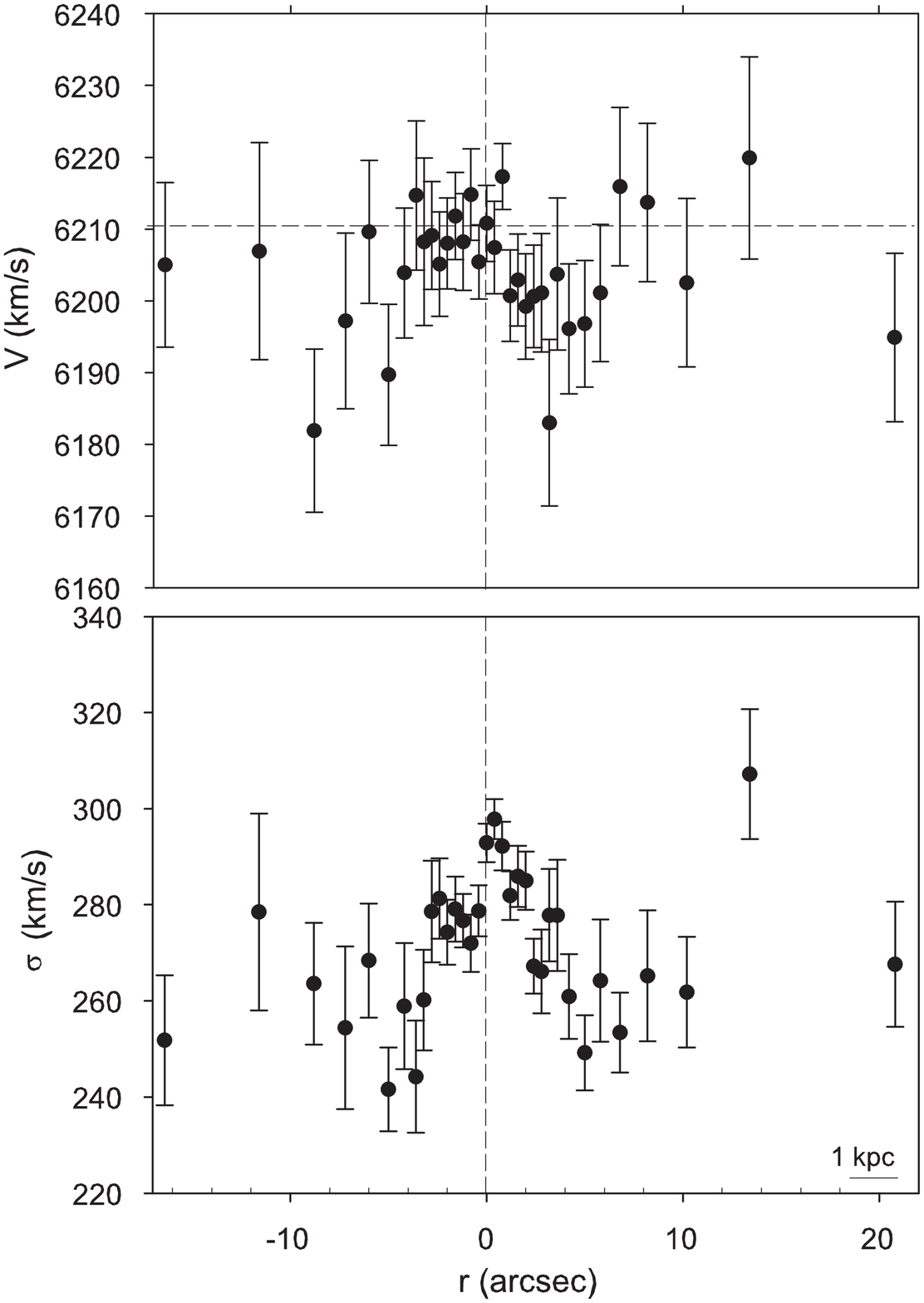}}\quad
         \subfigure[NGC4839.]{\includegraphics[scale=0.23]{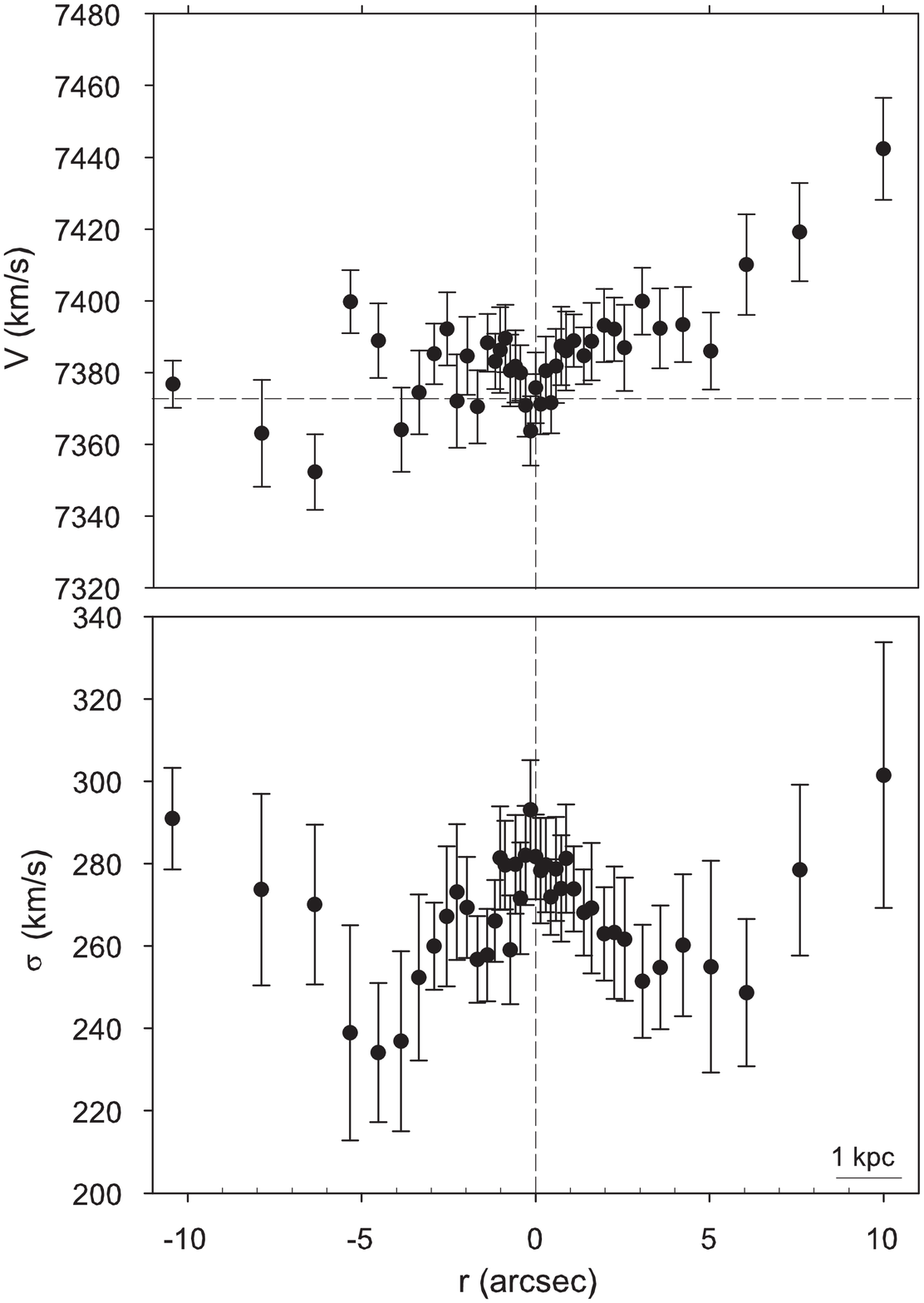}}\quad
         \subfigure[NGC4874.]{\includegraphics[scale=0.23]{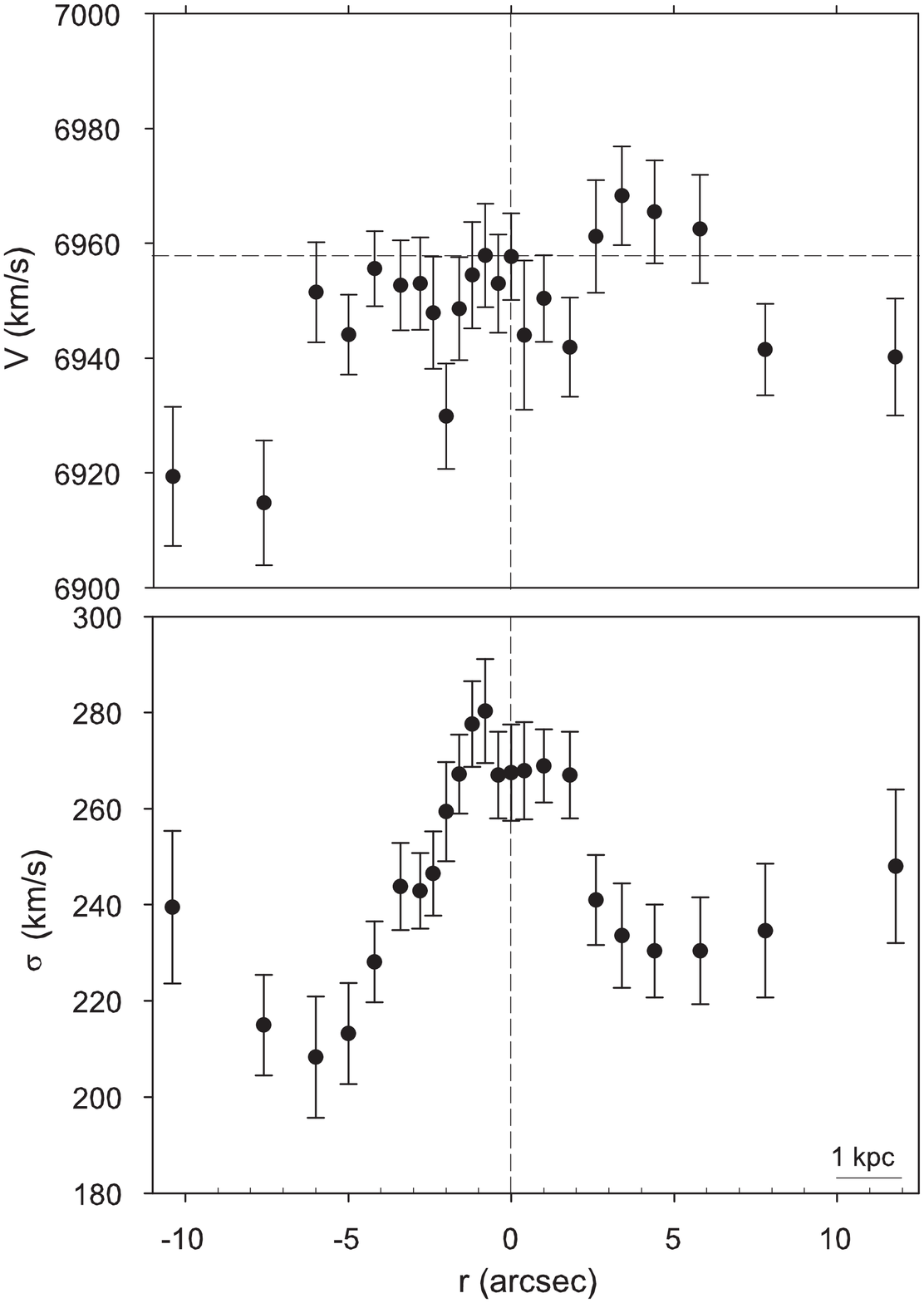}}}
   \mbox{\subfigure[NGC4889.]{\includegraphics[scale=0.23]{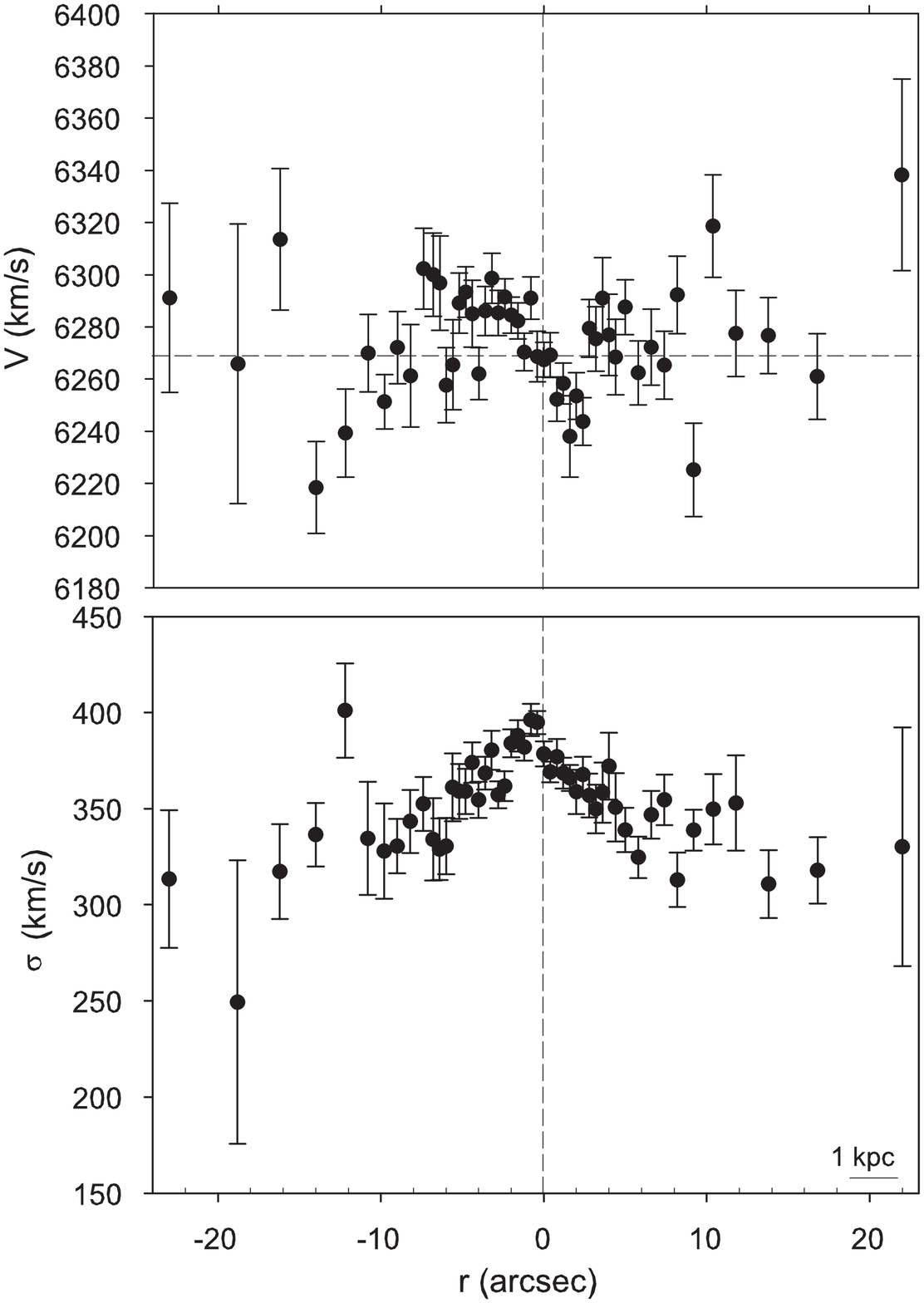}}\quad
         \subfigure[NGC4946.]{\includegraphics[scale=0.23]{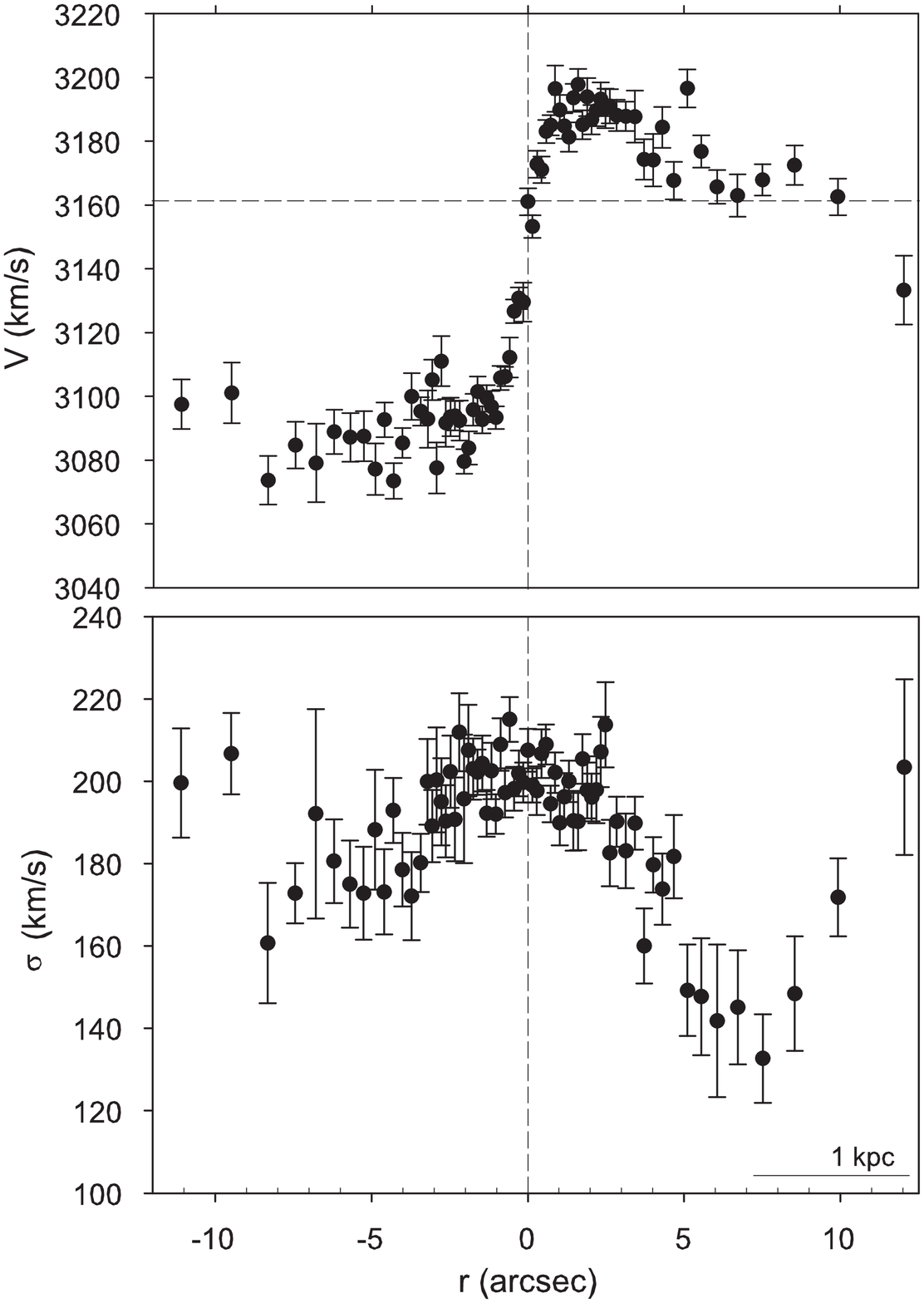}}\quad
         \subfigure[NGC6034.]{\includegraphics[scale=0.23]{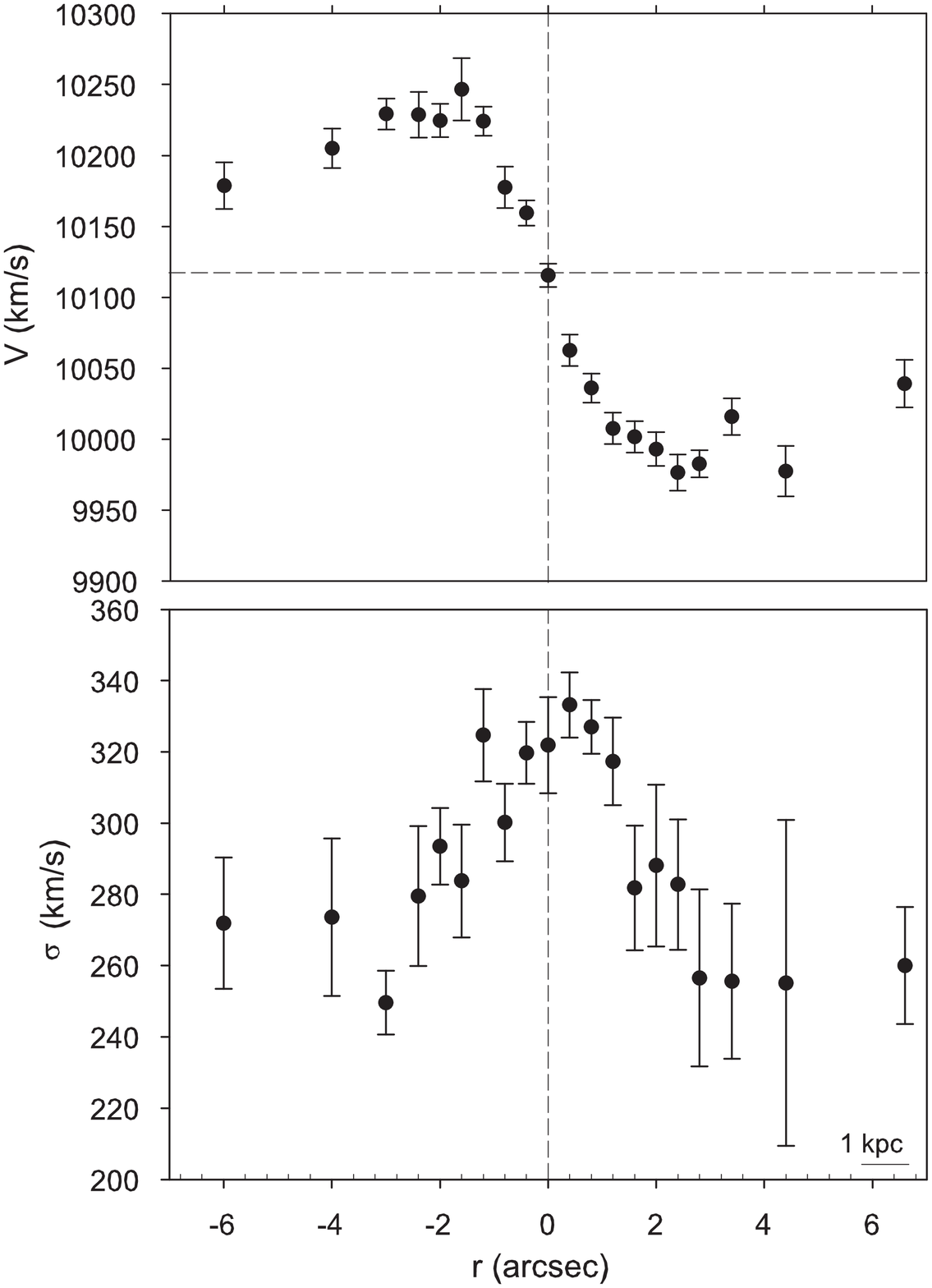}}}
   \mbox{\subfigure[NGC6047.]{\includegraphics[scale=0.23]{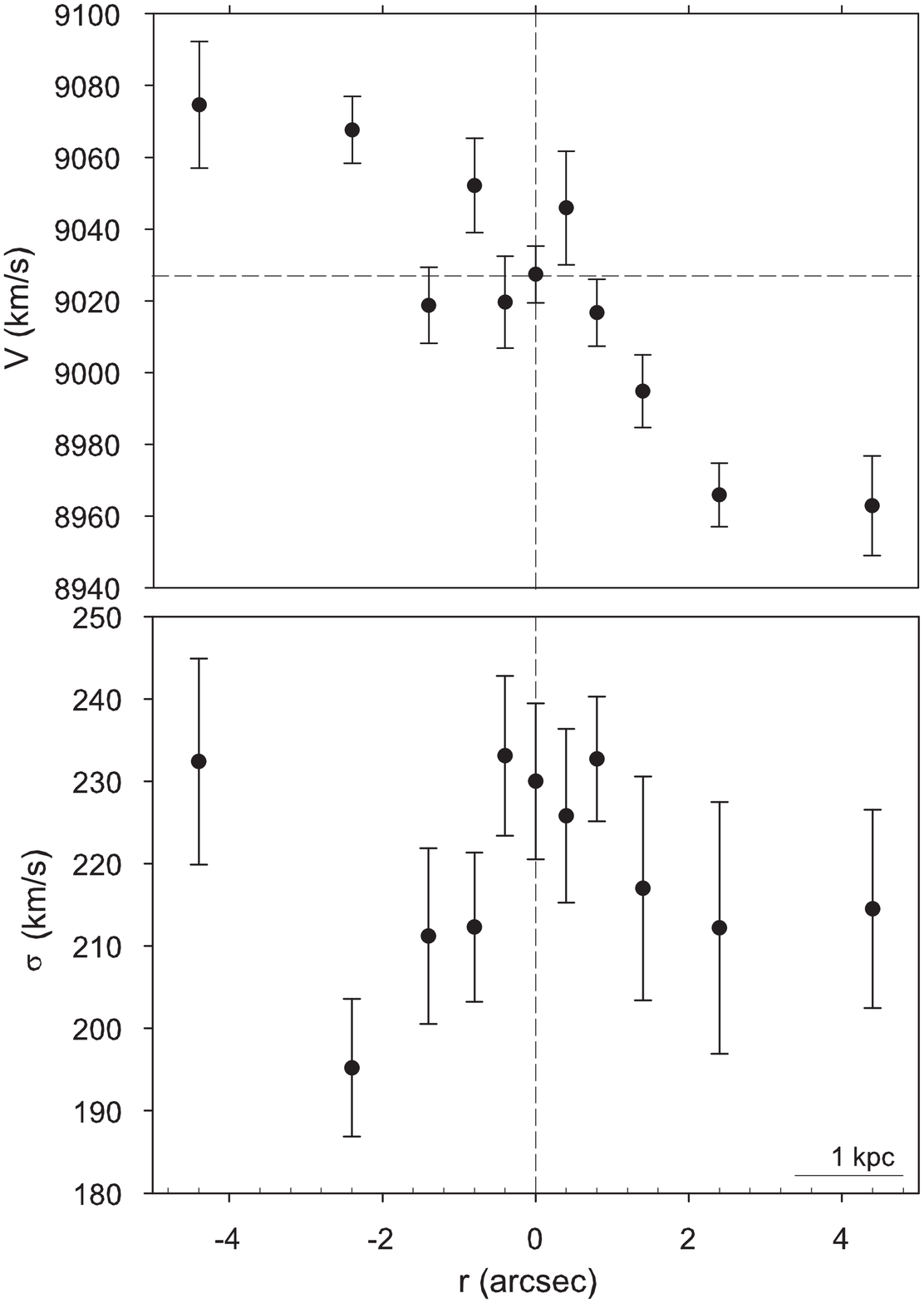}}\quad
         \subfigure[NGC6086.]{\includegraphics[scale=0.23]{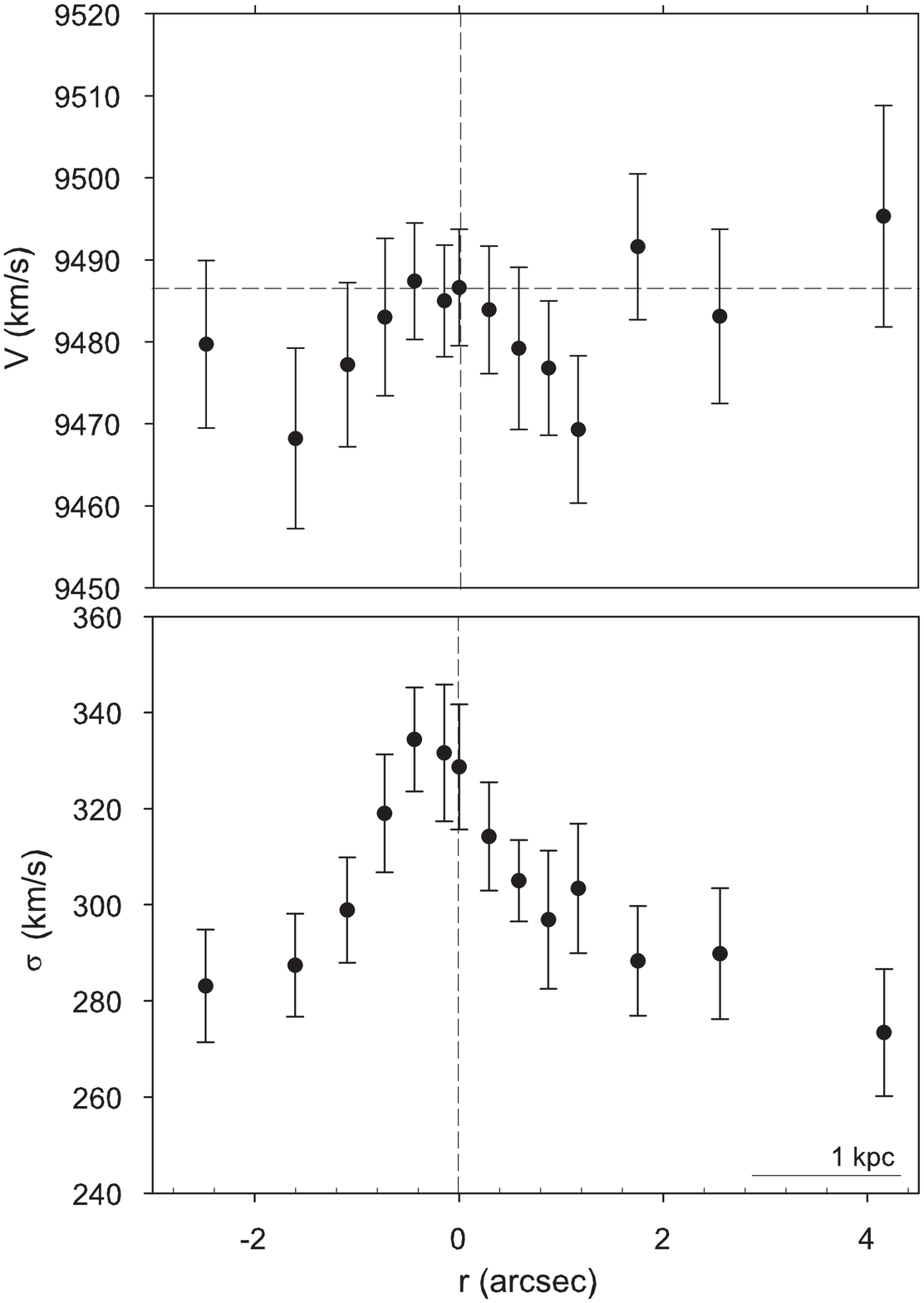}}\quad
         \subfigure[NGC6160.]{\includegraphics[scale=0.23]{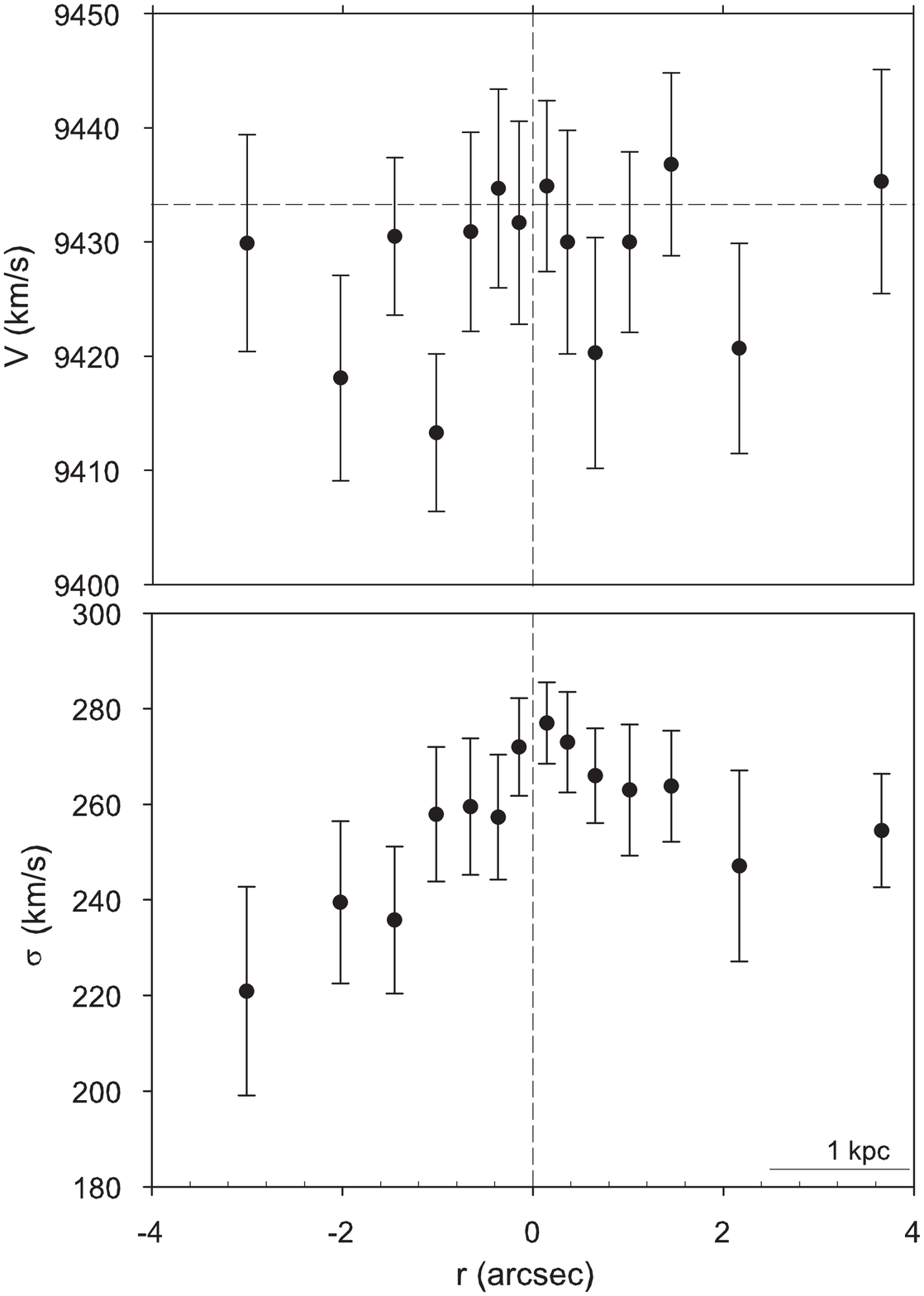}}}
\caption{Radial profiles of velocities (V) and velocity dispersion ($\sigma$). The dashed lines are as in Figure \ref{fig:NGC3842}.}
\label{fig:IC1633}
\end{figure*}

\begin{figure*}
   \centering
   \mbox{\subfigure[NGC6166.]{\includegraphics[scale=0.23]{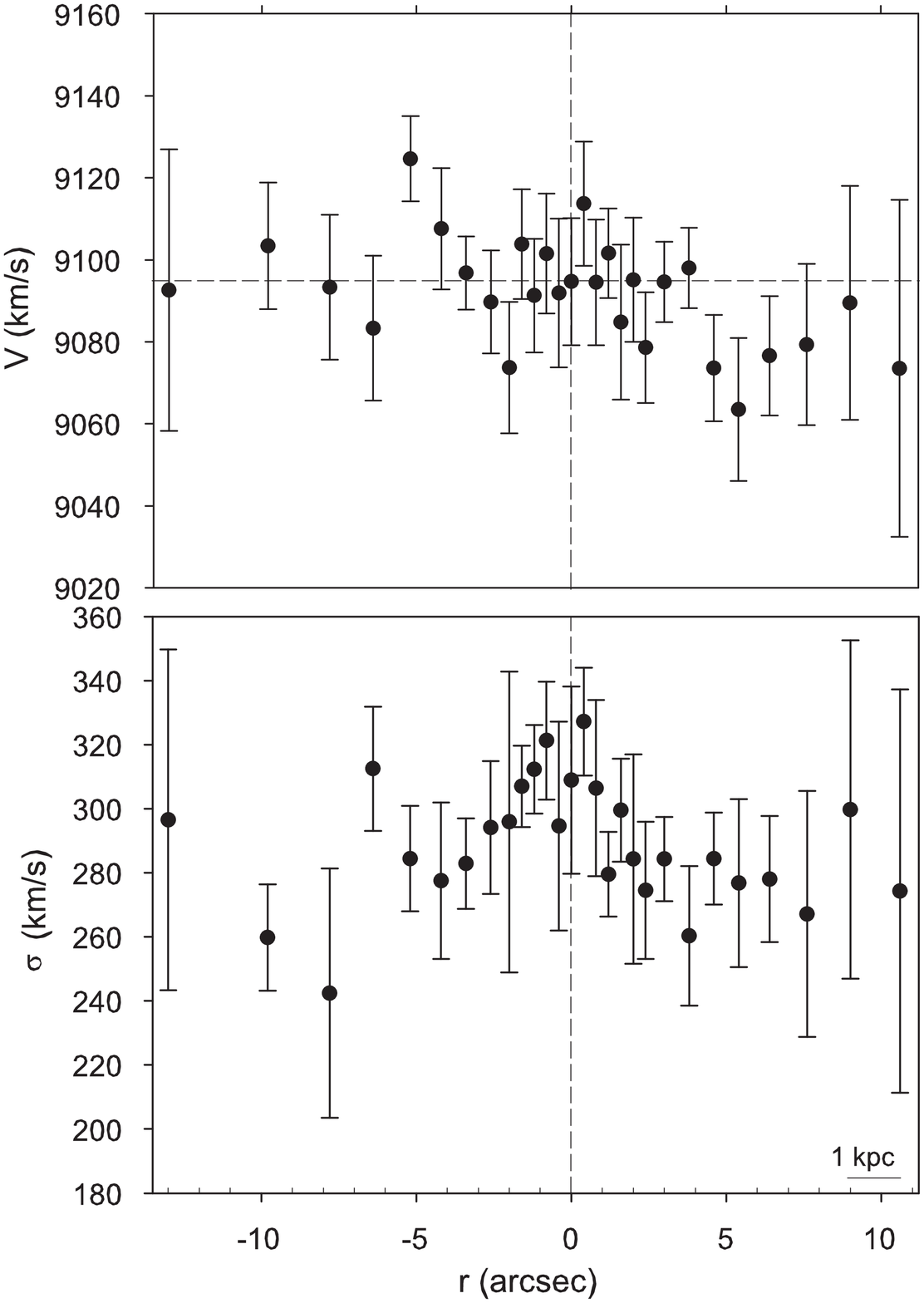}}\quad
         \subfigure[NGC6173.]{\includegraphics[scale=0.23]{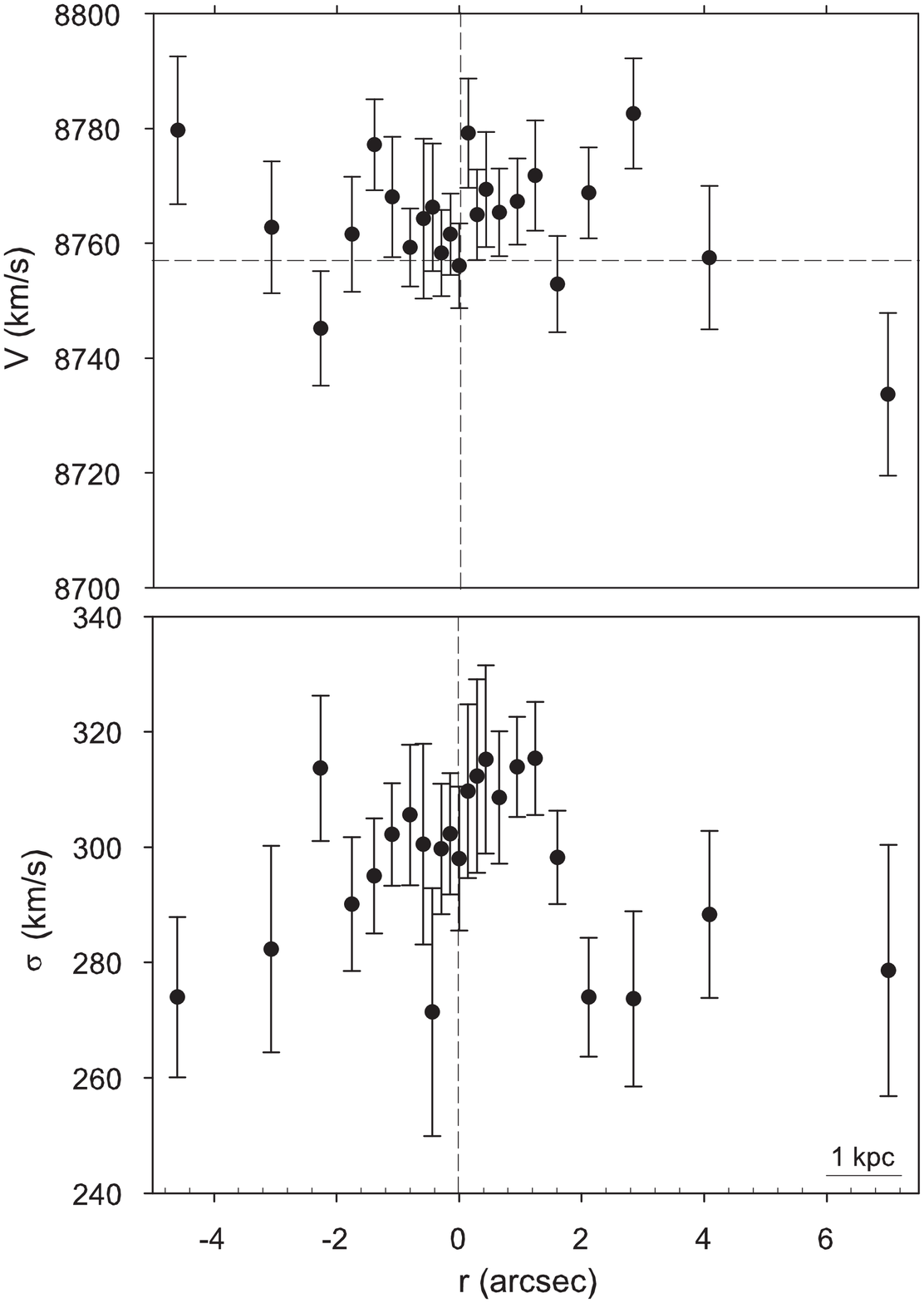}}\quad
         \subfigure[NGC6269.]{\includegraphics[scale=0.23]{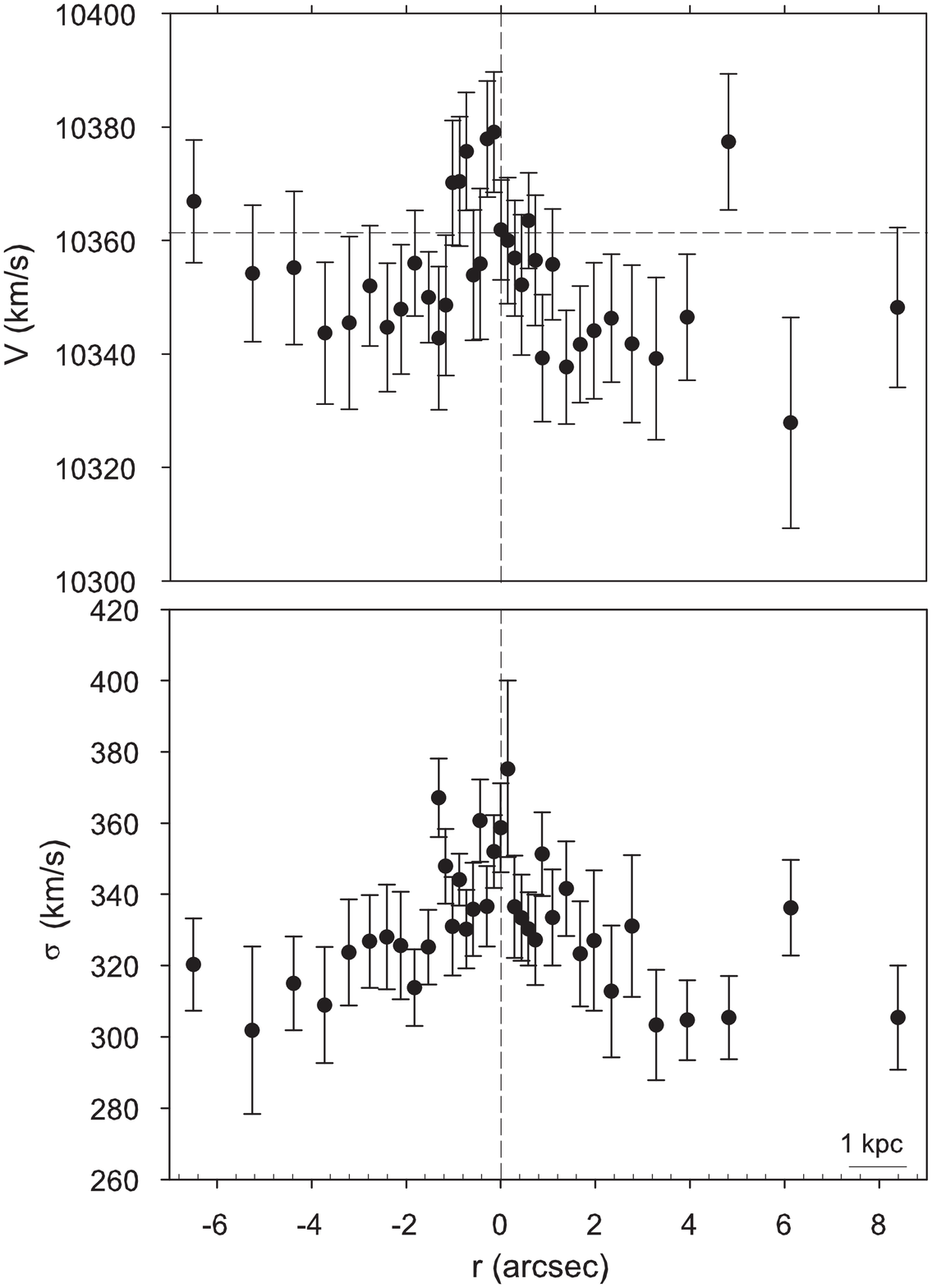}}}
   \mbox{\subfigure[NGC7012.]{\includegraphics[scale=0.23]{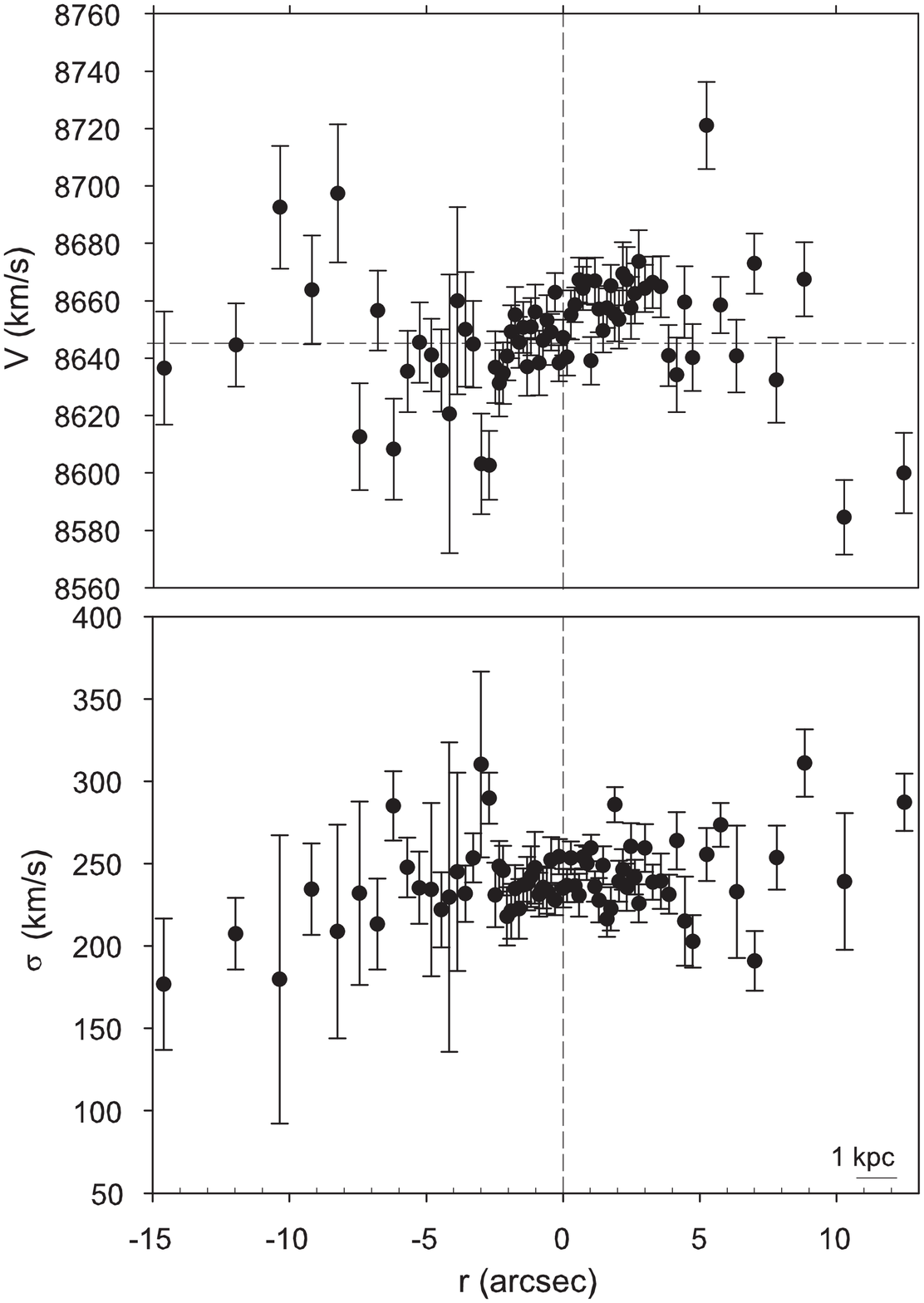}}\quad
         \subfigure[NGC7597.]{\includegraphics[scale=0.23]{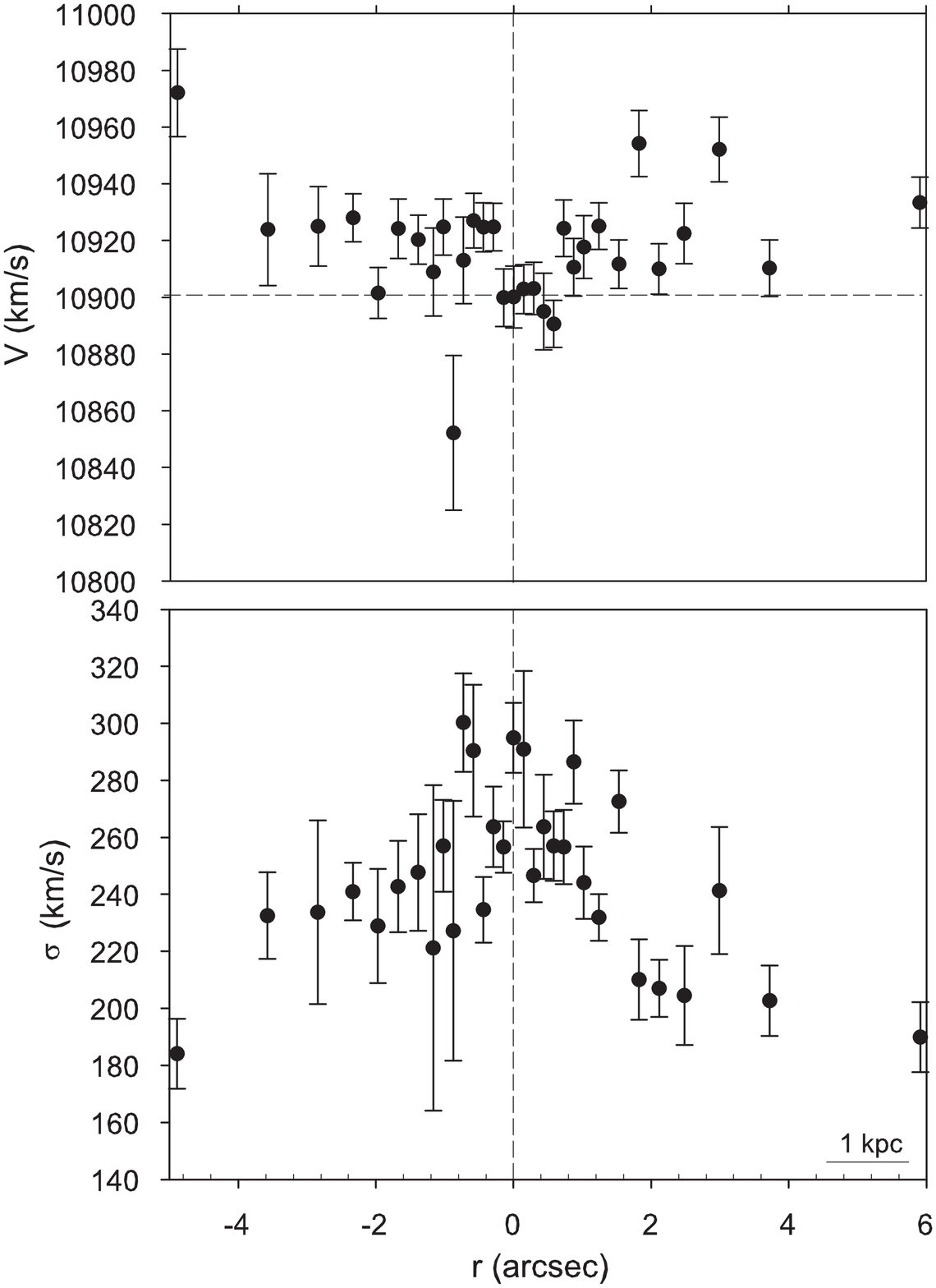}}\quad
         \subfigure[NGC7647.]{\includegraphics[scale=0.23]{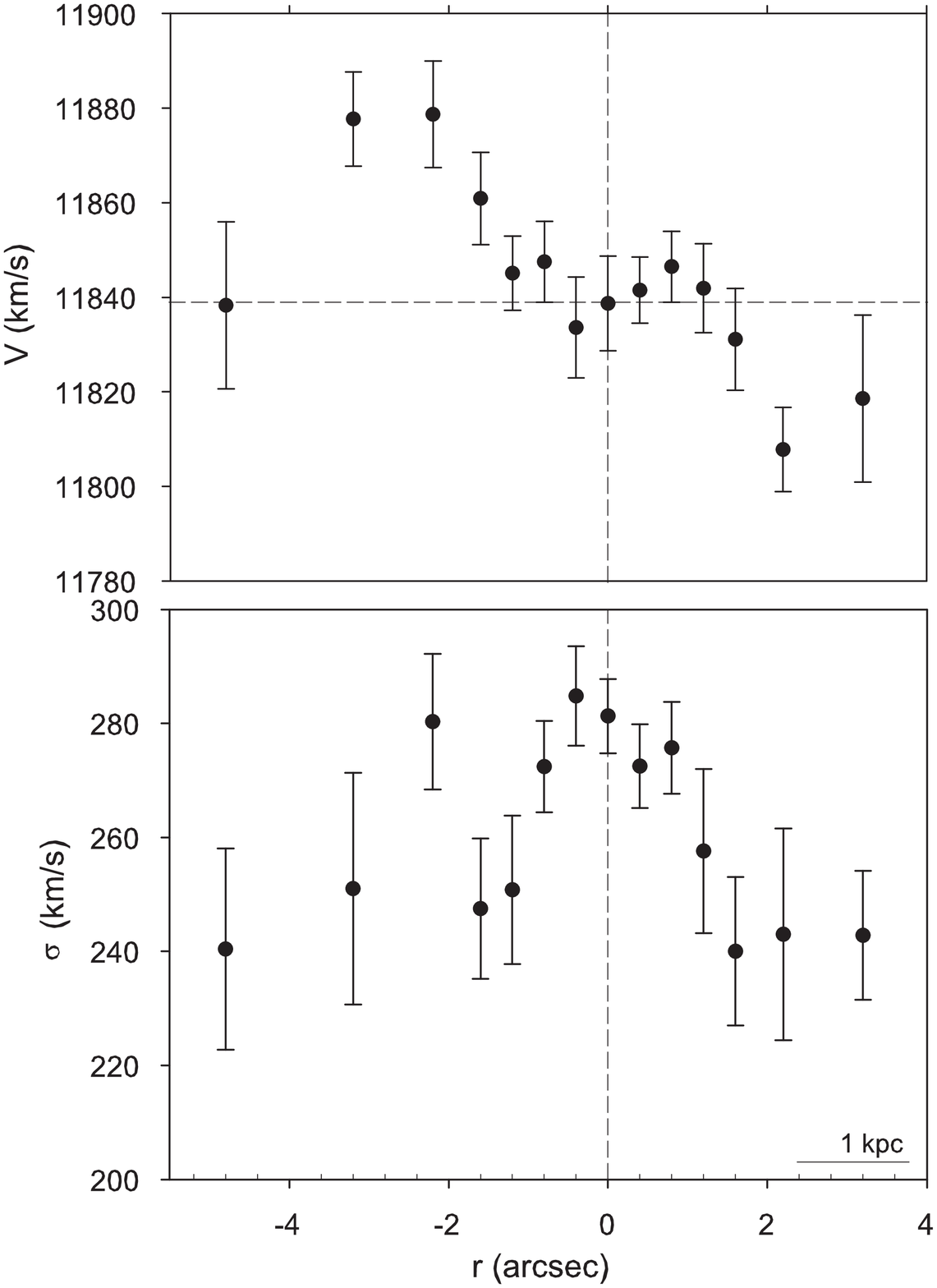}}}
   \mbox{\subfigure[NGC7649.]{\includegraphics[scale=0.23]{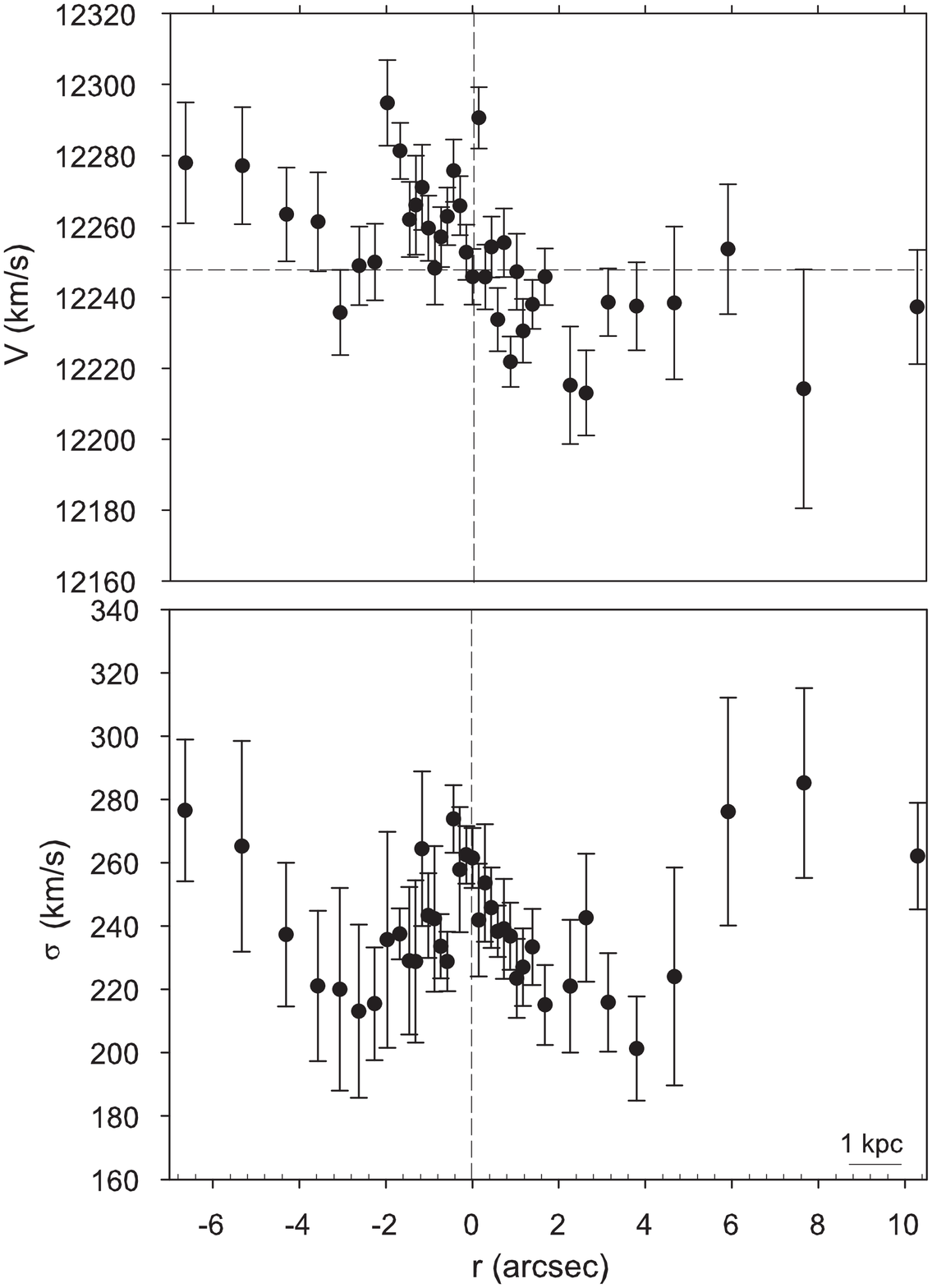}}\quad
         \subfigure[NGC7720.]{\includegraphics[scale=0.23]{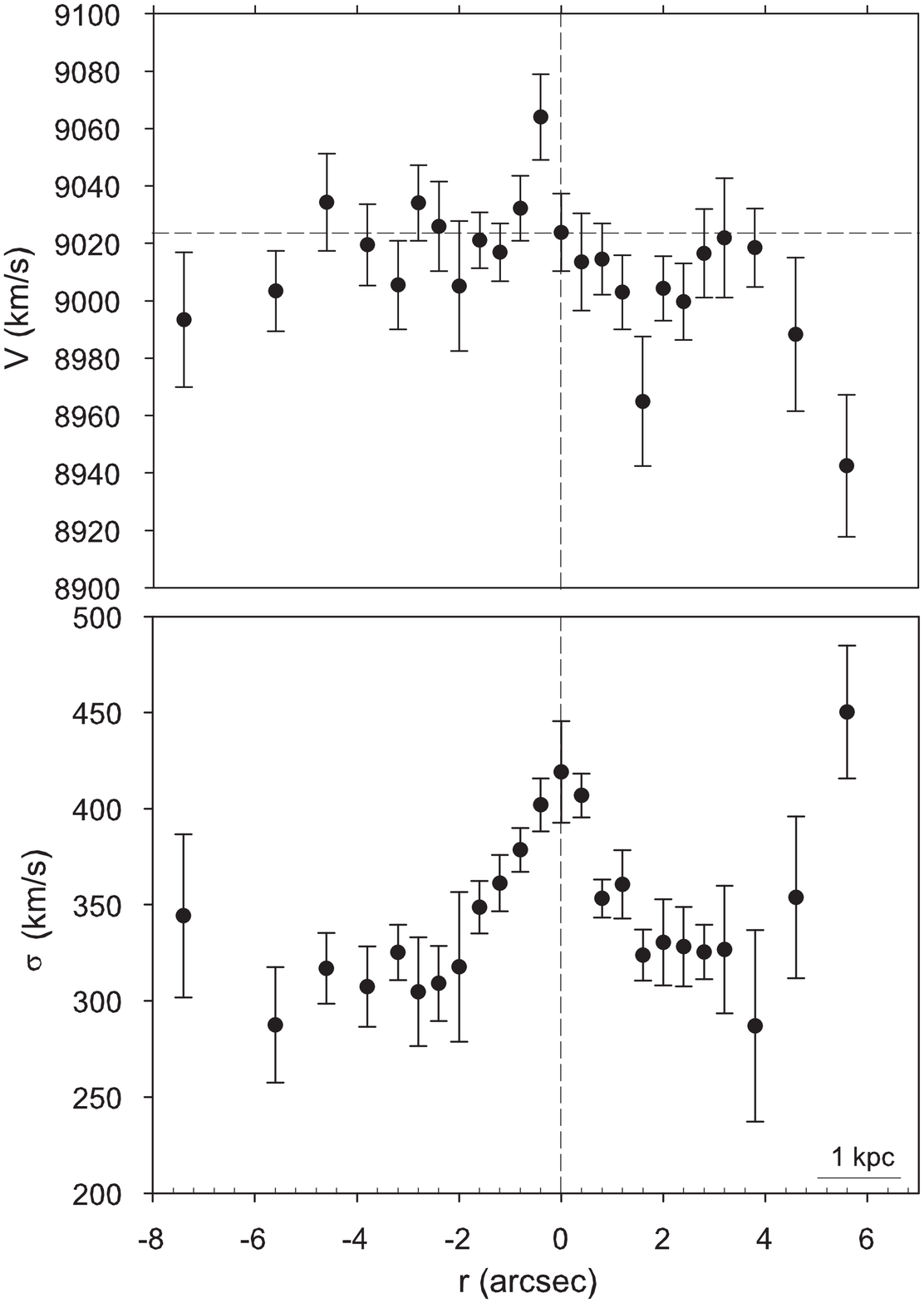}}\quad
         \subfigure[NGC7768.]{\includegraphics[scale=0.23]{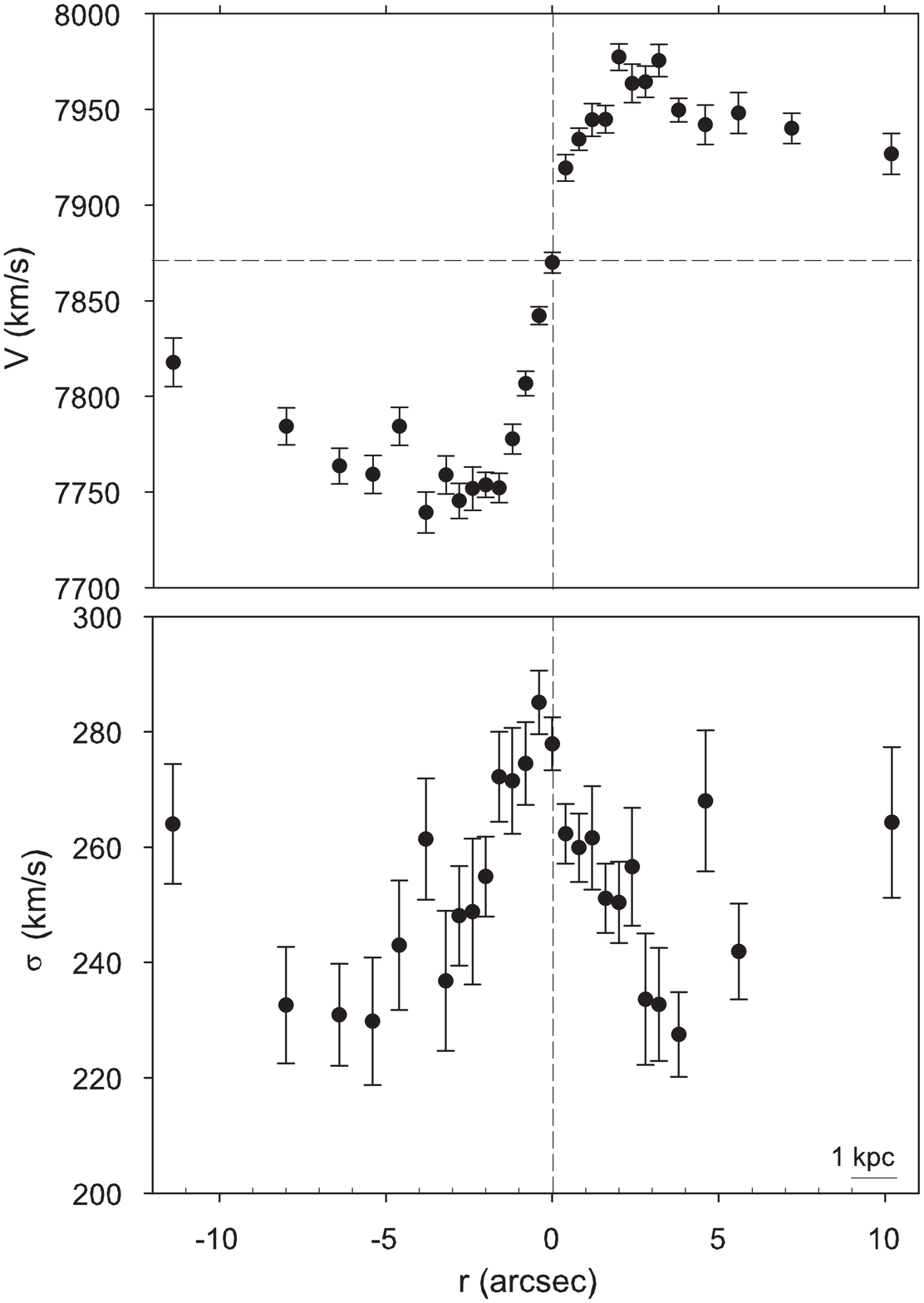}}}
\caption{Radial profiles of velocities (V) and velocity dispersion ($\sigma$). The dashed lines are as in Figure \ref{fig:NGC3842}.}
\label{fig:NGC4839}
\end{figure*}

\begin{figure*}
   \centering
   \mbox{\subfigure[PGC026269.]{\includegraphics[scale=0.23]{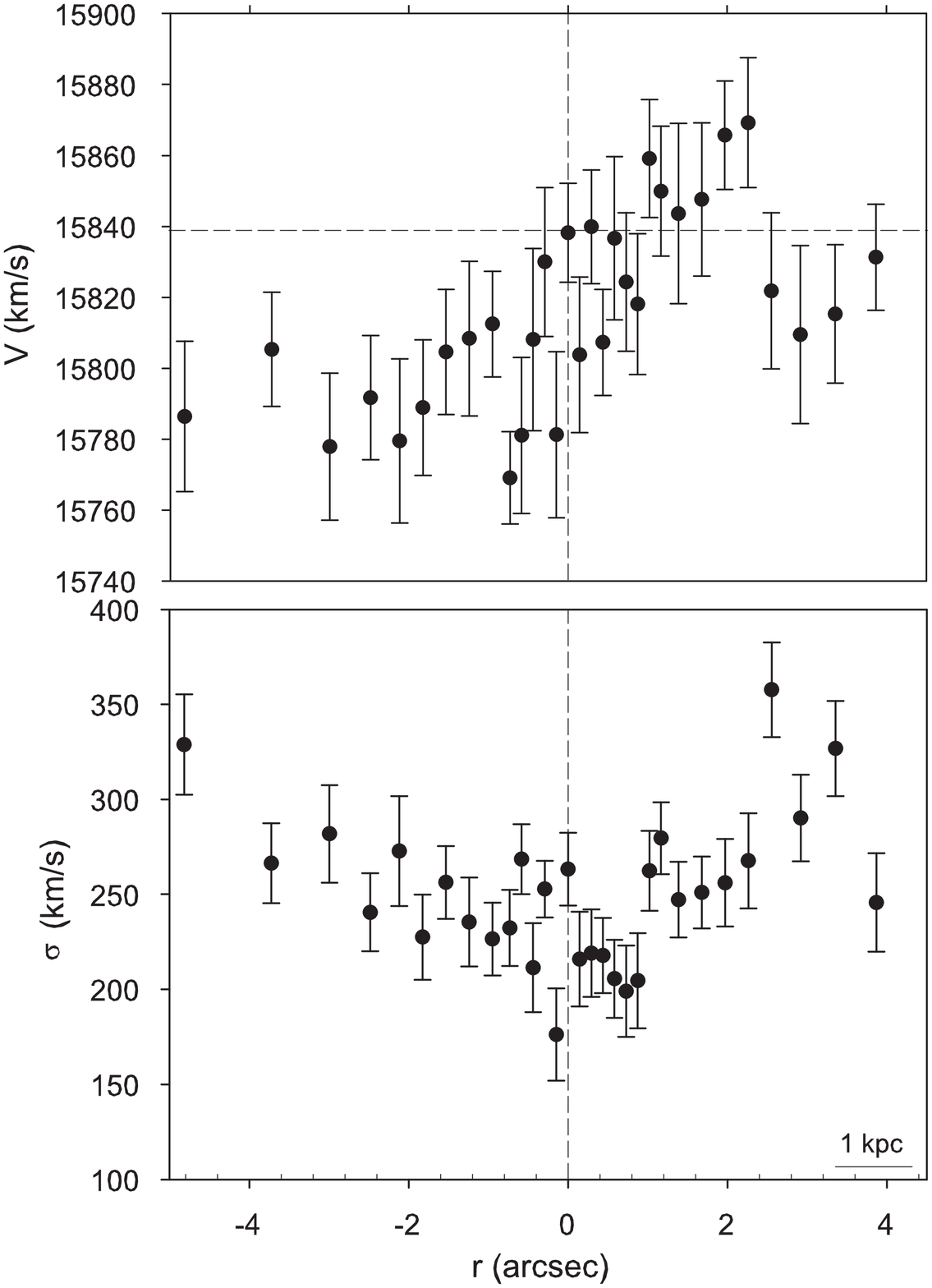}}\quad
         \subfigure[PGC044257.]{\includegraphics[scale=0.23]{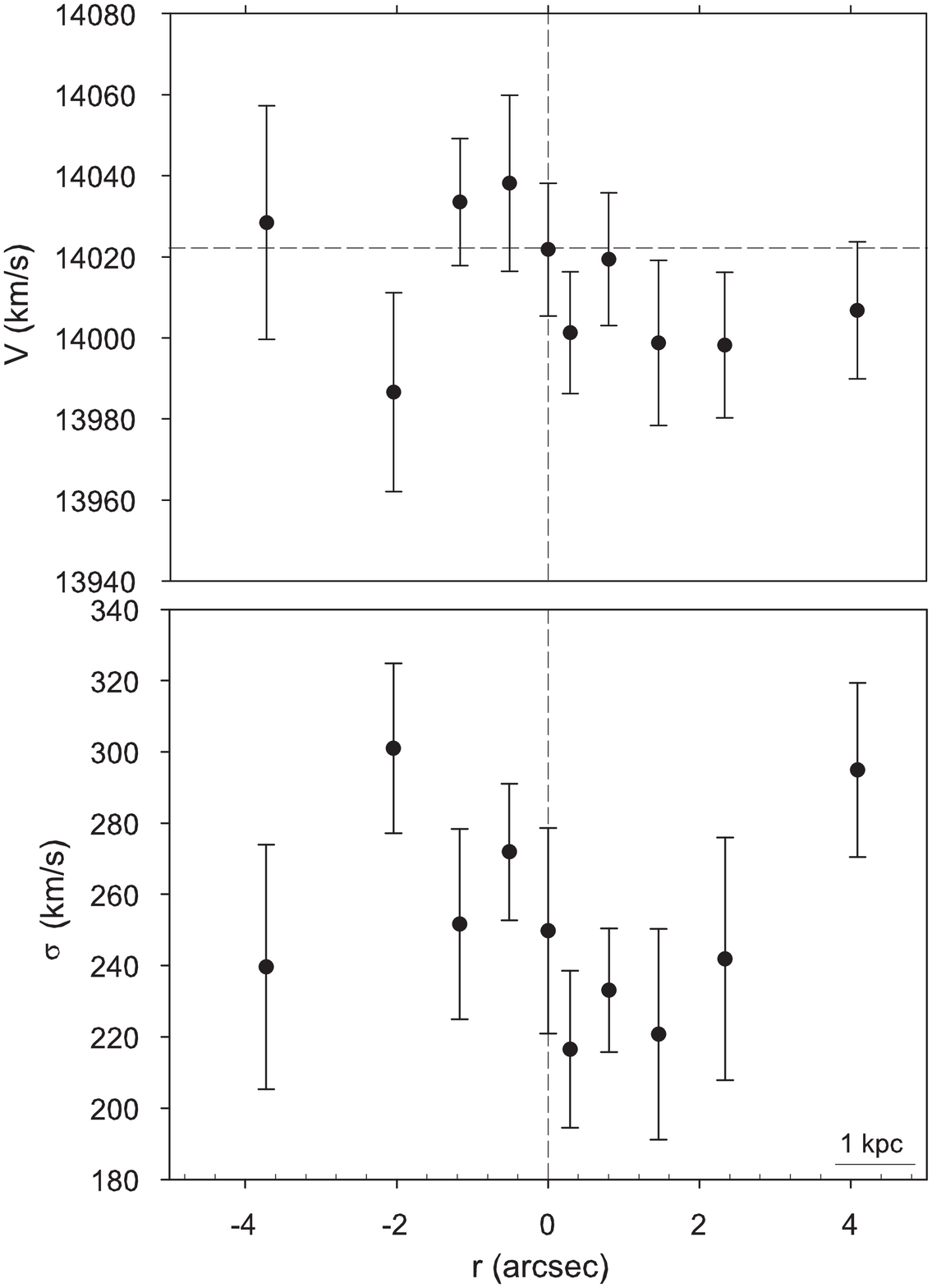}}\quad
         \subfigure[PGC071807.]{\includegraphics[scale=0.23]{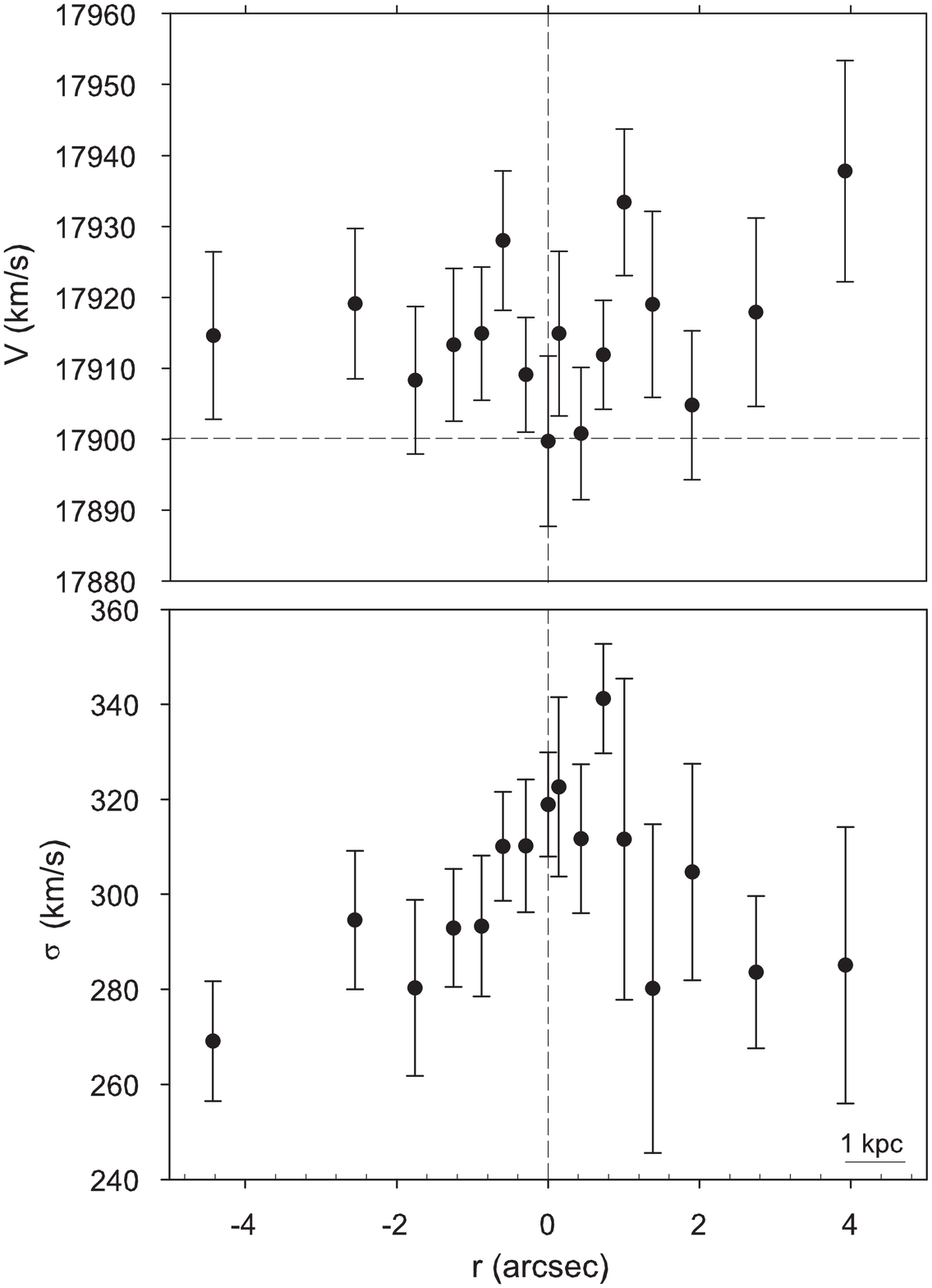}}}
   \mbox{\subfigure[PGC072804.]{\includegraphics[scale=0.23]{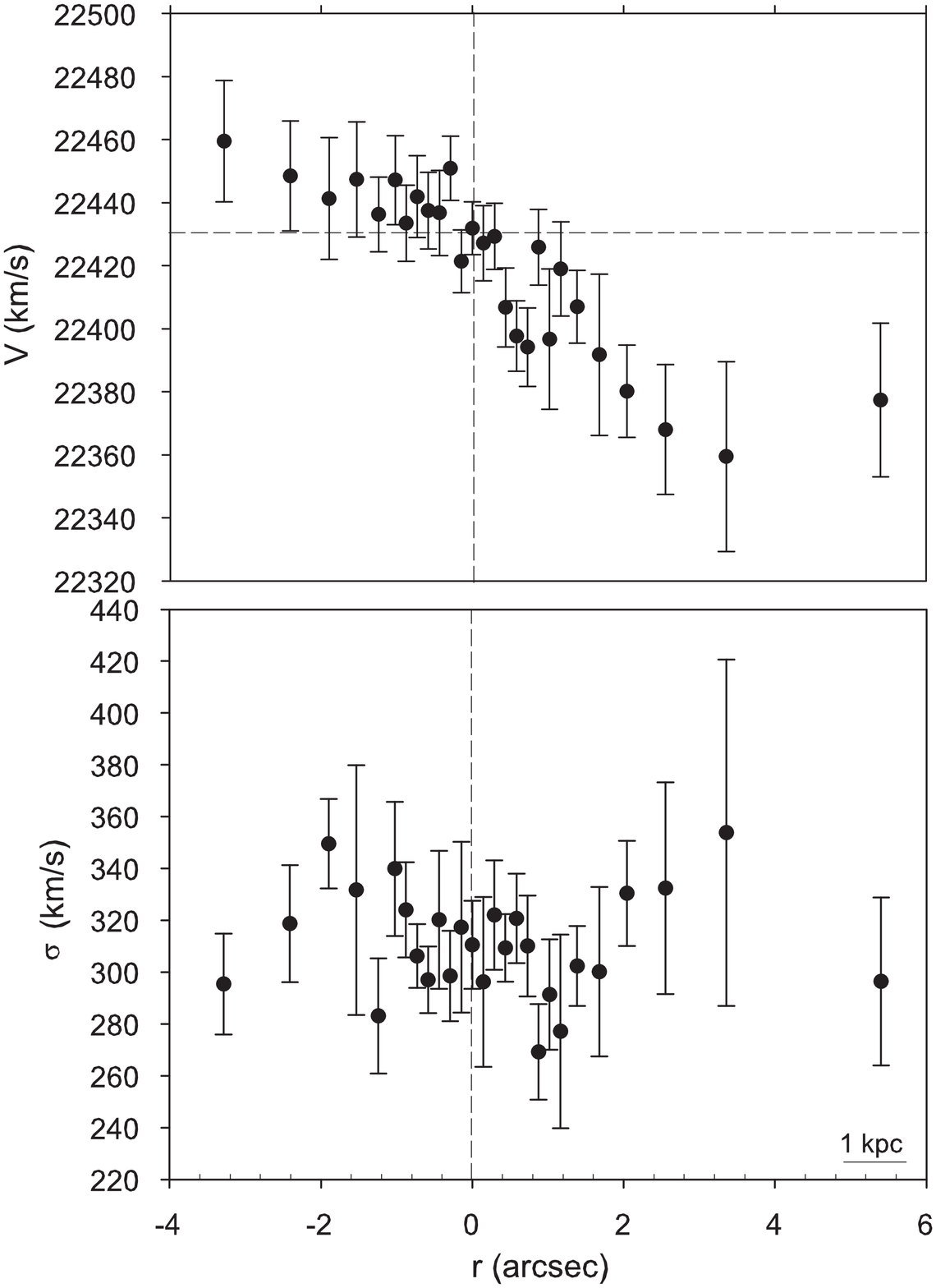}}\quad
         \subfigure[UGC02232.]{\includegraphics[scale=0.23]{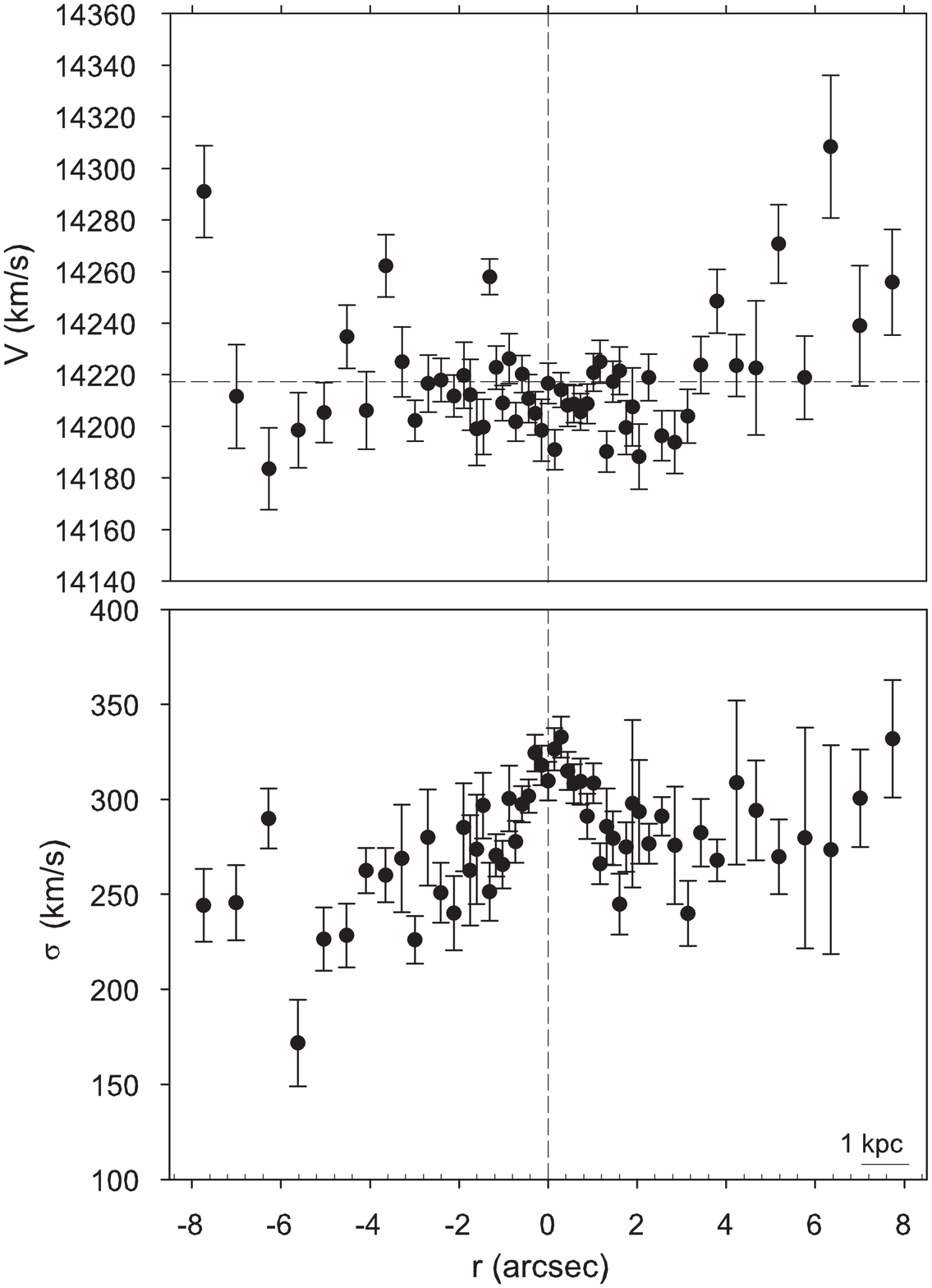}}\quad
         \subfigure[UGC05515.]{\includegraphics[scale=0.23]{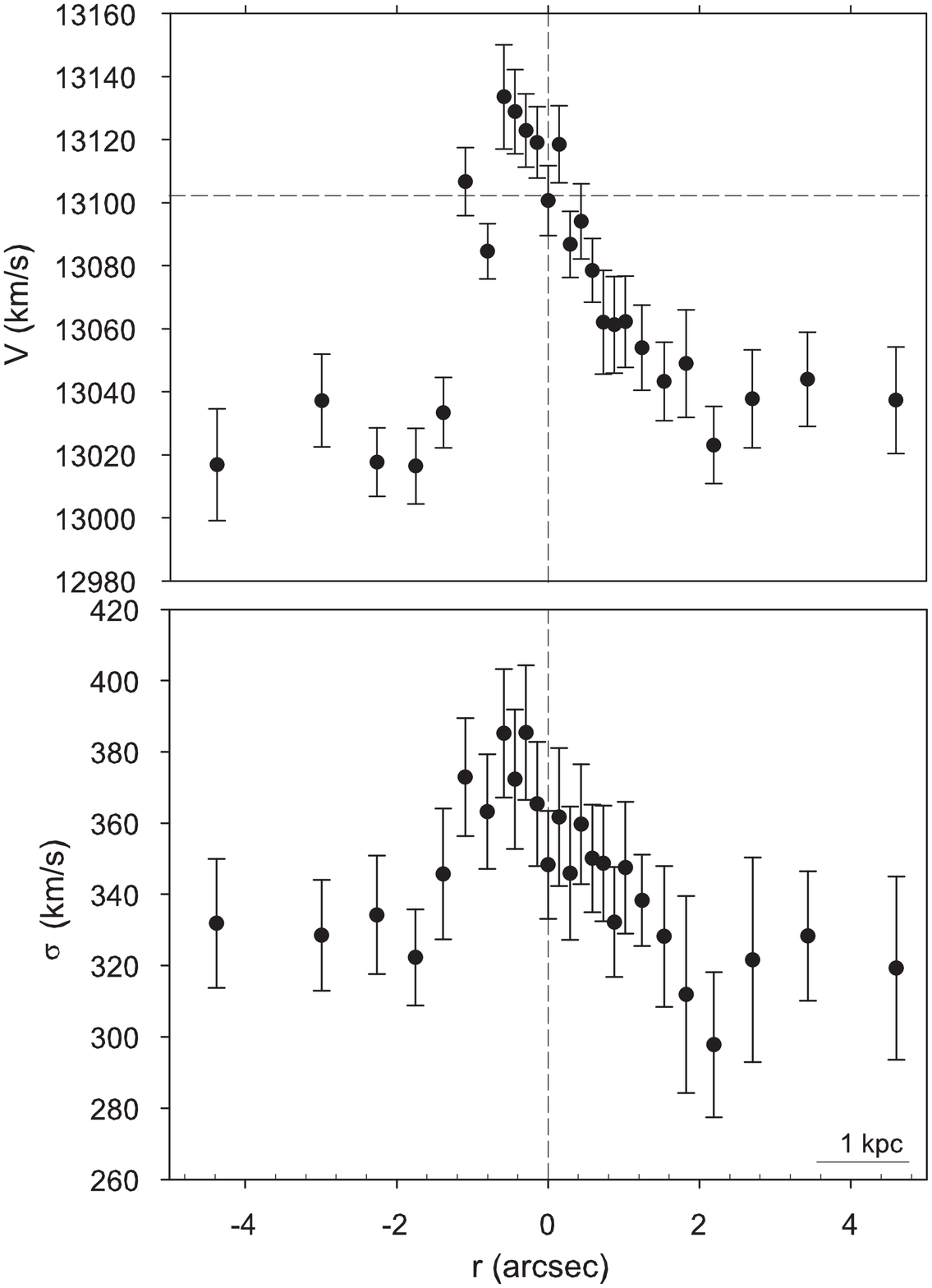}}}
   \mbox{\subfigure[UGC10143.]{\includegraphics[scale=0.23]{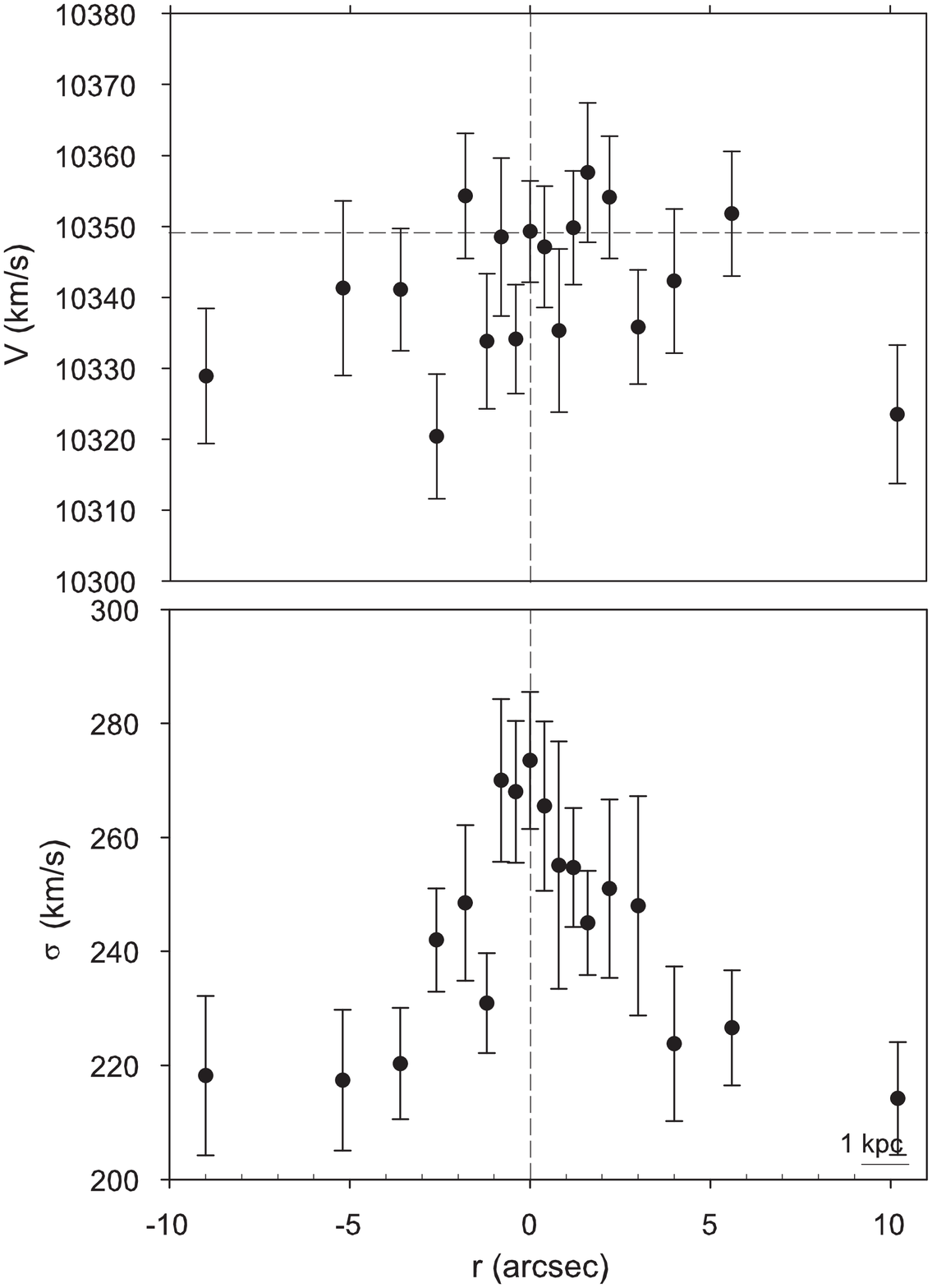}}}
\caption{Radial profiles of velocities (V) and velocity dispersion ($\sigma$). The dashed lines are as in Figure \ref{fig:NGC3842}.}
\label{fig:PGC026269}
\end{figure*}

\subsection{Notes on individual objects}
\label{Individual}
This section describes the individual objects and their radial kinematic profiles as shown in Figures \ref{fig:NGC3842}, \ref{fig:ESO146}, \ref{fig:IC1633}, \ref{fig:NGC4839} and \ref{fig:PGC026269}. We refer to results published prior to this study, as well as the results obtained here. The radial kinematic profiles of NGC4839, NGC4889, NGC6166 and IC1101 are compared to previous measurements from the literature in Appendix \ref{Appendix}. 

ESO146-028: 
The radial velocity of this galaxy is mostly flat and the velocity dispersion profile is decreasing outwards.

ESO303-005: 
The radial kinematic profiles of this galaxy show no significant rotation or substructure. 

ESO346-003:
The radial velocity profile shows clear rotation of the order of 51 km s$^{-1}$.

ESO349-010:
HST imaging of this galaxy was published by Laine et al.\ (2003). The radial velocity profile of this galaxy shows no significant rotation and the velocity dispersion profile has a positive gradient.

ESO444-046:
No clear rotation or velocity substructure is visible in the kinematic profiles.

ESO488-027:
The kinematic profiles show a KDC in the centre of this galaxy.

ESO552-020:
The radial velocity profile is mostly flat and the velocity dispersion profile shows a small dip south-west of the centre of the galaxy which might indicate a colder component.

GSC555700266:
Slow rotation but no clear substructure can be seen in the kinematic profiles.

IC1101:
This galaxy is one of the few known BCGs that were previously found to have a rising velocity dispersion profile (Fisher et al.\ 1995a). This result cannot be confirmed here since the derived velocity dispersion profile is not measured to the same radius as in the above mentioned study. No significant rotation is detected in the radial velocity profile.

IC1633:
The surface brightness profile was published by Schombert (1986), and HST imaging by Laine et al.\ (2003). The radial velocity profile reveals some evidence for substructure at the centre of this galaxy. 

IC4765:
The radial velocity profile shows rotation in the centre of the galaxy which is possibly rotational substructure.

IC5358:
Both the radial kinematic profiles show the clear presence of a KDC in this galaxy.

LEDA094683:
No significant rotation but some evidence of substructure is visible from the radial velocity profile.

MCG-02-12-039:
This BCG is a weak line-emitter. The surface brightness profile was published by Schombert (1986), and HST imaging by Laine et al.\ (2003). A KDC is detected in the radial kinematic profiles.

NGC1399:
The surface brightness profile was published by Schombert (1986). The velocity dispersion in the central region of this galaxy was found to be 388 km s$^{-1}$ decreasing to 200 km s$^{-1}$ at 10 arcseconds from the centre by Longo et al.\ (1994). The central radial velocity dispersion of this galaxy was measured in this study as 371 $\pm$ 3 km s$^{-1}$. Thus, the central velocity dispersion measured in this study corresponds to that measured by Longo et al.\ (1994), and the steep decreasing velocity dispersion profile is confirmed as well. This galaxy has a flat radial velocity profile. 

NGC1713:
The radial velocity profile of this galaxy reveals a definite KDC in the centre of this galaxy.

NGC2832:
This galaxy is thought to be tidally interacting with its companion, NGC2831. The surface brightness profile were published by Schombert (1986) and Jord\'an et al.\ (2004), and HST imaging by Laine et al.\ (2003). The radial velocity profile shows the presence of a KDC in the centre of this galaxy. 

NGC3311:
HST imaging was published by Laine et al.\ (2003). No clear rotation or velocity substructure can be seen in the radial kinematic profiles, but the velocity dispersion profile has a positive gradient.

NGC3842:
This galaxy belongs to a cluster which forms part of the Coma supercluster. The surface brightness profile was published by Schombert (1986), and HST imaging by Laine et al.\ (2003). The radial velocity profile of this galaxy reveals the presence of a kinematically decoupled core (KDC) in the centre.

NGC4839:
The location of this cD galaxy far from the centre of the Coma cluster is unusual. Surface brightness profiles were published by Schombert (1986), Oemler (1976) as well as Jord\'an et al.\ (2004), who all confirm the presence of a very prominent cD envelope. Rotation of the order 44 km s$^{-1}$ and a KDC in the centre of the galaxy are detected.

NGC4874:
This galaxy has a large, extended envelope. It is the second brightest galaxy of the famous pair of cDs at the centre of the Coma cluster. The surface brightness profile was published by Peletier et al.\ (1990). No significant rotation or velocity substructure is detected for this galaxy.

NGC4889:
A large cD galaxy with a very extended envelope. This is the brightest galaxy of the Coma cluster. The surface brightness profile was published by Peletier et al.\ (1990), and HST imaging by Laine et al.\ (2003). The radial velocity profile clearly shows a KDC.

NGC4946:
This is an ordinary elliptical galaxy. The radial velocity profile of this galaxy reveals rotation of the order of 62 km s$^{-1}$ along the major axis.

NGC6034:
The surface brightness profile was published by Schombert (1986). The radial velocity profile of this galaxy shows significant rotation of the order of 134 $\pm$ 15 km s$^{-1}$ along the major axis.

NGC6047: 
The surface brightness profile was published by Schombert (1986), and Schombert (1987) classified it as an E/SO galaxy. The radial velocity profile of this galaxy shows rotation of 59 km s$^{-1}$ (with the slit placement 29 degrees from the major axis).

NGC6086:
The surface brightness profile was published by Schombert (1986), and HST imaging by Laine et al.\ (2003). Carter et al.\ (1999) found this galaxy to have a KDC. Even though some evidence of substructure can be seen, this core cannot be confirmed since the profiles measured here do not cover the same radial extend as those in Carter et al.\ (1999). 

NGC6160:
The radial velocity profile of this galaxy reveals no significant rotation or substructure.

NGC6166:
This is a classic multiple nucleus cD galaxy in a rich cluster. The surface brightness profile was published by Schombert (1986). Carter et al.\ (1999) found the velocity dispersion of this galaxy to increase from 325 km s$^{-1}$ at the centre to 450 km s$^{-1}$ at 35 arcsec along the major axis. They also found the galaxy to show modest major-axis rotation (45 km s$^{-1}$ at 40 arcsec). The profiles derived here do not extend out to the radii necessary to confirm the positive velocity dispersion slope found by Carter et al.\ (1999). Rotation of the order of 31 km s$^{-1}$ is found in the centre of the galaxy although this is not a very clear rotation curve.

NGC6173:
The surface brightness profile was published by Schombert (1986), and HST imaging by Laine et al.\ (2003). The derived radial kinematic profiles of this galaxy show no significant rotation or substructure. 

NGC6269:
Surface brightness profiles were published by Schombert (1986) and by Malumuth $\&$ Kirshner (1985). The radial velocity profile shows some velocity substructure in the centre of this galaxy. This galaxy was observed at an angle of 46 degrees away from the major axis.

NGC7012:
The radial velocity profile shows a small amount of rotation in the centre which might be a KDC.

NGC7597:
The surface brightness profile was published by Schombert (1986). The radial kinematic profiles show possible evidence for a KDC.

NGC7647: 
This is a cD galaxy with a red envelope and a large number of small companions (Vitores et al.\ 1996). The surface brightness profile was published by Schombert (1986), and HST imaging by Laine et al.\ (2003). The radial velocity profile of this galaxy shows a clear KDC in the centre.

NGC7649:
The surface brightness profile was published by Schombert (1986), and HST imaging by Laine et al.\ (2003). A KDC is clearly visible in the centre of this galaxy.

NGC7720: 
This is a multiple nuclei galaxy and has a massive close companion galaxy, also an elliptical. The surface brightness profile was published by Lauer (1988), and HST imaging by Laine et al.\ (2003). The radial velocity profile reveals no significant rotation but there might be some substructure in this galaxy.

NGC7768:
Surface brightness profiles were published by Schombert (1986), Malumuth $\&$ Kirshner (1985) and Jord\'an et al.\ (2004) and HST imaging was published by Laine et al.\ (2003). The radial velocity profile of this galaxy shows significant rotation of the order of 114 $\pm$ 11 km s$^{-1}$ along the major axis. This agrees with the 101 $\pm$ 5 km s$^{-1}$ and the 99 km s$^{-1}$ rotation found previously by Fisher et al.\ (1995a) and Prugniel $\&$ Simien (1996), respectively.

PGC026269:
Also known as Hydra A. The radial kinematic profiles of this galaxy reveal rotation of the order of 51 km s$^{-1}$ along the major axis, as well as an increasing velocity dispersion profile.

PGC044257:
The data indicate that the velocity dispersion might increase with radius, although the large scatter and the small number of bins make it difficult to assess the slope.

PGC071807: 
No significant rotation or substructure can be seen in this galaxy.

PGC072804:
The surface brightness profile was published by Malumuth $\&$ Kirshner (1985). The radial velocity profile of this galaxy shows rotation of the order of 50 km s$^{-1}$ along the major axis.

UGC02232:
This galaxy has an extended envelope and a large number of objects in the vicinity. HST imaging was published by Laine et al.\ (2003). No significant rotation or clear substructure is detected.

UGC05515:
The surface brightness profile was published by Schombert (1986). The radial velocity profile shows a KDC in the centre of this galaxy.

UGC10143:
This galaxy has a chain of companions. The surface brightness profile was published by Schombert (1986), and HST imaging by Laine et al.\ (2003). No significant rotation or substructure can be inferred from the radial kinematic profiles of this galaxy.

\section{Discussion}
\label{kinematicanalysis}

(i) \textit{Five out of 41 BCGs (ESO349-010, ESO444-046, ESO552-020, NGC3311, PGC026269) were found to have a positive velocity dispersion gradient.} 

The radial kinematic studies done so far on early-type galaxies are mostly limited to normal ellipticals, for which flat or decreasing velocity dispersion profiles are found. The majority of the results previously obtained for very small samples of BCGs are similar to those of normal ellipticals. Fisher et al.\ (1995a) found one galaxy (IC1101) in their sample of 13 BCGs with a positive velocity dispersion gradient. Carter et al.\ (1999) found one (NGC6166) of their sample of three BCGs to have a positive velocity dispersion gradient, although Fisher et al.\ (1995a) did not find it for this galaxy. Brough et al.\ (2007) found negative velocity dispersion gradients in five out of their sample of six brightest cluster and group galaxies (the other one had a zero velocity dispersion gradient). 

Both IC1101 and NGC6166 form part of the sample of BCGs studied here, but in both cases the measured velocity dispersion profiles do not reach the radius achieved in the much smaller samples in the above mentioned studies. Thus, the positive velocity dispersion gradient could not be confirmed for IC1101 or NGC6166 (see Figures \ref{fig:NGC6166kin} and \ref{fig:IC1101kin}). However, five other BCGs were found to have a positive velocity dispersion gradient, though admittedly the slope is marginally positive in some cases. If these positive velocity dispersion gradients are not caused by systematic errors in the various data reduction or velocity dispersion measurements by different authors, then they imply a rising mass-to-light ratio.

\medskip (ii) \textit{At least 12 BCGs (NGC3842, NGC4889, NGC7647, ESO488-027, IC5358, MCG-02-12-039, NGC1713, NGC2832, NGC4839, NGC6269, NGC7649 and UGC05515) show clear velocity substructure in their profiles.}

From studies of elliptical galaxies in high density environments (e.g. Koprolin $\&$ Zeilinger 2000), the incidence of KDCs is observed to be about 33 per cent, rising to 50 per cent when projection effects are considered. Hau $\&$ Forbes (2006) finds KDCs in 40 per cent of their isolated elliptical galaxies. For BCGs, we have found at least 12 of the 41 BCGs show clear velocity substructure, amounting to 29 per cent of the sample (intermediate and minor axis data included).

KDCs can be the result of a merger event (Koprolin $\&$ Zeilinger 2000), but can also occur when the galaxy is triaxial and supports different orbital types in the core and main body (Statler 1991). The fact that the incidence of KDCs in BCGs compares with that found for normal elliptical galaxies in high density environments suggests that the two classes share the same fraction of galaxies with triaxial shapes.

\medskip (iii) \textit{NGC6034 and NGC7768 possess significant rotation (134 and 114 km s$^{-1}$ respectively) along the major axis. Several other BCGs show rotation that is $>$ 40 km s$^{-1}$ and more than three times the standard error: ESO346-003, GSC555700266 and NGC4839 (major axis spectra); ESO488-027, IC5358 and UGC05515 (intermediate axis spectra), as do the two elliptical galaxies NGC4946 (major axis spectra) and NGC6047 (intermediate axis spectra).}

Carter et al.\ (1999) found small rotation along the major axis at large radii (30 -- 40 arcsec) for their sample of three BCGs, which is consistent with the nearly complete lack of rotation found near the centres of the sample of 13 BCGs by Fisher et al.\ (1995a). According to Fisher at al.\ (1995a), the lack of rotation found in samples of BCGs are in agreement with the expectation of declining importance of rotation with increasing luminosity for elliptical galaxies. The lack of rotation is also compatible with the idea that these objects formed through dissipationless mergers (Boylan-Kolchin et al.\ 2006). The remnants left by mergers with or without dissipation are expected to differ in their kinematical structure. In a merger which involves gas-rich galaxies, the gas will form a disk. After the gas has been removed from the system at the end of the merger (through ejection and converted into stars), the remnant will show rotation (Bournaud et al. 2005). Whereas in a merger where dissipationless processes dominate, the remnant will show little or no rotation (Naab $\&$ Burkert 2003; Cox et al. 2006). In this study, clear rotation above 100 km s$^{-1}$ was found for NGC6034 and NGC7768, while most BCGs showed little or no rotation. This kinematical differentiation (the existence of slow and fast rotators) in early-type galaxies is also clearly visible in the SAURON data presented by Emsellem et al.\ (2006). 

The amount of flattening that is expected due to rotation in a galaxy depends on the balance between ordered and random motions, and this can be quantified using the anisotropy parameter. The rotation of elliptical galaxies is conventionally expressed as the anisotropy parameter, defined as $(V_{\rm max}/\sigma_{0})^{\ast} =$ ($V_{\rm max}/\sigma_{0}$)/$\sqrt{\epsilon/1-\epsilon}$\ \ (Kormendy 1982), where the rotational velocity $V_{\rm max}$ is measured as in Section \ref{Sec:kinematics}, and the central velocity dispersion $\sigma_{0}$ is taken as the measurement for the central velocity dispersion in Table \ref{table:GalVel}. A value of $(V_{\rm max}/\sigma_{0})^{\ast} \approx 1$ would be expected if a galaxy is flattened by rotation. The anisotropy parameter $(V_{\rm max}/\sigma_{0})^{\ast}$ can be used to separate galaxies that are rotationally supported from those that are supported by $\sigma$ anisotropy where the value is substantially less than unity. The division occurs at about $(V_{\rm max}/\sigma_{0})^{\ast} = 0.7$ (Bender, Burstein $\&$ Faber 1992). 

\begin{figure}
   \centering
   \includegraphics[scale=0.32]{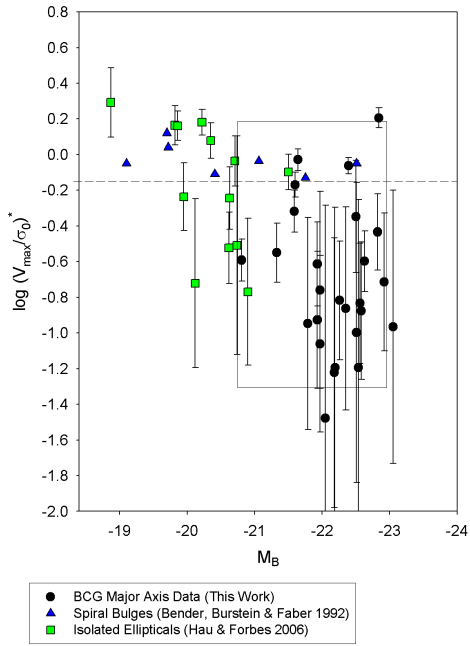}
   \caption[The Anisotropy-luminosity Diagram]{The anisotropy-luminosity diagram. The horizontal dashed line separates the rotationally supported galaxies from the anisotropic galaxies as described in the text. Only the BCGs for which major axis spectra were taken (within 10 degrees) are plotted. The box plotted in the figure outlines the region containing data on giant ellipticals from Bender et al.\ (1992).}
   \label{fig:Anisotropy}
\end{figure}

Figure \ref{fig:Anisotropy} shows the anisotropy parameter as a function of galaxy $B$-band luminosity, where the above mentioned division is indicated by the horizontal dashed line. Note that the errors indicated are only propagated from the errors on the velocity measurements taken to be the extreme radial velocity points, and the errors on the central velocity dispersion. They do not take into account the general uncertainties involved in determining the most extreme velocity measurements. Therefore, care has to be taken when interpreting individual points on the diagram. The only notable BCG data point that lies significantly above the dashed line is for the galaxy PGC026269, which possesses moderate rotation (51 km s$^{-1}$) and surprisingly low central velocity dispersion (222 km s$^{-1}$). However, the ellipticity of this galaxy is zero and it is not rotationally supported.

Another factor that complicates the dynamical interpretation of individual points is that the observed ellipticity is a global property of the galaxy, whereas the kinematic measurements taken here only reflect the kinematics along the axis where the slit was placed, and only close to the centre of the galaxy. For example, a disk component may dominate the measured kinematics but will have little effect on the ellipticity, making the galaxy appear to rotate faster than its global ellipticity would suggest (Merrifield 2004). 

For comparison, the sample of isolated ellipticals from Hau $\&$ Forbes (2006) are also plotted in Figure \ref{fig:Anisotropy}. Of this sample, 11 galaxies were observed along the major axis, and the central velocity dispersions were derived from the bins closest to the galaxy cores. The BCGs show less rotational support than the isolated elliptical galaxies, as a class. Spiral bulge (typically rotationally supported) and giant elliptical data from Bender et al.\ (1992) are also plotted. Their central velocity dispersions were derived over the whole half-light radii of the galaxies. Even though large rotation velocities were found for a few individual cases, the BCGs are consistent with the general trend for very massive galaxies. 

\section{Summary}

In this paper, we have presented the largest optical spectroscopic sample of BCGs with radial information to date. We have derived the rotational velocity and velocity dispersion profiles for 41 BCGs. We find clear rotation curves for a number of galaxies for which we have obtained major axis spectra, and two galaxies with rotational velocities exceeding 100 km s$^{-1}$. However, because of the generally large central velocity dispersions, the BCG data are consistent with the trend for very massive elliptical galaxies on the anisotropy-luminosity diagram. At least 29 per cent of the galaxies show very clear velocity substructure. 

A future paper will be devoted to the stellar population properties of this galaxy sample. Despite the undeniably special nature of BCGs due to their extreme morphological properties and locations, the kinematic properties investigated here (rotation and incidence of velocity substructure) seem normal compared with their ordinary giant elliptical counterparts.

\section*{Acknowledgments}
SIL thanks the South African National Research Foundation and the University of Central Lancashire for a Stobie-SALT Scholarship. We gratefully acknowledge Bryan Miller (PI of the Gemini programs to collect the Lick star spectra) and J.J Gonzalez for useful discussions. We also thanks the anonymous referee for constructive comments which contributed to the improvement of this paper. PSB is supported by a Marie Curie Intra-European Fellowship within the 6th European Community Framework Programme.

Based on observations obtained on the WHT and Gemini North and South telescopes (Gemini program numbers GS-2006B-Q-71, GN-2006B-Q-88, GS-2007A-Q-73, GN-2007A-Q-123, GS-2007B-Q-43 and GN-2007B-Q-101). The WHT is operated on the island of La Palma by the Royal Greenwich Observatory at the Observatorio del Roque de los Muchachos of the Instituto de Astrof\'{\i}sica de Canarias. The Gemini Observatory is operated by the Association of Universities for Research in Astronomy, Inc., under cooperative agreement with the NSF on behalf of the Gemini Partnership: the National Science Foundation (USA), the Science and Technology Facilities Council (UK), the National Research Council (Canada), CONICYT (Chile), the Australian Research Council (Australia), CNPq (Brazil) and CONICET (Argentina). This research has made use of the NASA/IPAC Extragalactic Database (NED) which is operated by the Jet Propulsion Laboratory, California Institute of Technology and the HyperLeda catalogue HyperLEDA which is an extended version of the Lyon-Meudon Extragalactic Database operated by the Centre de Recherche Astronomique de Lyon.

\appendix
\section{Kinematic profiles: Comparison with literature}
\label{Appendix}

\begin{figure}
   \centering
   \includegraphics[scale=0.37]{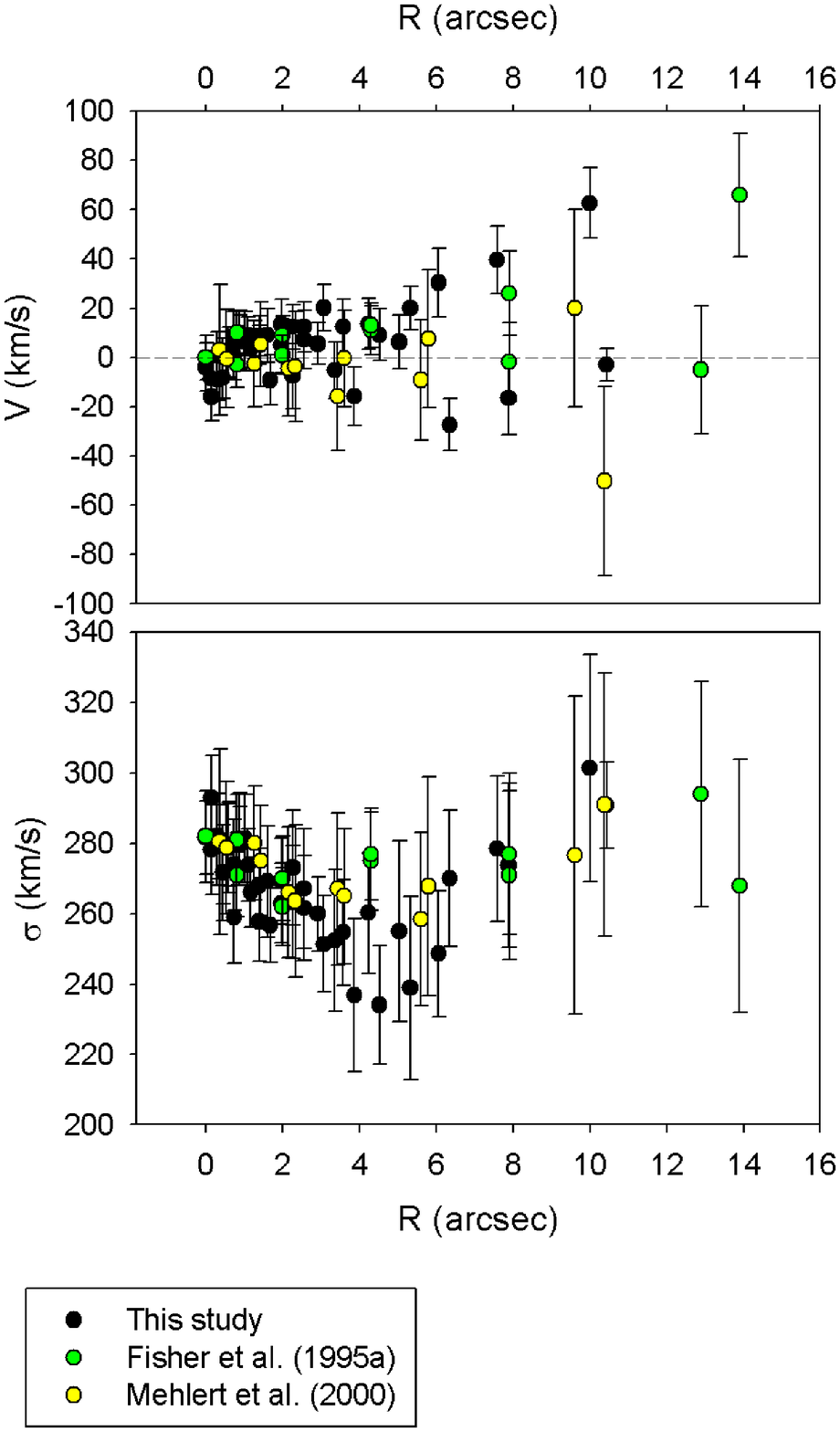}
\caption[]{NGC4839. Kinematic profile compared to previous literature. The data are folded with respect to the centre of the galaxy, and the radial velocities are given as relative to the central radial velocity.}
\label{fig:NGC4839kin}
\end{figure}

\begin{figure}
   \centering
   \includegraphics[scale=0.37]{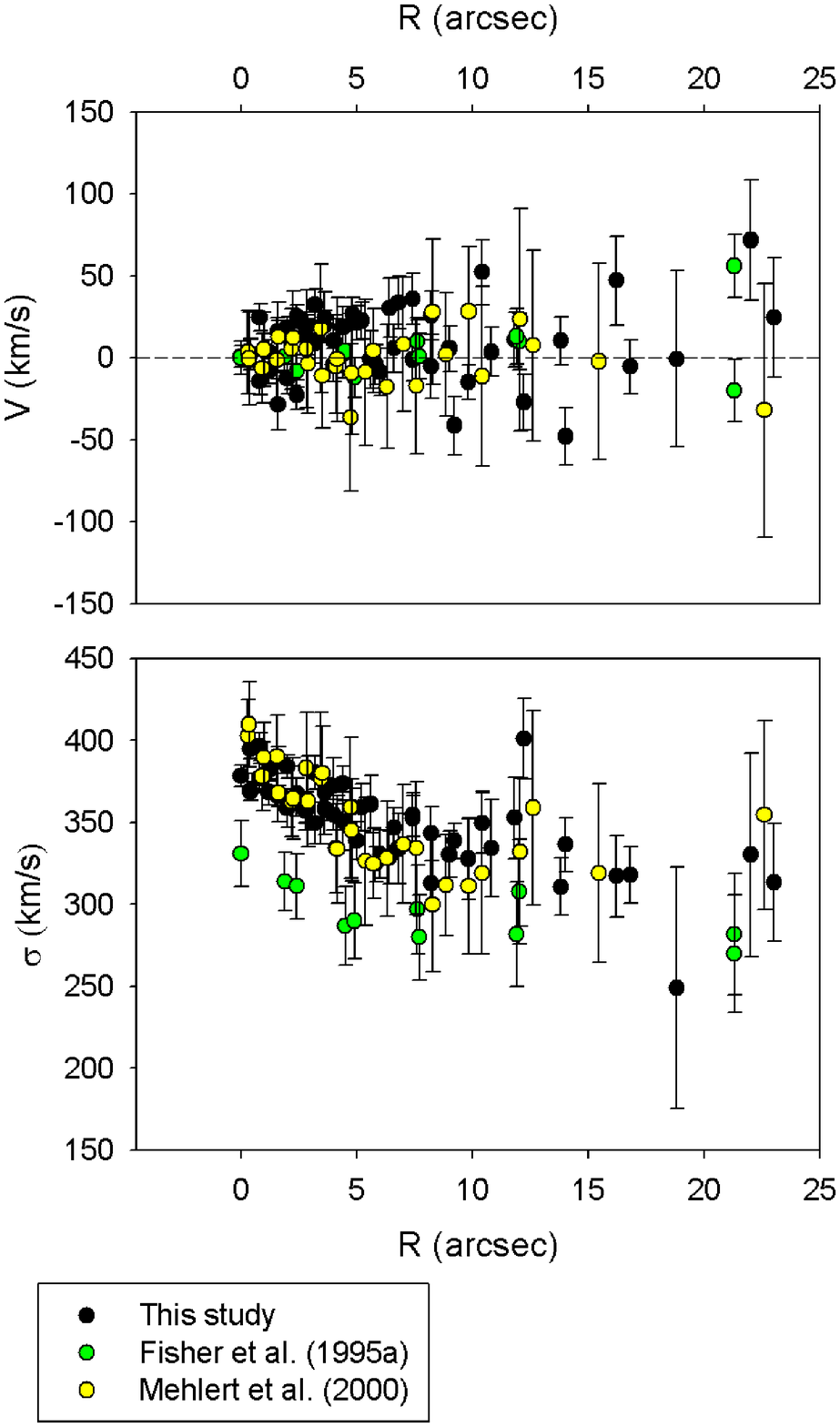}
\caption[]{NGC4889. Kinematic profile compared to previous literature.}
\label{fig:NGC4889kin}
\end{figure}

\begin{figure}
   \centering
   \includegraphics[scale=0.37]{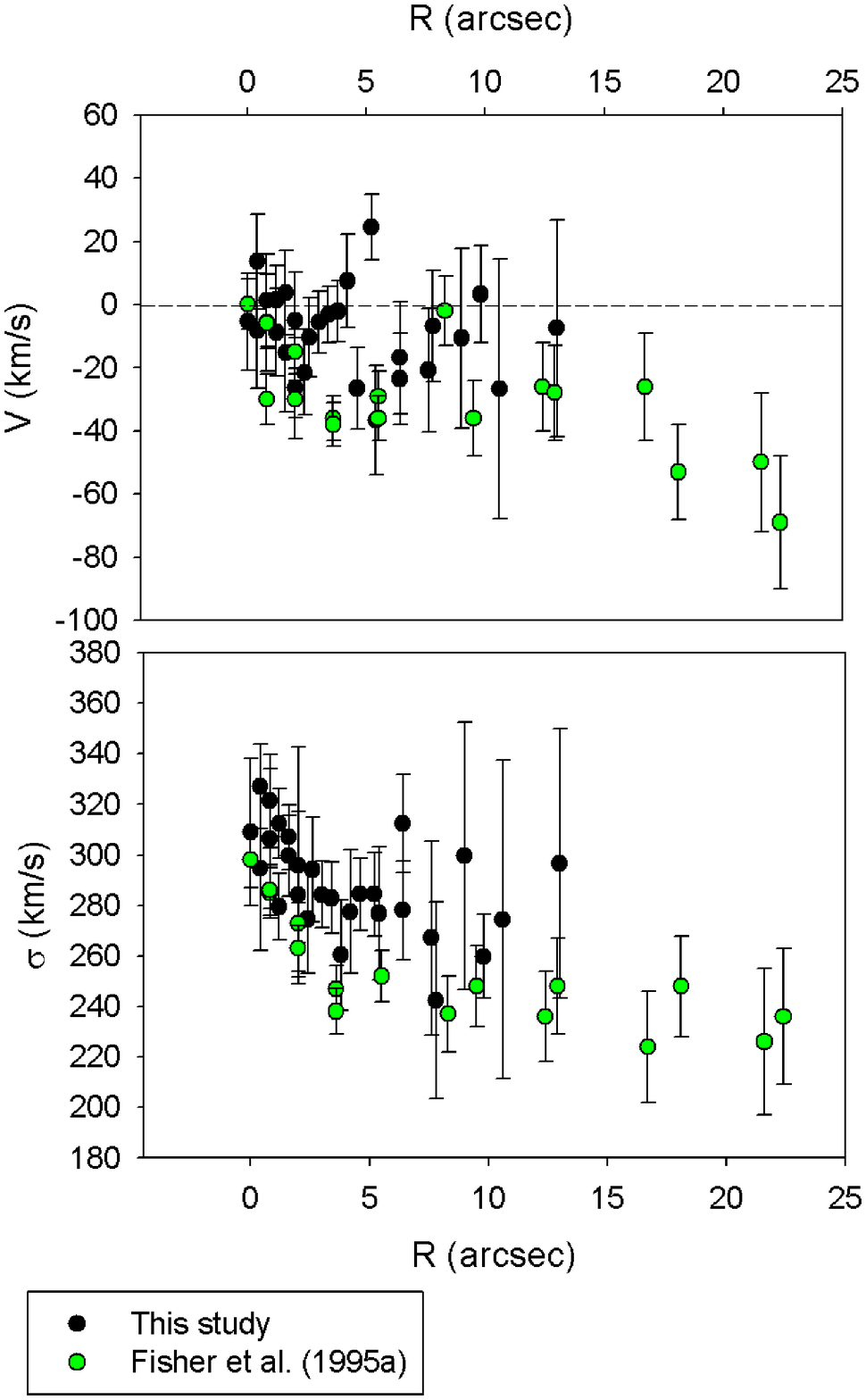}
\caption[]{NGC6166. Kinematic profile compared to previous literature.}
\label{fig:NGC6166kin}
\end{figure}

\begin{figure}
   \centering
   \includegraphics[scale=0.37]{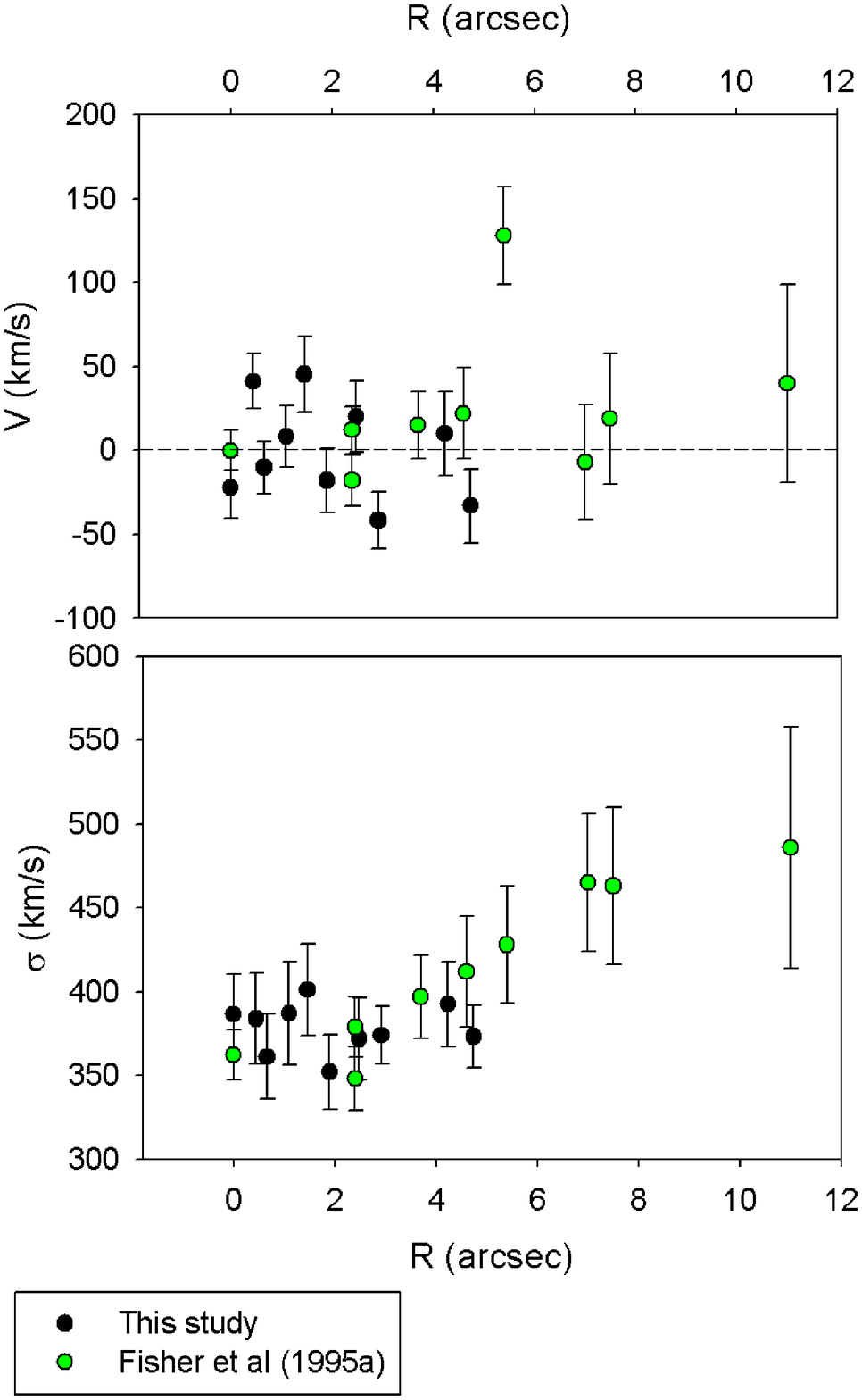}
\caption[]{IC1101. Kinematic profile compared to previous literature.}
\label{fig:IC1101kin}
\end{figure}

The kinematic radial profiles of some of the galaxies could be compared to previous measurements in the literature. Four comparisons are shown in Figures \ref{fig:NGC4839kin} to \ref{fig:IC1101kin}. Two of the Coma cluster cD galaxies (NGC4839 and NGC4889) were compared to data from Fisher et al.\ (1995a) and Mehlert et al.\ (2000). The third Coma cD (NGC4874) was not compared as it was observed at a different slit position angle than in the other studies. No systematic differences are found (within the error bars) for the measurements of this study compared to the previous studies. The velocity dispersion measurements of NGC4889 in Fisher et al.\ (1995a) are lower than the other studies. This discrepancy was already reported in Mehlert et al.\ (2000). NGC6166 was compared to Fisher et al.\ (1995a) and agree within the errors. IC1101 was compared to Fisher et al.\ (1995a), who found the velocity dispersion profile to be increasing. As can be seen from Figure \ref{fig:IC1101kin}, the velocity dispersion profile measured here does not extend out to the same radius as that from Fisher et al.\ (1995a) but the central measurements are in good agreement. 

\bsp

\label{lastpage}


\begin{thebibliography}{99}
\bibitem{Adam} Adami C., Biviano A., Durret F., Mazure A. 2005, A$\&$A, 443, 17
\bibitem{Andr} Andreon S., Davoust E., Michard R., Nieto J.L., Poulain P. 1996, A$\&$AS, 116, 429
\bibitem{Arag} Aragon-Salamanca A., Baugh C.M., Kauffmann G. 1998, MNRAS, 297, 427
\bibitem{Bai2} Baier F.W., Wipper H. 1995, AN, 316, 319
\bibitem{Baut} Bautz L.P., Morgan W.W. 1970, BAAS, 2, 294
\bibitem{Ben1} Bender R., Burstein D., Faber S.M. 1992, ApJ, 399, 462
\bibitem{Bro1} Brough S., Collins C.A., Burke D.J., Mann R.G., Lynam P.D. 2002, MNRAS, 329, 533
\bibitem{Brou} Brough S., Proctor R., Forbes D.A., Couch W.J., Collins C.A., Burke D.J., Mann R.G. 2007, MNRAS, 378, 1507
\bibitem{Bour} Bournaud F., Jog C.J., Combes F. 2005, A$\&$A, 437, 69
\bibitem{Boyl} Boylan-Kolchin M., Ma C., Quataert E. 2006, 369, 1081
\bibitem{Burs} Burstein D., Faber S.M., Gaskell C.M., Krumm N. 1984, ApJ, 287, 586
\bibitem{Car0} Cardiel N. 1999, Ph.D. thesis, Universidad Complutense de Madrid
\bibitem{Card1} Cardiel N., Gorgas J., Aragon-Salamanca A. 1998, MNRAS, 298, 977
\bibitem{Cart} Carter D., Metcalfe N. 1980, MNRAS, 191, 325
\bibitem{Car2} Carter D., Bridges T.J., Hau G.K.T. 1999, MNRAS, 307, 131
\bibitem{Coll} Collins C.A., Mann R.G. 1998, MNRAS, 297, 128
\bibitem{Cowi} Cowie L.L., Binney J. 1977, ApJ, 215, 723
\bibitem{Cox} Cox T.J., Dutta S.N., Di Matteo T., Hernquist L., Hopkins P.F., Robertson B., Springel V. 2006, ApJ, 650, 791
\bibitem{DeLu} De Lucia G., Blaizot J. 2007, MNRAS, 375, 2
\bibitem{DeL2} De Lucia G., Springer V., White S.D.M., Croton D., Kauffmann G. 2006, MNRAS, 366, 499
\bibitem{DeVo} De Vaucouleurs G. 1948, JO, 31, 113
\bibitem{Dre2} Dressler A. 1984, ApJ, 281, 512
\bibitem{Edwa} Edwards L.O.V., Hudson M.J., Balogh M.L., Smith R.J. 2007, MNRAS, 379, 100
\bibitem{Emse} Emsellem E., et al. 2007, MNRAS, 379, 401
\bibitem{Fabe} Faber S.M., Friel E.D., Burstein D., Gaskell C.M. 1985, ApJS, 57, 711
\bibitem{Fish} Fisher D., Illingworth G., Franx M. 1995a, ApJ, 438, 539
\bibitem{Fis2} Fisher D., Franx M., Illingworth G. 1995b, ApJ, 448, 119
\bibitem{Gall} Gallagher J.S., Ostriker J.P. 1972, AJ, 77, 288
\bibitem{Giac} Giacintucci S., Venturi T., Murgia M., Dallacasa D., Athreya R., Bardelli S., Mazzotta P., Saikia D.J. 2007, A$\&$A, 476, 99
\bibitem{Gonz} Gonz\'alez J.J. 1993, PhD thesis, Univ. California
\bibitem{Gorg} Gorgas J., Efstathiou G., Aragon-Salamanca A. 1990, MNRAS, 245, 217
\bibitem{Hau1} Hau G.K.T., Forbes D.A. 2006, MNRAS, 371, 633
\bibitem{Hill} Hill J.M., Oegerle W.R. 1992, AAS, 18, 1413
\bibitem{Hoes} Hoessel J.G., Gunn J.E., Thuan T.X. 1980, ApJ, 241, 486
\bibitem{Jarr} Jarrett T.H., Chester T., Cutri R., Schneider S.E., Huchra J.P. 2003, AJ, 125, 525
\bibitem{Jord} Jord\'an A., C\^ot\'e P., West M.J., Marzke R.O., Minniti D., Rejkuba M. 2004, AJ, 127, 24
\bibitem{Korm} Kormendy J. 1982, Morphology and Dynamics of Galaxies, p.\ 115, eds. Martinet L., Major M., Geneva Observatory
\bibitem{Krop} Koprolin W., Zeilinger W. 2000, A$\&$A, 145, 71
\bibitem{Lain} Laine S., van der Marel R.P., Lauer T.R., Postman M., O'Dea C.P., Owen F.N. 2003, AJ, 125, 478
\bibitem{Laue} Lauer T.R. 1988, ApJ, 325, 49
\bibitem{Liu} Liu F.S., Xia X.Y., Mao S., Wu H., Deng Z.G. 2008, MNRAS, 385, 23
\bibitem{Long} Longo G., Zaggio S.R., Brusarello G., Richter G. 1994, A$\&$AS, 105, 433
\bibitem{Malu} Malumuth E.M., Kirshner R.P. 1985, ApJ, 291, 8
\bibitem{Matt} Matthews T.A., Morgan W.W., Schmidt M. 1964, ApJ, 140, 35
\bibitem{MacN} McNamara B.R., O'Connell R.W. 1992, ApJ, 393, 579
\bibitem{Mac2} McNamara B.R., Wise M., Sarazin C.L., Jannuzi B.T., Elston R. 1996, ApJL, 466, 9
\bibitem{Mehl} Mehlert D., Saglia R.P., Bender R., Wegner G. 2000, A$\&$AS, 141, 449
\bibitem{Meri} Merrifield, M. 2004, MNRAS, 353, L13
\bibitem{Merr} Merritt D. 1983, ApJ, 264, 24
\bibitem{Naab} Naab T., Burkert A. 2003, ApJ, 597, 893
\bibitem{Neum} Neumann D.M. et al. 2001, A$\&$A, 365, 74
\bibitem{Oege} Oegerle, W.R., Hill, J.M. 2001, AJ, 122, 2858
\bibitem{Oeml} Oemler A. 1976, ApJ, 209, 693
\bibitem{Ostr} Ostriker J.P., Hausman M.A. 1977, ApJ, 217, 125
\bibitem{Ost0} Ostriker J.P., Tremaine S.D. 1975, ApJ, 202, 113
\bibitem{Pate} Patel P., Maddox S., Pearce F.R., Arag\'on-Salamanca A., Conway E. 2006, MNRAS, 370, 851
\bibitem{Pele} Peletier R.F., Davies R.L., Illingworth G.D., Davis L.E., Cawson M. 1990, AJ, 100, 1091
\bibitem{Post} Postman M., Lauer T.R. 1995, ApJ, 440, 28
\bibitem{Pro4} Proctor R.N., Forbes D.A., Forestell A., Gebhardt K. 2005, MNRAS, 362, 857
\bibitem{Prug} Prugniel Ph., Simien F. 1996, A$\&$A, 309, 749
\bibitem{Rine} Rines K., Geller M.J., Diaferio A., Mahdavi A., Mohr J.J., Wegner G. 2002, AJ, 124, 1266
\bibitem{Rome} Romeo A.D., Napolitano N.R., Covone G., Sommer-Larsen J., Antonuccio-Delogu V., Capacciolo M. 2008, arXiv:0804.1517
\bibitem{Rood} Rood H.J., Sastry G.N. 1971, PASP, 83, 313
\bibitem{Salo} Salom\'{e} P., Combes F. 2003, A$\&$A, 412, 657 
\bibitem{San1} S\'{a}nchez-Bl\'{a}zquez P., Gorgas J., Cardiel N., Gonz\'{a}lez J.J. 2006, A$\&$A, 457, 787 
\bibitem{Sara} Sarazin C.L. 1988, ``X-ray Emission From Clusters of Galaxies'', Cambridge University Press
\bibitem{Sarg} Sargent W., Schechter P., Boksenberg A., Shortridge K. 1977, ApJ, 212, 326
\bibitem{Scho} Schombert J.M. 1986, ApJS, 60, 603
\bibitem{Sch2} Schombert J.M. 1987, ApJS, 64, 643
\bibitem{Sch3} Schombert J.M. 1988, ApJ, 328, 475
\bibitem{Seig} Seigar M.S., Graham A.W., Jerjen, H. 2007, MNRAS, 378, 1575
\bibitem{Stat} Statler T.S. 1991, AJ, 102, 882
\bibitem{Stru} Struble M.F., Rood H.J. 1987, ApJS, 63, 555
\bibitem{Tonr} Tonry J.L. 1984, ApJ, 279, 13
\bibitem{Ton2} Tonry J.L. 1985, AJ, 90, 2431
\bibitem{Torl} Torlina L., De Propris R., West M.J. 2007, ApJ, 660L, 97
\bibitem{VanM} Van der Marel R.P., Franx M. 1993, ApJ, 407, 525
\bibitem{Vito} Vitores A.G., Zamorano J., Rego M., Alonso O., Gallego J. 1996, A$\&$AS, 118, 7
\bibitem{vond} Von der Linden A., Best P.N., Kauffmann G., White S.D.M. 2007, MNRAS, 379, 867
\bibitem{West} West M.J. 1989, ApJ, 344, 535
\bibitem{Whil} Whiley I.M. et al.\ 2008, arXiv:0804.2152
\bibitem{Wort} Worthey G., Faber S.M., Gonz\'alez J.J., Burstein D. 1994, ApJS, 94, 687
\bibitem{Wor3} Worthey G., Ottaviani D.L. 1997, ApJS, 111, 377
\bibitem{Yama} Yamada T. et al.\ 2002, ApJ, 577, 89
\end{thebibliography}
\end{document}